\documentclass[10pt,onecolumn,twoside]{article}

\usepackage[margin = 1.0in]{geometry}
\usepackage{mathtools}

\usepackage{amsmath}

\usepackage{graphicx,epsfig,color}
\usepackage{amsfonts}
\usepackage{subfigure}
\usepackage{bm}
\usepackage{amssymb}
\usepackage{amsthm}
\usepackage{epstopdf}
\usepackage{mathtools}  
\usepackage{tabulary}
\usepackage{booktabs}
\usepackage{cite}

\DeclareSymbolFont{bbold}{U}{bbold}{m}{n}
\DeclareSymbolFontAlphabet{\mathbbold}{bbold}
\newcommand{\vect}[1]{\mathbbold{#1}}
\newcommand{\vectorones}[1][]{\vect{1}_{#1}}
\newcommand{\vectorzeros}[1][]{\vect{0}_{#1}}

\def\dist{\mathcal{D}}
\def\diag{\operatorname{diag}}

\providecommand{\keywords}[1]{\textbf{\textit{Index terms:}} #1}

\newcommand{\until}[1]{\{1,\dots, #1\}}

\newcommand{\subscr}[2]{#1_{\textup{#2}}}

\newcommand{\argmin}{\operatorname{argmin}}
\newcommand{\argmax}{\operatorname{argmax}}

\renewcommand{\baselinestretch}{1.2}

\renewcommand{\theenumi}{\roman{enumi}}

\newcommand{\intDeltan}{\text{int}(\Delta_n)}

\newcommand{\vleft}{\subscr{\bm{v}}{left}}

\newtheorem{theorem}{Theorem}
\newtheorem{lemma}[theorem]{Lemma}
\newtheorem{remark}[theorem]{Remark}

\newtheorem{assumption}{Assumption}

\newtheorem{conjecture}[theorem]{Conjecture}

\renewcommand{\theenumi}{(\roman{enumi}}

\title{Dynamic Models of Appraisal Networks Explaining\\ Collective Learning\thanks{This material is based upon work supported by, or in part by, the U.~S.~Army Research Laboratory and the U.~S.~Army Research Office under grant numbers W911NF-15-1-0577 and W911NF-16-1-0005. The content of the information does not necessarily reflect the position or the policy of the Government, and no official endorsement should be inferred. A preliminary version~\cite{WM-NEF-KL-FB:15h} has been accepted to the 53rd IEEE Conference on Decision and Control. Compared with~\cite{WM-NEF-KL-FB:15h}, this paper proposes more generalized models, discusses numerous model variations, and provides all the mathematical proofs absent in~\cite{WM-NEF-KL-FB:15h}.}}
    
\author{\qquad Wenjun Mei \qquad Noah E. Friedkin \qquad Kyle Lewis \qquad Francesco Bullo \thanks{Wenjun Mei and
    Francesco Bullo are with the Department of Mechanical Engineering and
    with Center for Control, Dynamical Systems, and Computation, University
    of California, Santa Barbara, Santa Barbara, CA 93106, USA,
    \texttt{meiwenjunbd@gmail.com}, \texttt{bullo@engineering.ucsb.edu} Noah E. Friedkin is with the Department of Sociology and with Center for Control, Dynamical Systems, and Computation, University
    of California, Santa Barbara, Santa Barbara, CA 93106, USA, \texttt{ friedkin@soc.ucsb.edu} Kyle Lewis is with the Technology Management Program, University
    of California, Santa Barbara, Santa Barbara, CA 93106, USA, \texttt{klewis@tmp.ucsb.edu} }}

\begin{document}
\maketitle 
\begin{abstract}
  This paper proposes models of learning process in teams of
  individuals who collectively execute a sequence of tasks and whose
  actions are determined by individual skill levels and networks of
  interpersonal appraisals and influence. The closely-related proposed
  models have increasing complexity, starting with a centralized
  manager-based assignment and learning model, and finishing with a
  social model of interpersonal appraisal, assignments, learning, and
  influences. We show how rational optimal behavior arises along the
  task sequence for each model, and discuss conditions of
  suboptimality. Our models are grounded in replicator dynamics from
  evolutionary games, influence networks from mathematical sociology,
  and transactive memory systems from organization science.
\end{abstract}

 \keywords{collective learning, transactive memory systems, appraisal networks, influence networks, evolutionary games, replicator dynamics, multi-agent systems}
 
\section{Introduction}
\subsection{Motivation and problem description}
Researchers in sociology, psychology, and organization science have
long studied the inner functioning and performance of teams with
multiple individuals engaged in tasks. Extensive qualitative studies,
conceptual models and empirical studies in the laboratory and field
reveal some statistical features and various phenomena of
teams~\cite{DWL-RM-LA:95,KL:03,SW-BFJ-BU:07,AWW-CFC-AP-NH-TWM:10,WM-DJW:11,JMF-JLQ:11},
but only a few quantitative and mathematical models are
available~\cite{JGM:91,DL-AF:07,EGA-KL:13}. In this paper we build
mathematical models for the dynamics of team structure and performance.
Our work is based on the core idea that all the exhibited phenomena
and features of teams must result from some essential elements, that
is, individual (member) attributes and the team's inner structure. 
We aim to mathematically characterize these essential
elements of teams, investigate how they are related to team
performance and, most importantly, how they evolve with time.

%We aim to mathematically characterize these essential elements of teams, investigate how they are related to team performance and most importantly, how they evolve with time.

We consider a team of individuals with unknown skill levels who
complete a sequence of tasks. The team's basic inner structure
is characterized by the appraisal network. The appraisal network 
determines how each task is assigned, and is updated via performance 
feedback and its co-evolution with the team's influence network. 
We aim to build multi-agent dynamical models in which (i) the team 
as an entirety eventually achieves the optimal assignment of tasks to members; (ii)
 each individual's true relative skill level is asymptotically 
 learned by the team members, which is referred to as collective 
learning. We then investigate the model variations for what impairs
 collective learning.

%the 
%appraisal/influence networks evolve with time; (ii) as a result of the evolution, the team as an entirety 
%eventually learns the members' true skill levels, which is referred to
%as the collective learning process; (iii) the learning process leads 
%the team to a rational state in which the optimal task assignment is
%achieved. We then investigate variations of the baseline model for 
%what can go wrong and thus impair team learning.

\subsection{Literature review} 

Our work is deeply connected with a conceptual model of team learning and performance, a transactive memory system (TMS). A TMS is characterized by individuals' skills and knowledge, combined with members' collective understanding of which members possess what knowledge~\cite{DMW-TG-PTH:85}. When members share an understanding of who knows what on the team, tasks can be assigned to members most likely to possess the appropriate skills. As members observe the task performances of other members, their understanding of "who knows what" tends to become more accurate and more similar, leading to greater coordination and integration of members' knowledge. Empirical research across a range of team types and settings demonstrates a strong positive relationship between the development of a team TMS and team performance~\cite{DWL-RM-LA:95,KL:04,YCY-IC-KE:10}.

%Our work is deeply connected with a conceptual models of teams, the \emph{transactive memory system} (TMS), which generates a shared division of cognitive labor with respect to the encoding, storage, retrieval and communication of information~\cite{DMW-TG-PTH:85}. The TMS of a team is linked with the individual and team performance~\cite{DWL-RM-LA:95,KL:04,YCY-IC-KE:10}. Lewis~\cite{KL:03} %and Lewis et. al.~\cite{KL-BH:11}  describe the  behavioral indicators that a TMS is operating in a team: the degree to which members specialize in complementary yet distinct aspects of  the team's tasks, the extent that members rely on  the expertise of other members, and evidence of coordinated interdependent activity.

%Our work is deeply connected with a conceptual model of teams, the
%\emph{transactive memory system} (TMS) theory. A TMS is the shared
%division of cognitive labor with respect to the encoding, storage,
%retrieval and communication of information from different domains.
%Initially conceptualized by Wegner et. al.~\cite{DMW-TG-PTH:85}, TMS
%theory links the collective performance of teams with the group and
%individual cognition and has been studied in various types of
%teams~\cite{DWL-RM-LA:95,KL:04,KASJ-KK-JACB-ES:09,YCY-IC-KE:10}.

 % maybe clarify appraisal of what: expertise and responsibilities?
 
In our models, collective learning arises as the result of the co-evolution
of interpersonal appraisals and influence networks. Related
previous work includes social comparison theory~\cite{LF:54},
averaging-based social learning~\cite{AJ-AS-ATS:10}, opinion dynamics
on influence
networks~\cite{MHDG:74,NEF-ECJ:03,JL-DAL:10,AM-FB:11f}, reflected
appraisal mechanisms~\cite{NEF:11,PJ-AM-NEF-FB:13d,XC-JL-ZX-TB:15},
dynamic balance
theory~\cite{SAM-JK-RDK-SHS:11,VAT-PVD-PDL:13}, and the
combined evolution of interpersonal appraisals and influence
networks~\cite{PJ-NEF-FB:13n}.

In the modeling and analysis of the evolution of appraisal and
influence networks, we also build an insightful connection between our
model and the well-known replicator dynamics studied in evolutionary
game theory; see the textbook~\cite{WHS:10}, some control
applications~\cite{JRM-HPY-GA-JSS:09,MJF-BT-JSS:15}, and the
recent contributions~\cite{RC-YT:14,DM-CM:15}.
%% \wmmargin{build a fruitful connection... and the replicator dyn.}
%% \fbmargin{this seems ok to me}

%% for the journal version, to understand if the analysis of the time-varying replicator model is novel

\subsection{Contribution}
Firstly, based on a few natural assumptions, we propose three novel
models with increasing complexity for the dynamics of teams: the
manager dynamics, the assign/appraise dynamics, and the
assign/appraise/influence dynamics. Our work integrates three
well-established types of dynamics: the replicator dynamics, the dynamics of
appraisal networks, and the opinion dynamics on influence networks. To the
best of our knowledge, this is the first time that such an integration
has been proposed.  Our models provide an insightful perspective on
the connection between team performance and the interpersonal appraisal networks.  In
our models, the performances of a team of individuals, with fixed skill
levels, are determined by how a task is assigned. For the baseline
manager dynamics, task assignment is adjusted by an
outside authority, according to the replicator dynamics, with individuals' performances as the feedback signals. The assign/appraise dynamics elaborates the baseline model by assuming
that, instead of an outside authority, the team members' interpersonal appraisals as the basic inner structure determine the task assignments. In the assign/appraise/influence dynamics model, we further
elaborate the model by considering the co-evolution
of appraisal and influence networks.

Secondly, theoretical analysis is presented on the dynamical properties of the models we propose. We prove that, for the assign/appraise and the assign/appraise/influence dynamics, task assignments determined by the interpersonal appraisals satisfy the replicator dynamics in a generalized form. Moreover, results on the models' asymptotic behavior relate collective learning with the connectivity property of the observation network, which defines the heterogeneous feedback signals each individual observes. We find that, for the assign/appraise dynamics with the initial appraisal network that is strongly connected and ha a self loop for each node, the team achieves rational task assignment, if the observation network is strongly connected. For the assign/appraise/influence dynamics with strongly connected and aperiodic initial appraisal network, the team achieves collective learning, if the observation network has a globally reachable node. Our theoretical results on the asymptotic behavior can be interpreted as the exploration of the most relaxed condition for asymptotic optimal task assignment. In addition, the assign/appraise/influence dynamics describes an emergence process by which team members' perception of ``who knows what'' become more similar over time, a fundamental feature of TMS~\cite{ETP:05,JYL-DGB-KL:14}.

%In addition, theoretical results on assign/appraise/influence dynamics deepens the connection between our work and the TMS theory. As Palazzolo~\cite{ETP:05} and Lee et al.~\cite{JYL-DGB-KL:14} point out, transitive triads are positively correlated with mature TMS. Along assign/appraise/influence dynamics, appraisal conflicts are resolved and the team performance is asymptotically optimized, which leads to an appraisal network in which all the triads are transitive (for an appropriate notion of transitivity).    

Thirdly, besides the models in which the team eventually learns the individuals' true relative skill levels, we propose one variation in each of the three phases of the assign/appraise/influence dynamics: the assignment rule, the update of appraisal network based on feedback signal, and the opinion dynamics for the interpersonal appraisals. The variations reflect some sociological and psychological mechanisms known to prevent the team from learning. We investigate by simulation numerous possible causes of failure to learn.
%% \fbmargin{ask Kyle for references on these known mechanisms}

%\textcolor{blue}{Finally, we remark that our models have a natural connection with the
%TMS theory.  One important aspect of TMS is members’ shared
%(collective) understanding about who possesses what
%expertise. For the case of a one-dimension skill, this collective
%knowledge is approximately characterized by the appraisal network and
%the evolution of the interpersonal appraisals can thus be interpreted
%as the development of a TMS.  Moreover, according to the work by
%Palazzolo~\cite{ETP:05} and Lee et al.~\cite{JYL-DGB-KL:14},
%transitive triads are positively correlated with mature TMS.
%% In other words, internal cognitive consistence is good for a
%% team.\fbmargin{talk with Kyle about this sentence} Our third model
%% with opinion dynamics captures this feature\fbmargin{which feature?
%%   likely rewrite here} of TMS.
%
%Similarly, in our assign/appraise/influence model, individuals
%exchange their opinions to resolve their appraisal conflicts, leading
%eventually to an appraisal network consisting of consensus appraisal
%values, transitive triads (for an appropriate notion of transitivity)
%and thus high performance.
%}

\subsection{Organization}

The rest of this paper is organized as follows: the next subsection introduces
some preliminaries on evolutionary games and replicator dynamics; Section II proposes our problem set-up and centralized manager model; Section III introduces the assign/appraise dynamics; Section IV is the assign/appraise/influence model; Section V discusses some causes of failure to learn; Section VI provides some further discussions and conclusion. 

\subsection{Preliminaries}
\emph{Evolutionary games} apply game theory to evolving populations adopting different strategies. Consider a game with finite pure strategies $\bm{e}_1,\bm{e}_2,\dots,\bm{e}_n$ and mixed strategies $\bm{w}\in \intDeltan$. The expected payoff for mixed strategy $\bm{v}$ against mixed strategy $\bm{w}$ is defined as the \emph{payoff function} $\pi(\bm{v},\bm{w})=\sum_{i=1}^n v_i\pi_i(\bm{w})$, with $\pi_i(\bm{w})=\pi(\bm{e}_i,\bm{w})$ for simplicity. A strategy $\hat{\bm{w}}$ is an \emph{evolutionary stable strategy} (ESS) if any mutant strategy $\bm{v}\neq \hat{\bm{w}}$, adopted by an $\epsilon$-fraction of the population, brings less expected payoff than the majority strategy $\hat{\bm{w}}$, as long as $\epsilon$ is sufficiently small. A necessary and sufficient condition for a local ESS is stated as follows: there exists a neighborhood $U(\hat{\bm{w}})$ such that, for any $\bm{v}\in U(\hat{\bm{w}})\setminus \{\hat{\bm{w}}\}$, $\pi(\hat{\bm{w}},\bm{v})>\pi(\bm{v},\bm{v})$.

\emph{Replicator dynamics}, given by equation~\eqref{eq:replicator-dyn}, models the evolution of sub-population distribution $\bm{w}(t)\in \Delta_n$. Each sub-population $i$, with fraction $w_i(t)$ at time $t$, is using strategy $\bm{e}_i$ and has the growth rate proportional to its \emph{fitness}, defined as the expected payoff $\pi_i(\bm{w}(t))$.
\begin{equation}\label{eq:replicator-dyn}
\dot{w}_i = w_i \Big( \pi_i (\bm{w}) - \sum_{k=1}^n w_k\pi_k (\bm{w})  \Big). 
\end{equation} 
The time index $t$ is omitted for simplicity. %It is easy to check that $\bm{w}(t)\in \Delta_n$ for any $t\ge 0$. 
There is a simple connection between the ESS and the replicator dynamics~\cite{WHS:10}: if the payoff function $\pi(\bm{v},\bm{w})$ is linear to $\bm{w}$, then an ESS is a globally asymptotically stable equilibrium for system~\eqref{eq:replicator-dyn}; if $\pi(\bm{v},\bm{w})$ is nonlinear to $\bm{w}$, then the ESS is locally asymptotically stable. 

\section{Problem Set-up and Manager Dynamics}
In this section we introduce some basic formulations and a baseline centralized system on the evolution of a team. Frequently used notations are listed in Table~\ref{table:notations}.
\subsection{Model assumptions and notations}
\emph{a) Team, tasks and assignments: } The basic assumption on the individuals and the tasks are given below.
\begin{assumption}[Team, task type and assignment]\label{asmp-team-task-assign}
Consider a team of $n$ individuals characterized by a fixed but unknown vector $\bm{x}=(x_1,\dots,x_n)^{\top}$ satisfying $\bm{x}\succ \vectorzeros[n]$ and $\bm{x}^{\top}\vectorones[n]=1$, where each $x_i$ denotes the \emph{skill level} of individual $i$. %The individual skill levels are fixed but unknown to the team. 
The tasks being completed by the team are assumed to have the following properties:
\begin{enumerate}
\item The total workload of each task is characterized by a positive scalar and is fixed as $1$ in this paper;
\item The task can be arbitrarily decomposed into $n$ sub-tasks according to the \emph{task assignment}\\ $\bm{w}=(w_1,\dots,w_n)^{\top}$, where each $w_i$ is the sub-task workload assigned to individual $i$. The task assignment satisfies $\bm{w}\succ \vectorzeros[n]$ and $\bm{w}^{\top}\vectorones[n]=1$. The sub-tasks are executed simultaneously.
\end{enumerate}
\end{assumption}
\smallskip
The selection of scalar values for skill levels and task assignments
can be simply interpreted as the assumption that the type of tasks
considered in this paper only requires some one-dimension
skill. Alternatively, the skill levels can be
considered more generally as the individuals' overall abilities of contributing to the
completion of tasks, and the task assignments are the individuals'
relative responsibilities to the team.
%There are two ways of interpreting the scalar settings of the task workload and the individual skill level.
%The task workload and the individual skill levels are characterized by scalars. There are two ways of interpreting this set-up. 
%The first is to assume that the tasks only require some one-dimension skill. The second way is to think of the workload as some abstract measure of the general difficulty, and the skill levels as the individuals' overall abilities of contributing to the completion of the task.  

\emph{b) Individual performance: } With fixed skill levels $\bm{x}$, each individual $i$'s performance is assumed to depend only on the task assignment $\bm{w}$, denoted by $p_i(\bm{w})$. The measure of individual performance is defined below.
\begin{assumption}[Individual performance]\label{asmp:indiv-perf}
Given fixed skill levels, each individual $i$'s performance, with the assignment $\bm{w}$, is measured by $p_i(\bm{w})=f(x_i/w_i)$, where $f:[0,+\infty)\to [0,+\infty)$ is a concave, continuously differentiable and monotonically increasing function.
\end{assumption}
\smallskip
The function $f$ is assumed to be concave in that, it is widely adopted that the relation between the performance $f$ and individual ability $x$ obeys the power law, i.e., $f(x)\sim x^\gamma$, with $0<\gamma<1$~\cite{EGA-KL:13}. Despite the specific form $f(x_i/w_i)$ as in Assumption~\ref{asmp:indiv-perf}, the measure of individual performance can be quite general by adopting difference measures of $x_i$ and $w_i$.

\emph{c) Optimal assignment: }It is reasonable to claim that, in a well-functioning team, individuals' relative responsibilities, characterized by the task assignment in this paper, should be proportional to their actual abilities. Define the measure of the mismatch between task assignment and individual's true skill levels as $\mathcal{H}_1(\bm{w})= \sum_{i=1}^n | w_i/x_i-1 |$. For fixed $\bm{x}$, the \emph{optimal assignment} $\bm{w}^*=\bm{x}$ minimizes $\mathcal{H}_1(\bm{w})$. 

%That is, the team performance is maximized at the task assignment $\bm{w}^*=\bm{x}$. There are various cost functions or measurements of team performance taking $\bm{w}^*=\bm{x}$ as the optimal assignment. For example, the mismatch $\mathcal{H}_1(\bm{w})= \sum_{i=1}^n | w_i/x_i-1 |$, for fixed $\bm{x}$, is minimized at $\bm{w}^*=\bm{x}$, and the minimal individual performance $\mathcal{H}_2(\bm{w})=\min_i f(x_i/w_i)$ is maximized at $\bm{w}^*=\bm{x}$.  

\begin{table}[htbp]\caption{Notations frequently used in this paper}\label{table:notations}
\begin{center}
\vspace{-0.4cm}
\begin{tabular}{r p{13cm} }

\toprule
$\succ$~($\prec$ resp.) & entry-wise greater than (less than resp.).\\
$\succeq$~($\preceq$ resp.) & entry-wise no less than (no greater than resp.).\\
$\vectorones[n]$ ($\vectorzeros[n]$ resp. ) & $n$-dimension column vector with all entries equal to $1$ ($0$ resp.) \\  
$\bm{x}$ & vector of individual skill levels, with $\bm{x}=(x_1,x_2,\dots,x_n)^{\top}\succ \vectorzeros[n]$ and $\bm{x}^{\top}\vectorones[n]=1$. \\
$\bm{w}$ & task assignment. $\bm{w}\succ \vectorzeros[n]$ and $\bm{w}^{\top}\vectorones[n]=1$\\ 
$f$ & a concave, continuously differentiable and increasing function $f$: $[0,+\infty)\to [0,+\infty)$\\ 
$\bm{p}(\bm{w})$ & vector of individual performances. $\bm{p}(\bm{w})=\big( p_1(\bm{w}),\dots,p_n(\bm{w}) \big)^{\top}$, where $p_i(\bm{w})=f(w_i/x_i)$ is the performance of individual $i$.\\
%$\bm{\phi}(\bm{w})$ & vector of individual relative performances. $\phi_i(\bm{w})=p_i(\bm{w})-\sum_k w_k p_k(\bm{w})$.\\
$A$ & appraisal matrix. $A=(a_{ij})_{n\times n}$, where $a_{ij}$ is individual $i$'s appraisal of $j$'s skill level. \\
$W$ & influence matrix. $W=(w_{ij})_{n\times n}$, where $w_{ij}$ is the weight individual $i$ assigns to $j$'s opinion.\\
$\Delta_n$ & $n$-dimension simplex $\{\bm{y}\in \mathbb{R}_{\ge 0}^{n}\,|\,\bm{y}^{\top}\vectorones[n]=1\}$\\
$\intDeltan$ & the interior of $\Delta_n$. \\
$\vleft(A)$ & the left dominant eigenvector of the non-negative and irreducible matrix $A$, i.e., the normalized entry-wise positive left eigenvector associated with the eigenvalue equal to $A$'s spectral radius. \\
$G(B)$ & the directed and weighted graph associated with the adjacency matrix $B\in \mathbb{R}^{n\times n}$.\\
\bottomrule
\end{tabular}
\end{center}
\end{table} 

\subsection{Centralized manager dynamics}
\begin{figure*}
\begin{center}
\subfigure[manager]{\label{fig:diagram-manager} \includegraphics[width=.205\linewidth]{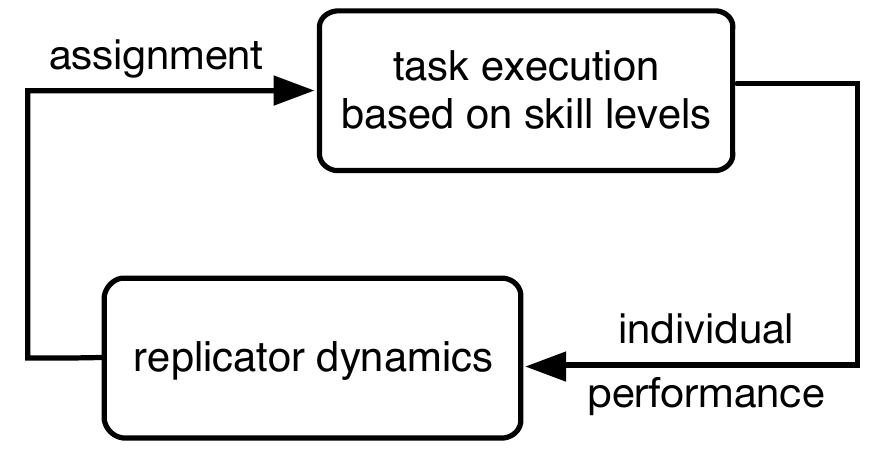}}
\subfigure[assign/appraise]{\label{fig:diagram-ass/app} \includegraphics[width=.37\linewidth]{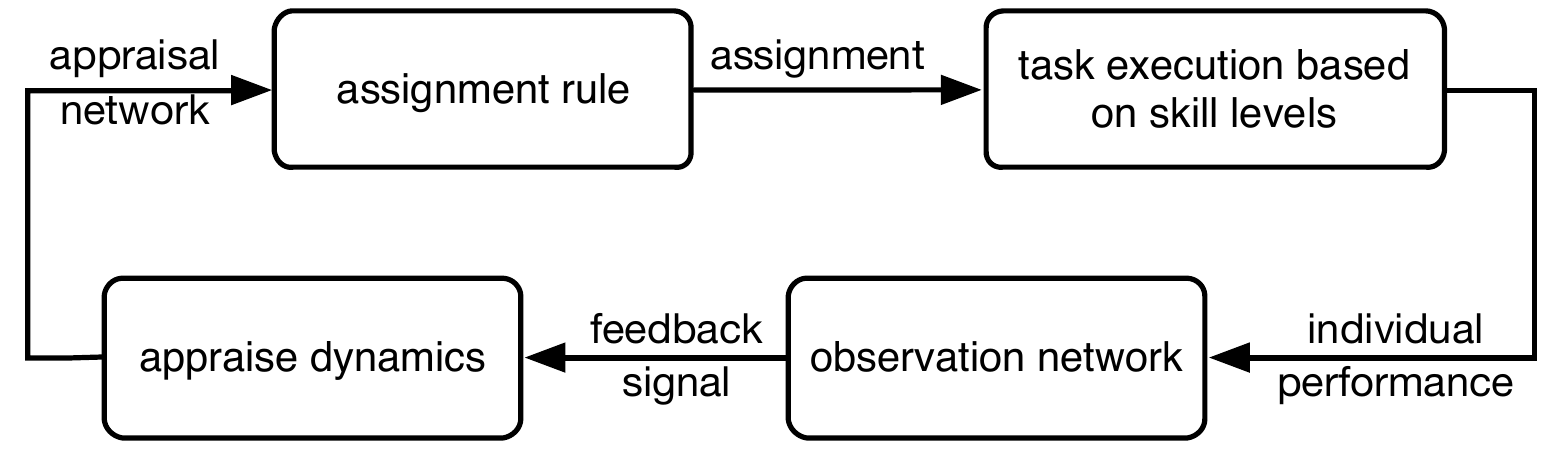}}
\subfigure[assign/appraise/influence]{\label{fig:diagram-ass/app/inf} \includegraphics[width=.37\linewidth]{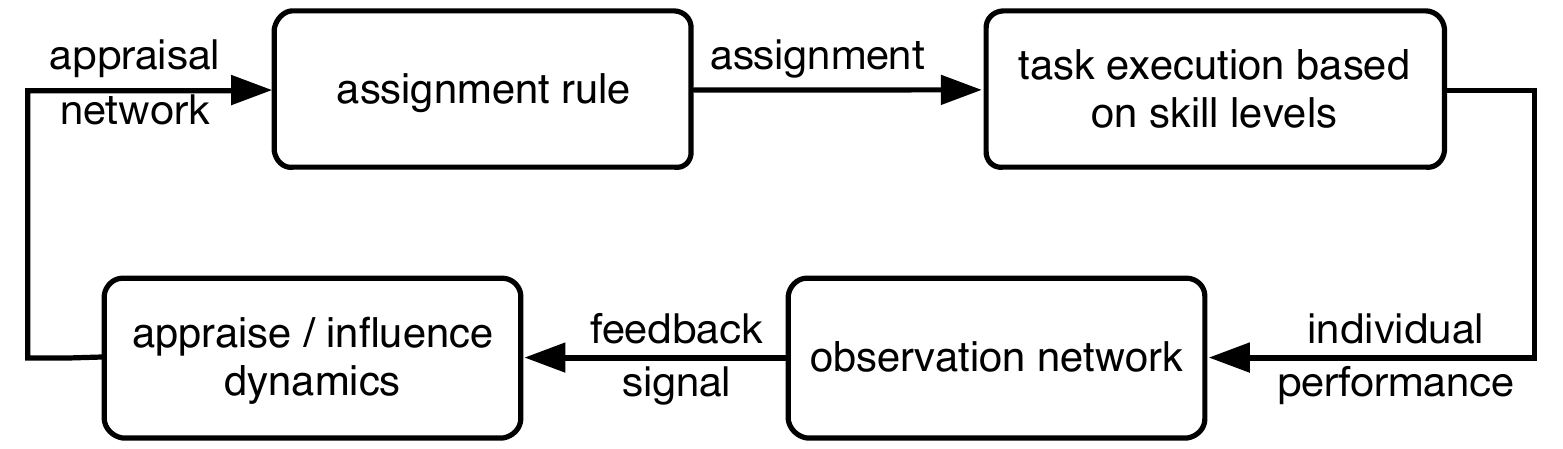}}
\caption{Diagram illustrations of manager dynamics, assign/appraise dynamics, and assign/appraise/influence dynamics.}
\label{fig:diagrams}
\end{center}
\end{figure*}

In this subsection we introduce a continuous-time centralized model on the evolution of task assignment, referred to as the \emph{manager dynamics}. The diagram illustration is given by Figure~\ref{fig:diagram-manager}. Suppose that, at each time $t$, a team of $n$ individuals is completing a task based on the assignment $\bm{w}(t)$, which is determined and adjusted along $t$ by an outside manager. The manager observes the individuals' real-time performance $\bm{p}\big( \bm{w}(t) \big)$ and adjust the task assignment $\bm{w}(t)$ according to the dynamics:
\begin{equation}\label{eq:manager-dyn}
\dot{w}_i = w_i \Big( p_i(\bm{w}) - \sum_{k=1}^n w_k p_k(\bm{w}) \Big),
\end{equation} 
for any $i\in \{1,\dots,n\}$. The following theorem states the asymptotic behavior of the manager dynamics.
\begin{theorem}[Manager dynamics]\label{thm:manager-dyn}
  Consider the manager dynamics~\eqref{eq:manager-dyn} for the
  task assignment as in Assumption~\ref{asmp-team-task-assign} with performance as in
  Assumption~\ref{asmp:indiv-perf}. Then
  \begin{enumerate}
  \item the set $\intDeltan$ is invariant;
  \item the optimal assignment $\bm{w}^*=\bm{x}$ is the ESS for the evolutionary game defined by the payoff function $\pi(\bm{v},\bm{w})=\sum_{i=1}^n v_i f(x_i/w_i)$, and is thus a locally asymptotically stable equilibrium for equation~\eqref{eq:manager-dyn} as a replicator dynamics; 
  \item for any $\bm{w}(0)\in \intDeltan$, the
    manager's assignment $\bm{w}(t)$ converges to $\bm{w}^*=\bm{x}$, as $t\rightarrow \infty$.
  \end{enumerate}
\end{theorem}
\smallskip

The proof is given in Appendix~\ref{proof:manager-dyn}. Equation~\eqref{eq:manager-dyn} takes the same form as the classic replicator dynamics~\cite{WHS:10}, with the nonlinear fitness function $\pi_i(\bm{w})=f(x_i/w_i)$. The same Lyapunov function is used in the proof for the asymptotic stability. Distinct from the classic result that the ESS, with nonlinear payoff function, can only lead to local asymptotic stability for the replicator dynamics, our model is a special case in which the ESS associated with a nonlinear payoff function is also a globally asymptotically stable equilibrium of the replicator dynamics.

\section{The Assign/Appraise Dynamics of the Appraisal Networks}
Despite the desired property on the convergence of the task assignment to optimality, the manager dynamics does not capture one of the most essential aspects of team dynamics: the evolution of the team's inner structures. In this section, we introduce a multi-agent system, in which task assignments are determined by the team members' interpersonal appraisals, rather than any outside authority, and the appraisal network is updated in a decentralized manner, driven by the heterogeneous feedback signals observed by each team member. 

% and is updated in a decentralized manner， based on individuals' heterogeneous feedback signals.  

\subsection{Model description and problem statement}
\emph{Appraisal network:} Denote by $a_{ij}$ the individual $i$'s evaluation of $j$'s skill levels and refer to $A=(a_{ij})_{n\times n}$ as the \emph{appraisal matrix}. Since the evaluations are in the relative sense, we assume $A\succeq \vectorzeros[n\times n]$ and $A\vectorones[n]=\vectorones[n]$. The directed and weighted graph $G(A)$, referred to as the \emph{appraisal network}, reflects the team's collective knowledge on the distribution of its members' abilities.

%\begin{figure*}
%\begin{center}
%\includegraphics[width=.75\linewidth]{diagram_assign_appraise}
%\caption{Diagram illustration for the assign/appraise dynamics}
%\label{fig:diagram-assign-appraise}
%\end{center}
%\end{figure*}

\emph{Assign/appraise dynamics:} We propose a multi-agent model on the evolution of the appraisal network. The model is referred to as the \emph{assign/appraise dynamics} and illustrated by the diagram in Figure~\ref{fig:diagram-ass/app}. We model three phases: the task assignment, the feedback signal and the update of the appraisal network, specified by the following three assumptions respectively.

\begin{assumption}[Assignment rule]\label{asmp:assign-rule}
At any time $t\ge 0$, the task is assigned according to the left dominant eigenvector of the appraisal matrix, i.e., $\bm{w}(t)=\vleft\big( A(t) \big)$.
\end{assumption}
\smallskip

Justification of Assumption~\ref{asmp:assign-rule} is given in Appendix~\ref{justification:assign-rule}. For now we assume $A(t)$ is row-stochastic and irreducible for all $t\ge 0$, so that $\vleft\big( A(t) \big)$ is always well-defined. This will be proved later in this section. 
%According to the Perron-Frobenius theorem, for any $A(t)$ row-stochastic and irreducible, $\bm{w}(t)$ is well-defined and satisfies $\bm{w}\succ \vectorzeros[n]$; $\bm{w}^{\top}(t)\vectorones[n]=1$; $\bm{w}^{\top}(t)A=\bm{w}^{\top}(t)$. Justification of Assumption~\ref{asmp:assign-rule} is given in Section~VII.

\begin{assumption}[Feedback signal]\label{asmp:perf-feedback}
%After executing the task assignment $\bm{w}$, each individual $i$ observes, with no noise, the difference between her own performance and the weighted average performance of some part of the whole task, in which each individual $j$ participates $m_{ij}$ fraction of it, with $m_{ij}\ge 0$ and $\sum_j m_{ij}=1$. The performance feedback signal for individual $i$ is thus given by $p_i\big(  \bm{w}(t)\big)-\sum_k m_{ik}p_k\big( \bm{w}(t) \big)$.
After executing the task assignment $\bm{w}$, each individual $i$
observes, with no noise, the difference between her own performance
and the quality of some part of the whole task, given by $\sum_k
m_{ik}p_k(\bm{w})$, in which $m_{ik}$ denotes the fraction of workload
individual $k$ contributes to this part of task. The matrix
$M=(m_{ij})_{n\times n}$ defines a directed and weighted graph $G(M)$,
referred to as the \emph{observation network}, and satisfies $M\succeq
\vectorzeros[n\times n]$ and $M\vectorones[n]=\vectorones[n]$ by
construction.
\end{assumption}
\smallskip

The topology of the observation network defines the individuals' feedback signal structure and influences the asymptotic behavior of assign/appraise dynamics. Notice that, the feedback signal for each individual $i$ is the difference $p_i\big(  \bm{w}(t)\big)-\sum_k m_{ik}p_k\big( \bm{w}(t) \big)$, while the matrix $M$ is not necessarily known to the individuals.

\begin{assumption}[Update of interpersonal appraisals]\label{asmp:update-app-net}
With the performance feedback signal defined as in Assumption~\ref{asmp:perf-feedback}, each individual $i$ increases her self appraisal and decreases the appraisals of all the other individuals, if $p_i(\bm{w})>\sum_k m_{ik}p_k(\bm{w})$, and vice versa. In addition, the appraisal matrix $A(t)$ remains row-stochastic.
\end{assumption}
\smallskip

The following dynamical system for the appraisal matrix, referred to as the \emph{appraise dynamics}, is arguably the simplest model satisfying Assumptions~\ref{asmp:perf-feedback} and~\ref{asmp:update-app-net}:
\vspace{-0.07cm}
\begin{equation}\label{eq:app-dyn}
\begin{cases}
\displaystyle\, \dot{a}_{ii}  =  a_{ii}(1  -  a_{ii})\Big( p_i(\bm{w}) -  \sum_{k=1}^n m_{ik} p_k(\bm{w}) \Big),\\
\displaystyle\, \dot{a}_{ij} =  -a_{ii}a_{ij}\Big( p_i(\bm{w}) -  \sum_{k=1}^n m_{ik} p_k(\bm{w}) \Big).
\end{cases}
\end{equation} 
The matrix form of the appraise dynamics, together with the assignment rule as in Assumption~\ref{asmp:assign-rule}, is given by
\vspace{-0.05cm}
\begin{equation}\label{eq:ass/app-dyn-matrix}
\begin{cases}
\displaystyle\, \dot{A}=\diag\!\big(\bm{p}(\bm{w})-M\bm{p}(\bm{w})\big)\subscr{A}{d} (I_n-A),\\
\displaystyle\, \bm{w}=\vleft(A),
\end{cases}
\end{equation}
and collectively referred to as the assign/appraise dynamics. Here $\subscr{A}{d}=\diag(a_{11},\dots,a_{nn})$.

\emph{Problem statement:} In Section~III.B, we investigate the asymptotic behavior of dynamics~\eqref{eq:ass/app-dyn-matrix}, including:
\begin{enumerate}
\item convergence to the optimal assignment, which means that the team as an entirety eventually learns all its members' relative skill levels, i.e., $\lim_{t\to +\infty}\bm{w}(t)=\bm{x}$;
\item \emph{appraisal consensus}, which means that the individuals asymptotically reach consensus on the appraisals of all the team members, i.e., $a_{ij}(t)-a_{kj}(t)\to 0$ as $t\to +\infty$, for any $i,j,k$.
\end{enumerate}
Collective learning is the combination of the convergence to optimal assignment and appraisal consensus.

\subsection{Dynamical behavior of the assign/appraise dynamics}
We start by establishing that the appraisal matrix $A(t)$, as the solution to equation~\eqref{eq:ass/app-dyn-matrix}, is extensible to all $t\in [0,+\infty)$ and the assignment $\bm{w}(t)$ is well-defined, in that $A(t)$ remains row-stochastic and irreducible. Moreover, some finite-time properties are investigated.

%\textcolor{blue}{We start by establishing that system~\eqref{eq:ass/app-dyn-matrix} is well-posed. By that we mean the assignment $\bm{w}(t)=\vleft\big( A(t) \big)$ is well-defined, and the solution to~\eqref{eq:ass/app-dyn-matrix}, i.e., $A(t)$, is extensible to all $t\in [0,+\infty)$. Moreover, some finite-time properties are investigated. All the relevant results are presented in the following theorem.}

\begin{theorem}[Finite-time properties of assign/appraise dynamics]\label{thm:ass/app-finite-time}
Consider the assign/appraise dynamics~\eqref{eq:ass/app-dyn-matrix}, based on Assumptions~\ref{asmp:assign-rule}-\ref{asmp:update-app-net}, describing a task assignment as in Assumption~\ref{asmp-team-task-assign}, with performance as in Assumption~\ref{asmp:indiv-perf}. For any observation network $G(M)$, and any initial appraisal matrix $A(0)$ that is row-stochastic, irreducible and has strictly positive diagonal,
\begin{enumerate}
\item The appraisal matrix $A(t)$, as the solution to~\eqref{eq:ass/app-dyn-matrix}, is extensible to all $t\in [0,+\infty)$. Moreover, $A(t)$ remains row-stochastic, irreducible and has strictly positive diagonal for all $t\ge 0$;  
%\item the assignment $\bm{w}(t)$ is well-defined, and the solution to~\eqref{eq:ass/app-dyn-matrix}, i.e., the matrix $A(t)$, is extensible to all $t\in [0,+\infty)$. For all $t\ge 0$, $A(t)$ is row-stochastic, irreducible and has strictly positive diagonal;
\item there exists a row-stochastic irreducible matrix $C\in
   \mathbb{R}^{n\times n}$ with zero diagonal such that
   \begin{equation}\label{eq:assign/appraise-struc-A(t)}
     A(t)= \diag\!\big(\bm{a}(t) \big) + \left( I_n-\diag\!\big( \bm{a}(t)
     \big) \right)C,
   \end{equation}
   for all $t\ge 0$, where $\bm{a}(t)=\big( a_{1}(t),\dots,a_{n}(t)\big)^{\top}$ and
   $a_{i}(t)=a_{ii}(t)$, for $i\in \{1,\dots,n\}$;
\item Define the \emph{reduced assign/appraise dynamics} as 
\vspace{-0.05cm}
\begin{equation}\label{eq:ass/app-reduced}
     \hspace{-0.33cm}\begin{cases}
       \displaystyle \dot{a}_i = a_i(1-a_i)\big( p_i(\bm{w})-\sum_{k=1}^n m_{ik} p_k(\bm{w}) \big),\\
       \displaystyle w_i  = \frac{c_i}{(1-a_i)} \Big/ \sum_{k=1}^n \frac{c_k}{(1-a_k)},
     \end{cases}
\end{equation}
where $\bm{c}=(c_1,\dots,c_n)^{\top}=\vleft(C)$. This dynamics is equivalent to system~\eqref{eq:ass/app-dyn-matrix} in the following sense: The matrix $A(t)$'s each diagonal entry $a_{ii}(t)$ satisfies the dynamics~\eqref{eq:ass/app-reduced} for $a_i(t)$, and, for any $t\ge 0$, $a_{ii}(t)=a_i(t)$ for any $i$, and $a_{ij}(t)=a_{ij}(0)\big( 1-a_i(t)\big)/\big( 1-a_i(0) \big)$ for any $i\neq j$;
\item The set $\Omega=\big{\{}\bm{a}\in [0,1]^n\big|0\le a_i\le 1-\zeta_i\big(\bm{a}(0)\big)\big{\}}$,
%\begin{equation*}
%\Omega=\big{\{}\bm{a}\in [0,1]^n\,\big|\,0\le a_i\le 1-\zeta_i\big(\bm{a}(0)\big)\big{\}},
%\end{equation*}
where $\zeta_i\big( \bm{a}(0) \big)=\frac{c_i}{x_i}\min_k \frac{x_k}{c_k}\big( 1-a_k(0) \big)$, is a compact positively invariant set for the reduced assign/appraise dynamics~\eqref{eq:ass/app-reduced};
\item the assignment $\bm{w}(t)$ satisfies the \emph{generalized replicator dynamics} with time-varying fitness function \\$a_i(t)\Big( p_i\big( \bm{w}(t) \big) - \sum_l m_{il} p_l\big( \bm{w}(t) \big) \Big)$ for each $i$:
   \begin{equation}\label{eq:ass/app-dyn-w(t)}
     \begin{split}
       \dot{w}_i = w_i\Big( a_i\big( p_i(\bm{w})-\sum_{l=1}^n m_{il} p_l(\bm{w}) \big) - \sum_{k=1}^n w_k a_k \big( p_k(\bm{w})-\sum_{l=1}^n m_{kl}p_l(\bm{w}) \big) \Big).
     \end{split}
   \end{equation}
\end{enumerate} 
\end{theorem}
\smallskip

The proof for Theorem~\ref{thm:ass/app-finite-time} is presented in Appendix~\ref{proof:ass/app-finite-time}. With the extensibility of $A(t)$ and the finite-time properties, we now present the main theorem of this section. 
\begin{theorem}[Asymptotic behavior of assign/appraise dynamics]\label{thm:ass-app-asym-behav}
Consider the assign/appraise dynamics~\eqref{eq:ass/app-dyn-matrix},
based on Assumptions~\ref{asmp:assign-rule}-\ref{asmp:update-app-net},
with the task assignment as in Assumption~\ref{asmp-team-task-assign}
and the performance as in Assumption~\ref{asmp:indiv-perf}. Assume the
observation network $G(M)$ is strongly connected.
%% For any initial appraisal network, with row-stochastic appraisal
%% matrix $A(0)$, that is strongly connected and has positive self
%% loops for all the individuals,
%% \wmmargin{How shall I include the condition that $A(0)$  is row-stochastic?}
For any initial appraisal matrix $A(0)$ that is row-stochastic,
irreducible and has positive diagonal,
\begin{enumerate}
\item the solution $A(t)$ converges, i.e., there exists $A^*\in \mathbb{R}^{n\times n}$ such that $\lim_{t\to \infty}A(t)=A^*$;
\item the limit appraisal matrix $A^*$ is row-stochastic and irreducible. Moreover, the task assignment satisfies $\lim_{t\to \infty}\bm{w}(t)=\vleft(A^*)=\bm{x}$.
\end{enumerate} 
\end{theorem}
\smallskip

\begin{figure}
\begin{center}
\subfigure[$t=0$]{ \includegraphics[width=.222\linewidth]{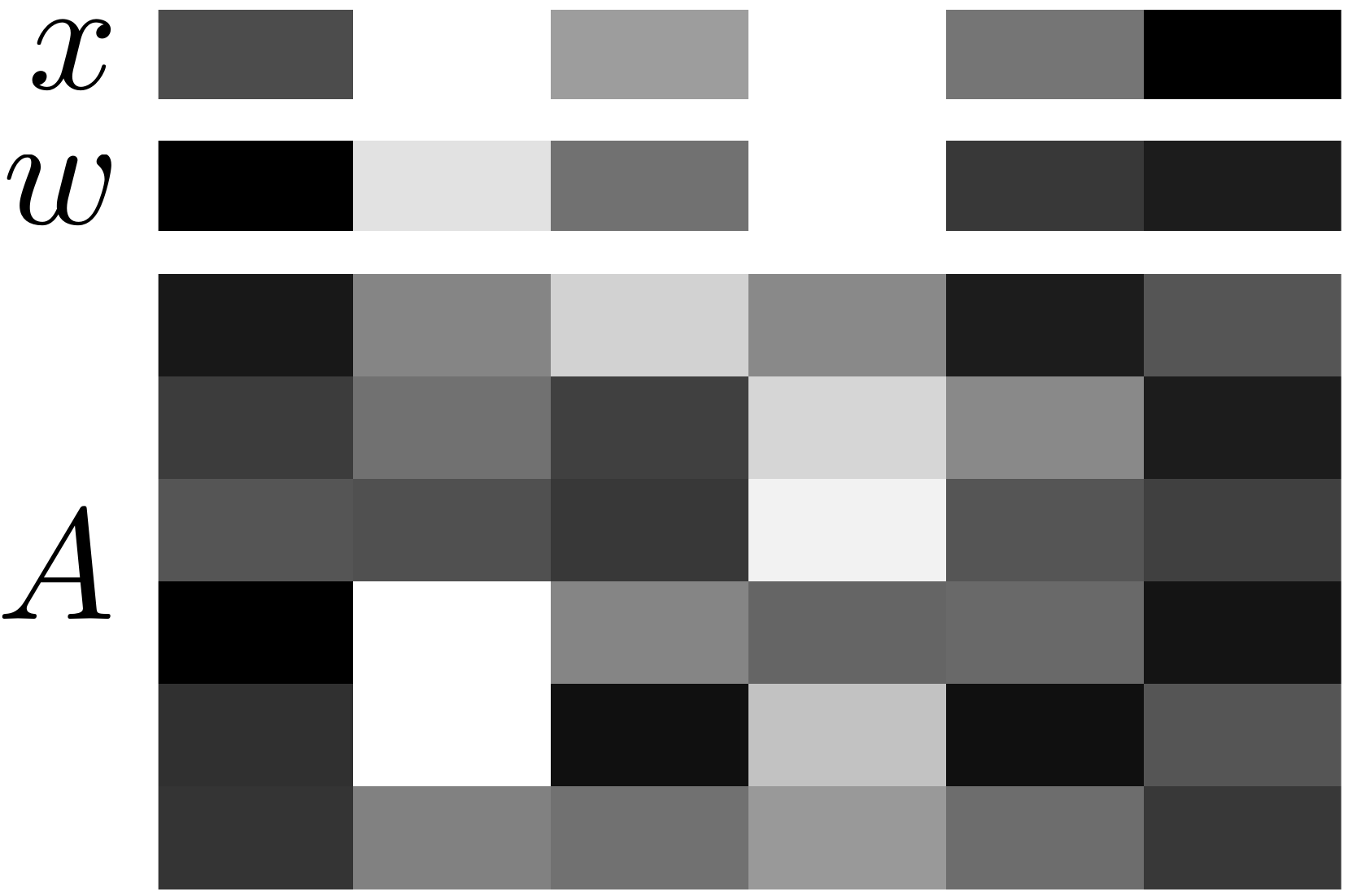}}
\subfigure[$t=2$]{ \includegraphics[width=.222\linewidth]{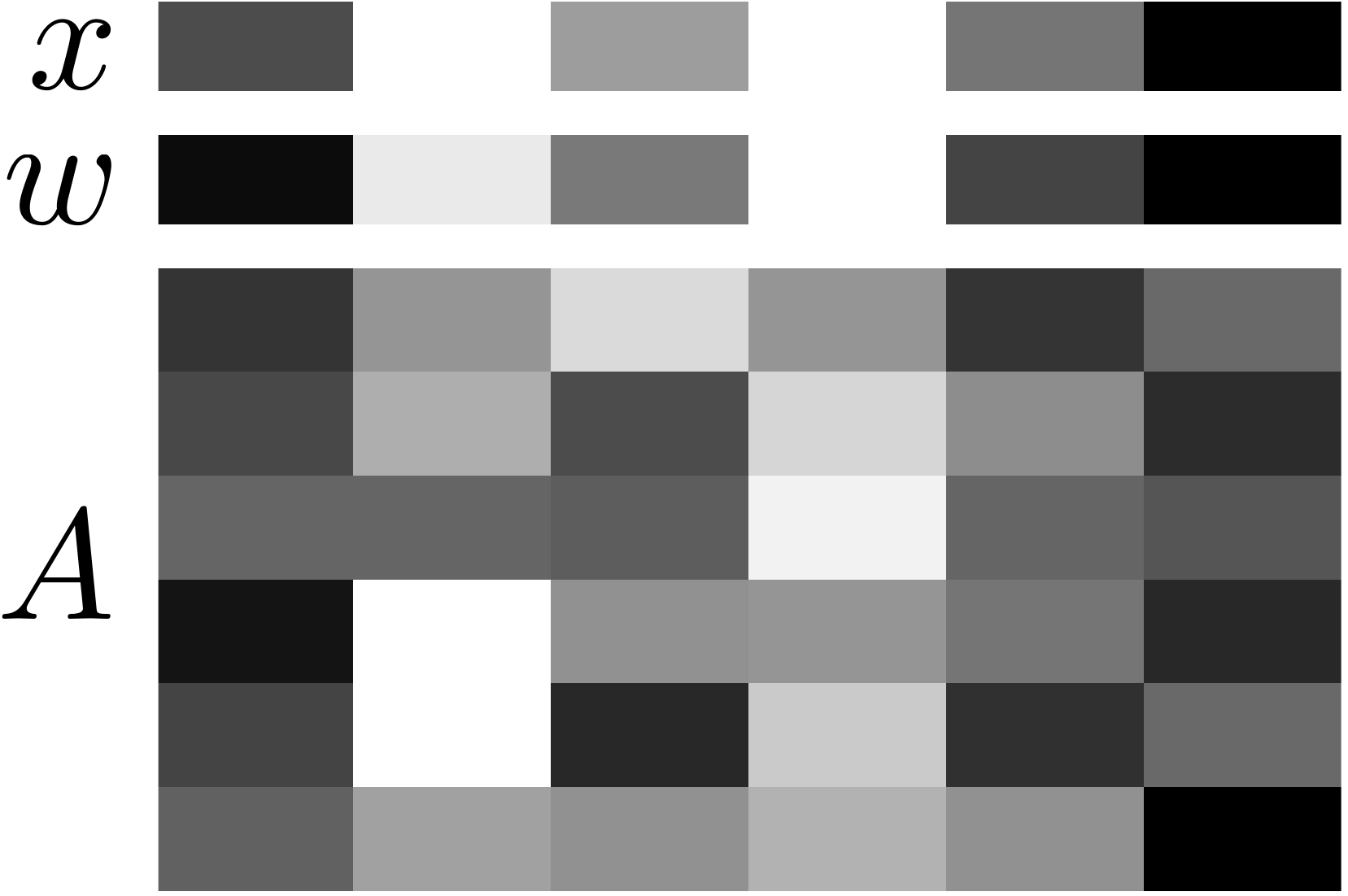}}
\subfigure[$t=10$]{ \includegraphics[width=.222\linewidth]{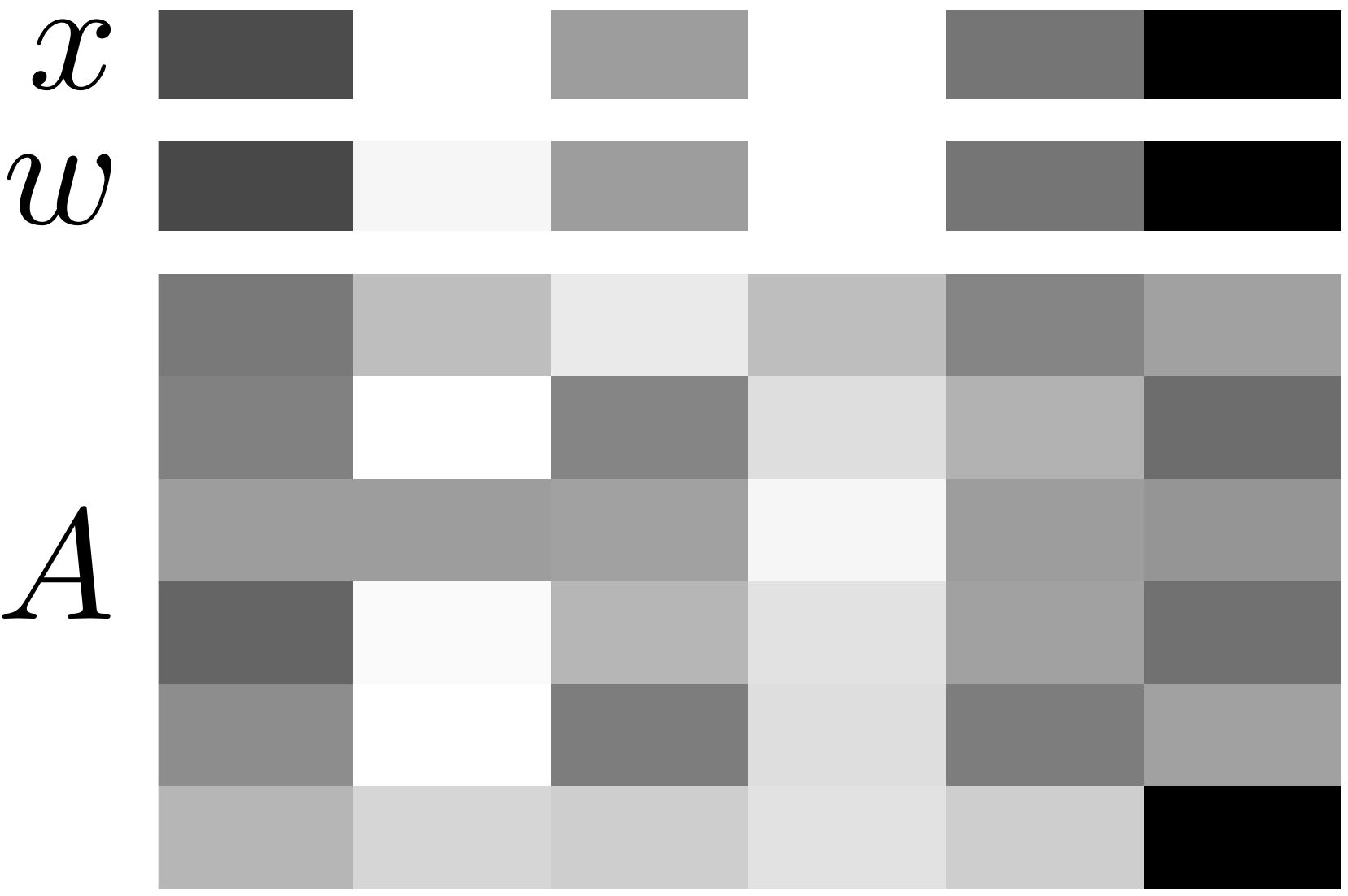}}
\subfigure[$t=30$]{ \includegraphics[width=.222\linewidth]{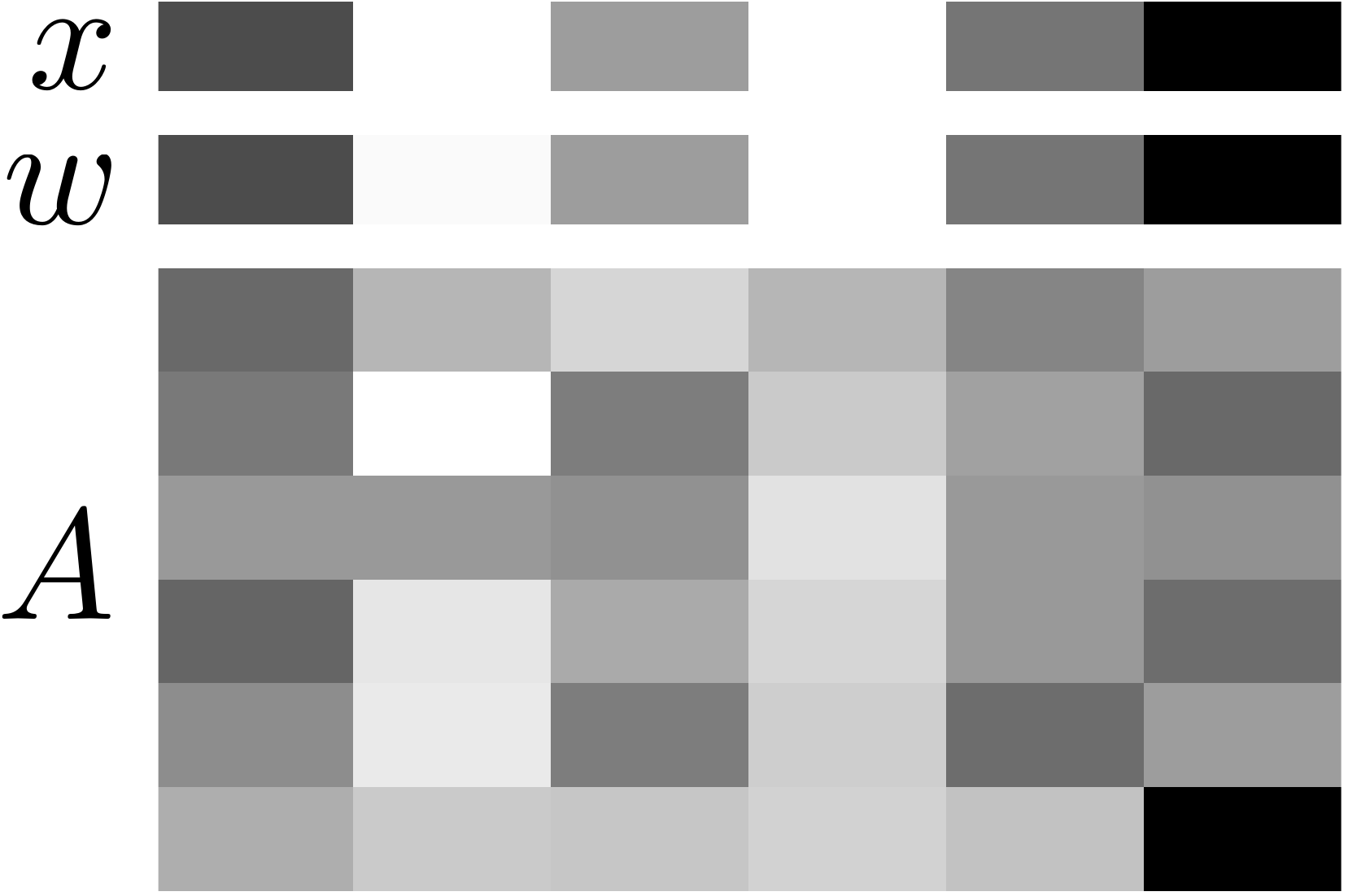}}
\caption{Visualization of the evolution of $A(t)$ and $\bm{w}(t)$ obeying the assign/appraise dynamics with $n=6$. The observation network is strongly connected. In these visualized matrices and vectors, the darker the entry, the higher value it has.}
\label{fig:visualization-assign-appraise}
\end{center}
\end{figure}

The proof is presented in Appendix~\ref{proof:ass-app-asym-behav}. Theorem~\ref{thm:ass-app-asym-behav} indicates that, the teams obeying the assign/appraise dynamics asymptotically achieves the optimal task assignment, but do not necessarily reach appraisal consensus. Figure~\ref{fig:visualization-assign-appraise} gives a visualized illustration of the asymptotic behavior of the assign/appraise dynamics.
%The convergence of $\bm{w}$ to the optimal assignment $\bm{x}$ indicates that the assign/appraise dynamics asymptotically achieves collective learning. However, since the ratio of any two off-diagonal entries in the same row of $A(t)$ remains constant, the  dynamics does not necessarily achieves appraisal consensus. 

\smallskip

\begin{remark}
From the proof for Theorem~\ref{thm:ass-app-asym-behav} we know that, the teams obeying the following dynamics
\begin{equation*}
\begin{cases}
\displaystyle \dot{a}_{ii} & \hspace{-0.3cm} =\gamma_i(t)a_{ii}(1-a_{ii})\big( p_i(\bm{w})-\sum_k m_{ik}p_k(\bm{w}) \big),\\
\displaystyle \dot{a}_{ij} & \hspace{-0.3cm}=-\gamma_i(t)a_{ii}a_{ij}\big( p_i(\bm{w})-\sum_k m_{ik}p_k(\bm{w}) \big),
\end{cases}
\end{equation*} 
also asymptotically achieve the optimal assignment, if each $\gamma_i(t)$ remains strictly bounded from $0$. This result indicates that our model can be generalized to the case of heterogeneous sensitivities to performance feedback. %\wmmargin{How to better rephrase Remark 10? Basically it means our model is quite general. As long as the individuals increase self appraises when above the average performance and vice versa, and the sensitivity to performance feedback is lower bounded, the collective learning is achieved.}
\end{remark}

\section{The Assign/appraise/influence Dynamics of the Appraisal Networks}
%In this section we introduce the influence network as another inner structure of teams, in addition to the appraisal network, and the team learning dynamics involving both structures.

In this section we further elaborate the assign/appraise dynamics by
assuming that the appraisal network is updated via not only the
performance feedback, but also its co-evolution with the team members' interpersonal influences. In other words, we include an opinion dynamics process
among the individuals who discuss and possibly reach consensus on the
values of interpersonal appraisals.

\subsection{Model description}
The new model, named the \emph{assign/appraise/influence
  dynamics}, is defined by three components: the assignment rule as in
Assumption~\ref{asmp:assign-rule}, the appraise dynamics based on
Assumptions~\ref{asmp:perf-feedback} and~\ref{asmp:update-app-net},
and the \emph{influence dynamics}, which is the opinion exchanges
among individuals on interpersonal appraisals. Denote by $w_{ij}$ the
weight individual $i$ assigns to $j$ (including self weight $w_{ii}$)
in the opinion exchange. The matrix $W=(w_{ij})_{n\times n}$ defines a
directed and weighted graph, referred to as the \emph{influence
  network}, is row-stochastic and possibly time-varying.

The diagram illustration of assign/appraise/influence dynamics is
presented in Figure~\ref{fig:diagram-ass/app/inf}, and the general
form is given as follows:
\begin{equation}\label{eq:ass/app/inf-dyn-general-form}
\begin{cases}
\displaystyle \dot{A} & \hspace{-0.25cm}= \frac{1}{\subscr{\tau}{ave}} \subscr{F}{ave}( A, W ) + \frac{1}{\subscr{\tau}{app}} \subscr{F}{app}( A,\bm{w} ),\\
\displaystyle \bm{w} & \hspace{-0.25cm}= \vleft( A ).
\end{cases}
\end{equation}
The time index $t$ is omitted for simplicity. The term $\subscr{F}{app}(A,\bm{w})$ corresponds to the appraise dynamics given by the right-hand side of the first equation in~\eqref{eq:ass/app-dyn-matrix}, while the term $\subscr{F}{ave}( A, W )$ corresponds to the influence dynamics specified by the assumption below. Parameters $\subscr{\tau}{ave}$ and $\subscr{\tau}{app}$ are positive, and relate to the time scales of influence dynamics and appraise dynamics respectively.

\begin{assumption}[Influence dynamics]\label{asmp:inf-dyn}
For the assign/appraise/influence dynamics, assume that, at each time $t\ge 0$, the influence network is identical to the appraisal network, i.e., $W(t)=A(t)$. Moreover, assume that the individuals obey the classic DeGroot opinion dynamics~\cite{MHDG:74} for the interpersonal appraisals, i.e., $\subscr{F}{ave}( W,A) = -( I_n-W )A.$
%\begin{equation*}
%\subscr{F}{ave}( W,A) = -( I_n-W )A.
%\end{equation*}
\end{assumption}
\smallskip

%\begin{figure*}
%\begin{center}
%\includegraphics[width=.75\linewidth]{diagram_assign_appraise_influence_dyn}
%\caption{Diagram illustration for the assign/appraise/influence dynamics}
%\label{fig:diagram-assign-appraise-influence}
%\end{center}
%\end{figure*}

Based on equation~\eqref{eq:ass/app/inf-dyn-general-form} and Assumptions~\ref{asmp:assign-rule}-\ref{asmp:inf-dyn}, the assign/appraise/influence dynamics is written as
\begin{equation}\label{eq:ass/app/inf-dyn-matrix-form}
\begin{cases}
\dot{A} & \hspace{-0.25cm}  =  \frac{1}{\subscr{\tau}{ave}}(A^2-A)+ \frac{1}{\subscr{\tau}{app}} \diag\!\big( \bm{p}(\bm{w})-M\bm{p}(\bm{w})\big)\subscr{A}{d} (I_n-A),\\
\bm{w} & \hspace{-0.25cm}  =  \vleft(A),
\end{cases}
\end{equation}
%For this model, the appraisal network and the influence network are both the team's basic inner structures, and the observation network defines the feedback signals received by each individual. In Section IV.B, 
In the next subsection, we relate the topology of the observation network $G(M)$ to the asymptotic behavior of the assign/appraise/influence dynamics, i.e., the convergence to optimal assignment and the appraisal consensus.%, i.e., the collective learning and the appraisal consensus.

\subsection{Dynamical behavior of the assign/appraise/influence dynamics}
The following lemma shows that, for the assign/appraise/influence dynamics, we only need to consider the all-to-all initial appraisal network.
\begin{lemma}[entry-wise positive for initial appraisal]\label{lem:ass/app/inf-initial->0}
Consider the assign/appraise/influence dynamics~\eqref{eq:ass/app/inf-dyn-matrix-form} based on Assumptions~\ref{asmp:assign-rule}-\ref{asmp:inf-dyn}, with the task assignment and performance as in Assumptions~\ref{asmp-team-task-assign} and~\ref{asmp:indiv-perf} respectively. For any initial appraisal matrix $A(0)$ that is primitive and row-stochastic, there exists $\Delta t>0$ such that $A(t)\succ \vectorzeros[n\times n]$ for any $t\in (0,\Delta t]$.
\end{lemma}
\smallskip

The proof is given in Appendix~\ref{proof:ass/app/inf-initial->0}. Before discussing the asymptotic behavior, we state a technical assumption. 
\begin{conjecture}[Strict lower bound of the interpersonal appraisals]\label{conj:ass/app/inf-lower-bound-A(t)}
Consider the assign/appraise/influence dynamics~\eqref{eq:ass/app/inf-dyn-matrix-form} based on Assumptions~\ref{asmp:assign-rule}-\ref{asmp:inf-dyn}, with the task assignment and performance as in Assumptions~\ref{asmp-team-task-assign} and~\ref{asmp:indiv-perf} respectively. For any $A(0)$ that is entry-wise positive and row-stochastic, there exists $a_{\min}>0$, depending on $A(0)$, such that $A(t)\succ a_{\min}\vectorones[n]\vectorones[n]^{\top}$ for any time $t\ge 0$, as long as $A(\tau)$ and $\bm{w}(\tau)$ are well-defined for all $\tau\in [0,t]$.
\end{conjecture}

Monte Carlo validation and a sufficient condition for Conjecture~\ref{conj:ass/app/inf-lower-bound-A(t)} are presented in Appendix~\ref{discussion:conjecture-pos}. Now we state the main results of this section.

\begin{theorem}[Assign/appraise/influence dynamical behavior]\label{thm:ass/app/inf-asym-behav}
Consider the assign/appraise/influence
dynamics~\eqref{eq:ass/app/inf-dyn-matrix-form} based on
Assumptions~\ref{asmp:assign-rule}-\ref{asmp:inf-dyn}, with the task
assignment and performance as in
Assumptions~\ref{asmp-team-task-assign} and
Assumption~\ref{asmp:indiv-perf} respectively. Suppose that
Conjecture~\ref{conj:ass/app/inf-lower-bound-A(t)} holds. Assume that
the observation network $G(M)$ contains a globally reachable node.
%% For any all-to-all initial appraisal network $G\big(A(0)\big)$,
%% \wmmargin{How shall I state the condition that $A(0)$ isrow-stochastic?}
For any initial appraisal matrix $A(0)$ that is entry-wise positive and row-stochastic, 
\begin{enumerate}
\item the solution $A(t)$ exists and $\bm{w}(t)=\vleft\big( A(t) \big)$ is well-defined for all $t\in [0,+\infty)$. Moreover, $A(t)\succ \vectorzeros[n\times n]$ and $A(t)\vectorones[n]=\vectorones[n]$ for any $t\ge 0$;
\item the assignment $\bm{w}(t)$ obeys the generalized replicator dynamics~\eqref{eq:ass/app-dyn-w(t)}, and $\xi_0 \vectorones[n] \preceq \bm{w}(t) \preceq \big( 1-(n-1)\xi_0 \big)\vectorones[n]$, where
\begin{equation*}
\xi_0 = \left( 1+(n-1)\frac{\max_k x_k}{\min_l x_l}\gamma_0 \right)^{-1},\quad \text{and}\quad 
\gamma_0 = \frac{\max_k x_k/w_k(0)}{\min_l x_l/w_l(0)};
\end{equation*}
%where 
%\begin{align*}
%\xi_0 & = \left( 1+(n-1)\frac{\max_k x_k}{\min_l x_l}\gamma_0 \right)^{-1},\quad \text{and}\\
%\gamma_0 & = \frac{\max_k x_k/w_k(0)}{\min_l x_l/w_l(0)};
%\end{align*}
\item as $t\to +\infty$, $A(t)$ converges to $\vectorones[n]\bm{x}^{\top}$ and thereby $\bm{w}(t)$ converges to $\bm{x}$.
\end{enumerate}
\end{theorem}  
\smallskip

\begin{figure}
\begin{center}
\subfigure[$t=0$]{ \includegraphics[width=.222\linewidth]{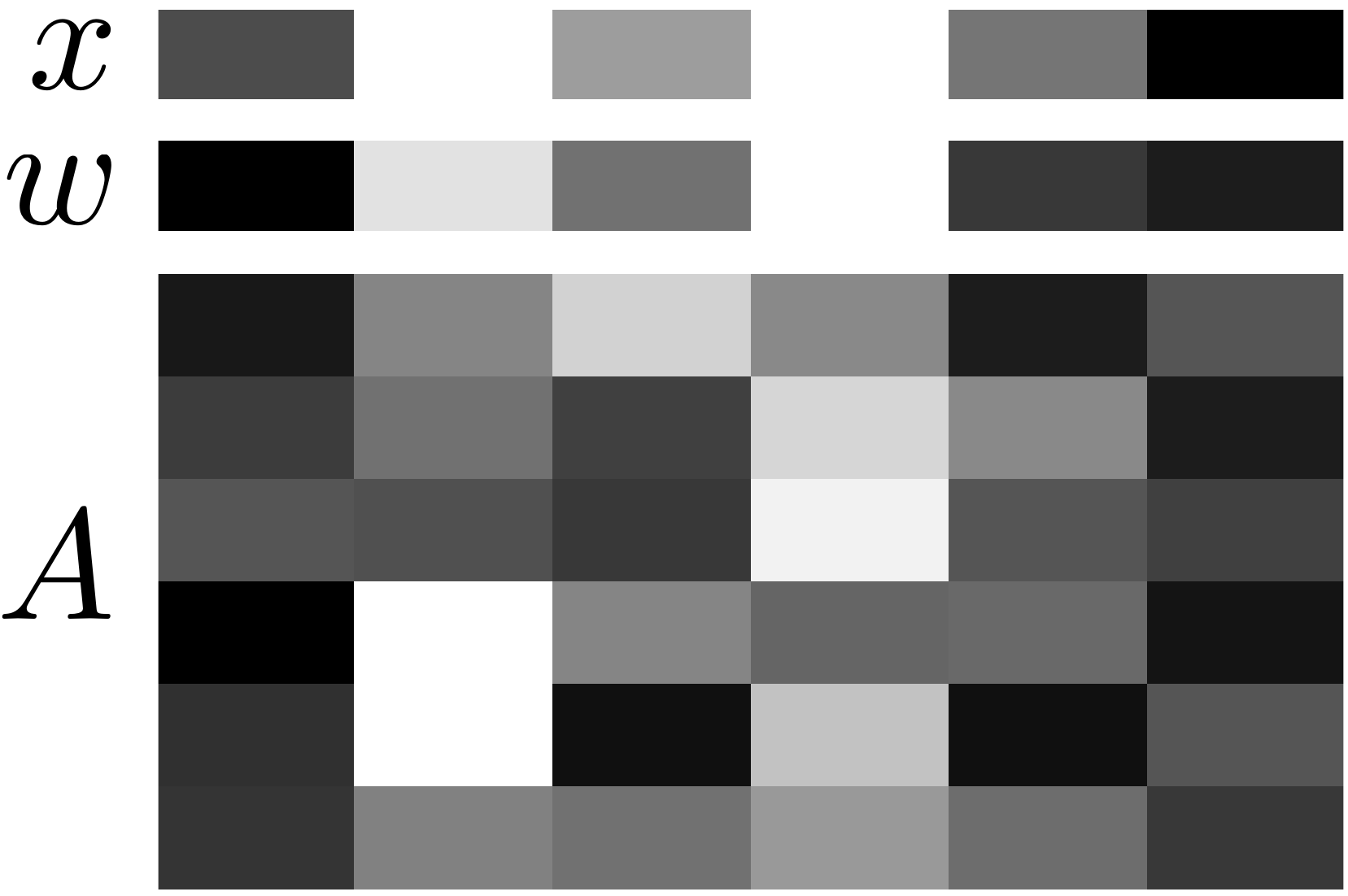}}
\subfigure[$t=2$]{ \includegraphics[width=.222\linewidth]{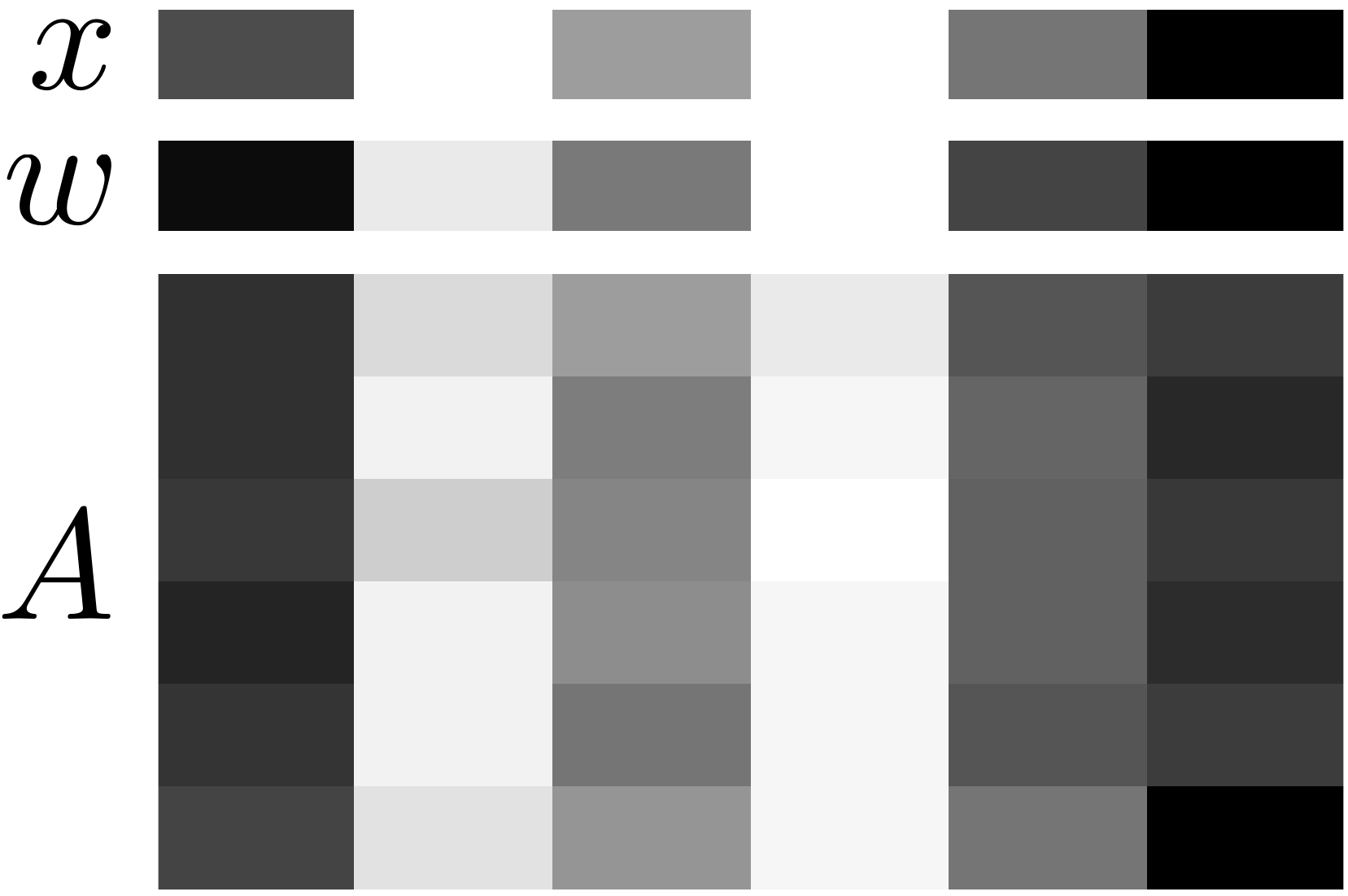}}
\subfigure[$t=10$]{ \includegraphics[width=.222\linewidth]{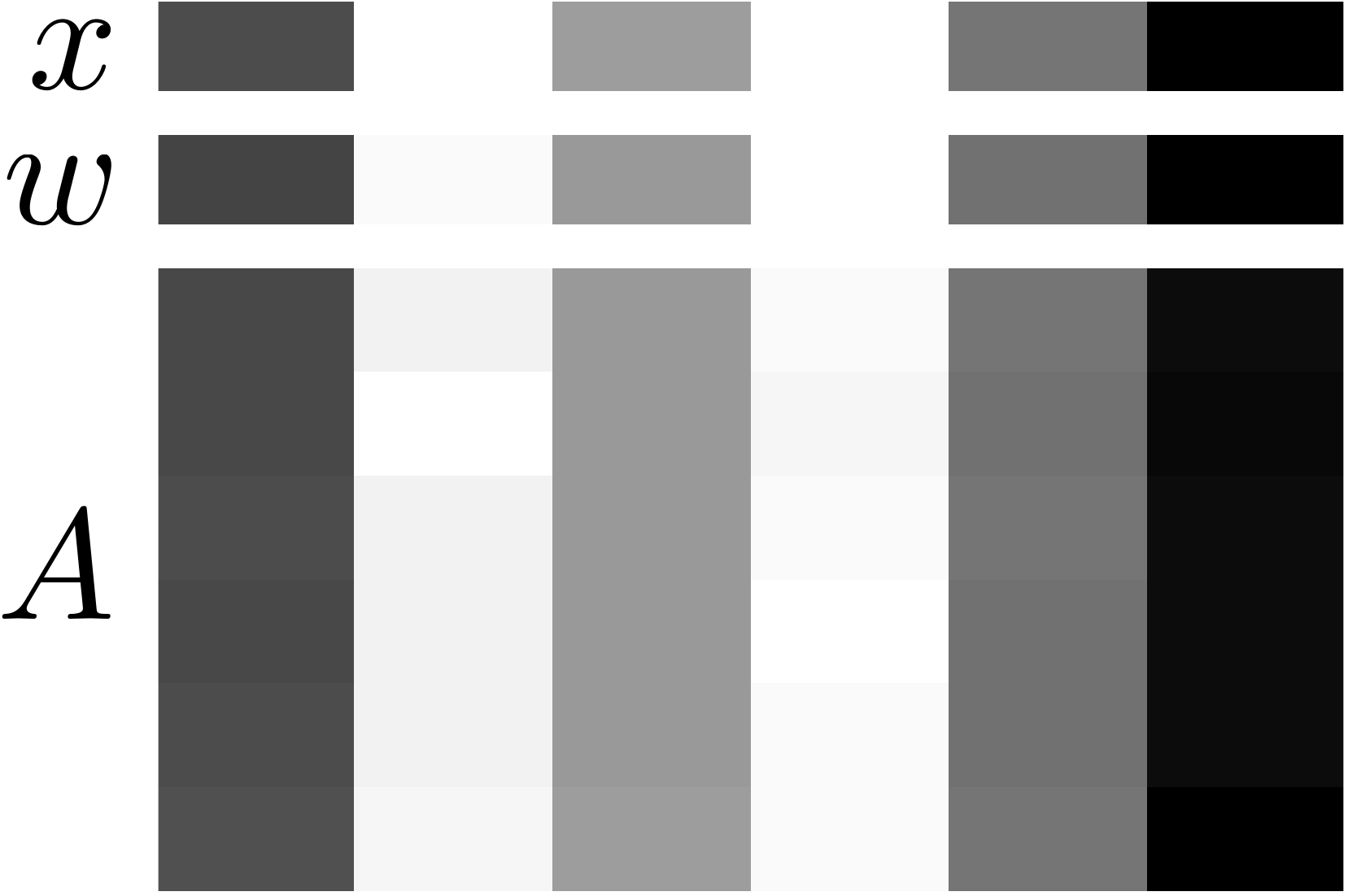}}
\subfigure[$t=30$]{ \includegraphics[width=.222\linewidth]{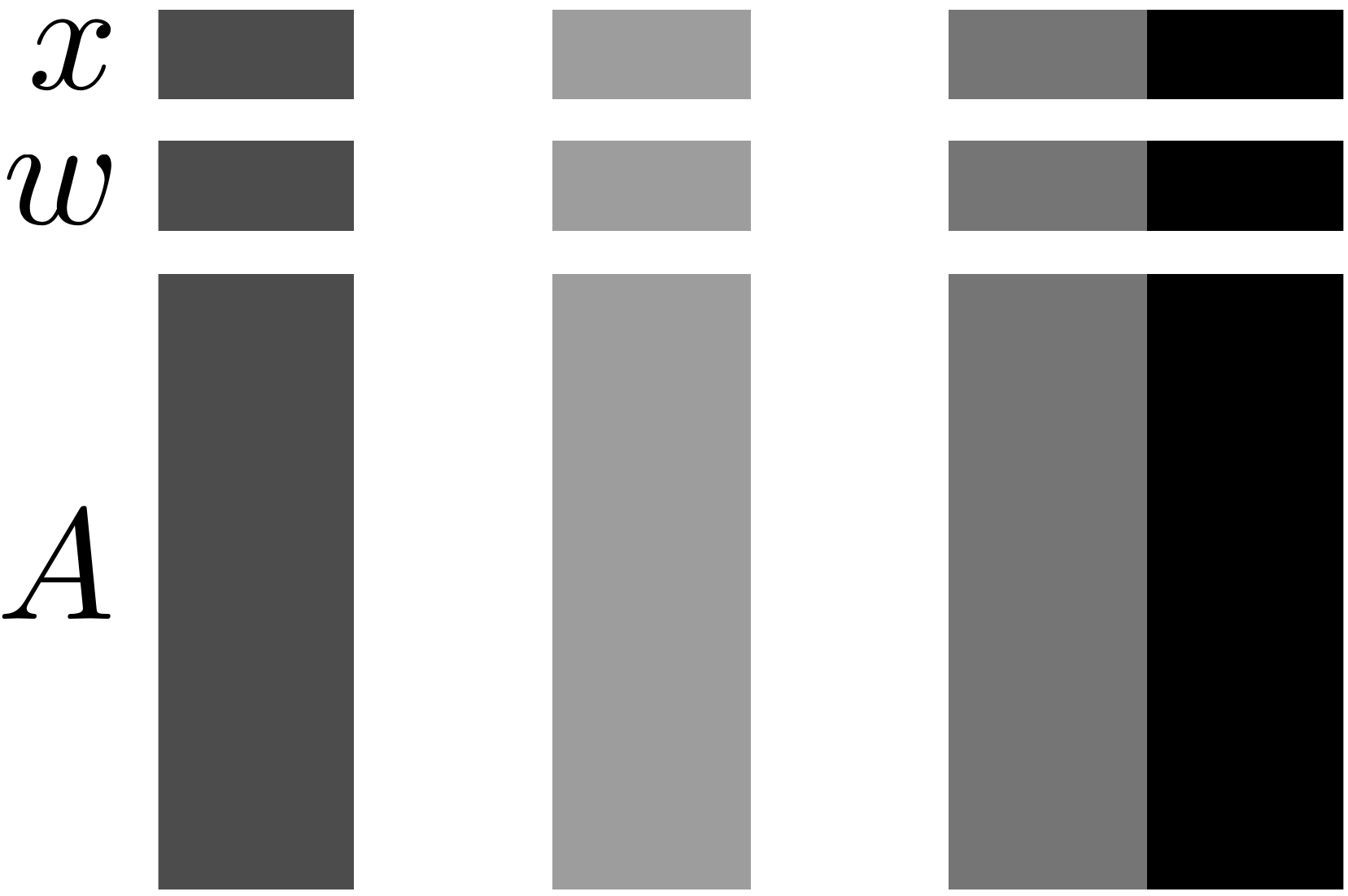}}
\caption{Visualization of the evolution of $A(t)$ and $\bm{w}(t)$ obeying the assign/appraise/influence dynamics with $n=6$. The observation network contains a globally reachable node. In these visualized matrices and vectors, the darker the entry, the higher value it has.}
\label{fig:visualization-ass/app/inf}
\end{center}
\end{figure}

The proof is given in Appendix~\ref{proof:ass/app/inf-asym-behav}. As Theorem~\ref{thm:ass/app/inf-asym-behav} indicates, the team obeying the assign/appraise/\\influence dynamics achieves collective learning. A visualized illustration of the dynamics is given by Figure~\ref{fig:visualization-ass/app/inf}. 

\section{Model Variations: Causes of Failure to Learn}
\label{sec:variations}
The baseline assign/appraise/influence dynamics~\eqref{eq:ass/app/inf-dyn-matrix-form} consists of three phases: the assignment rule, the appraise dynamics, and the influence dynamics. In this section, we propose one variation in each of the three phases, based on some socio-psychological mechanisms that may cause a failure in team learning. We investigate the behavior of each model variation by numerical simulation.

\begin{figure}
\begin{center}
\subfigure[no influence dynamics, $t=0$]{ \includegraphics[width=.222\linewidth]{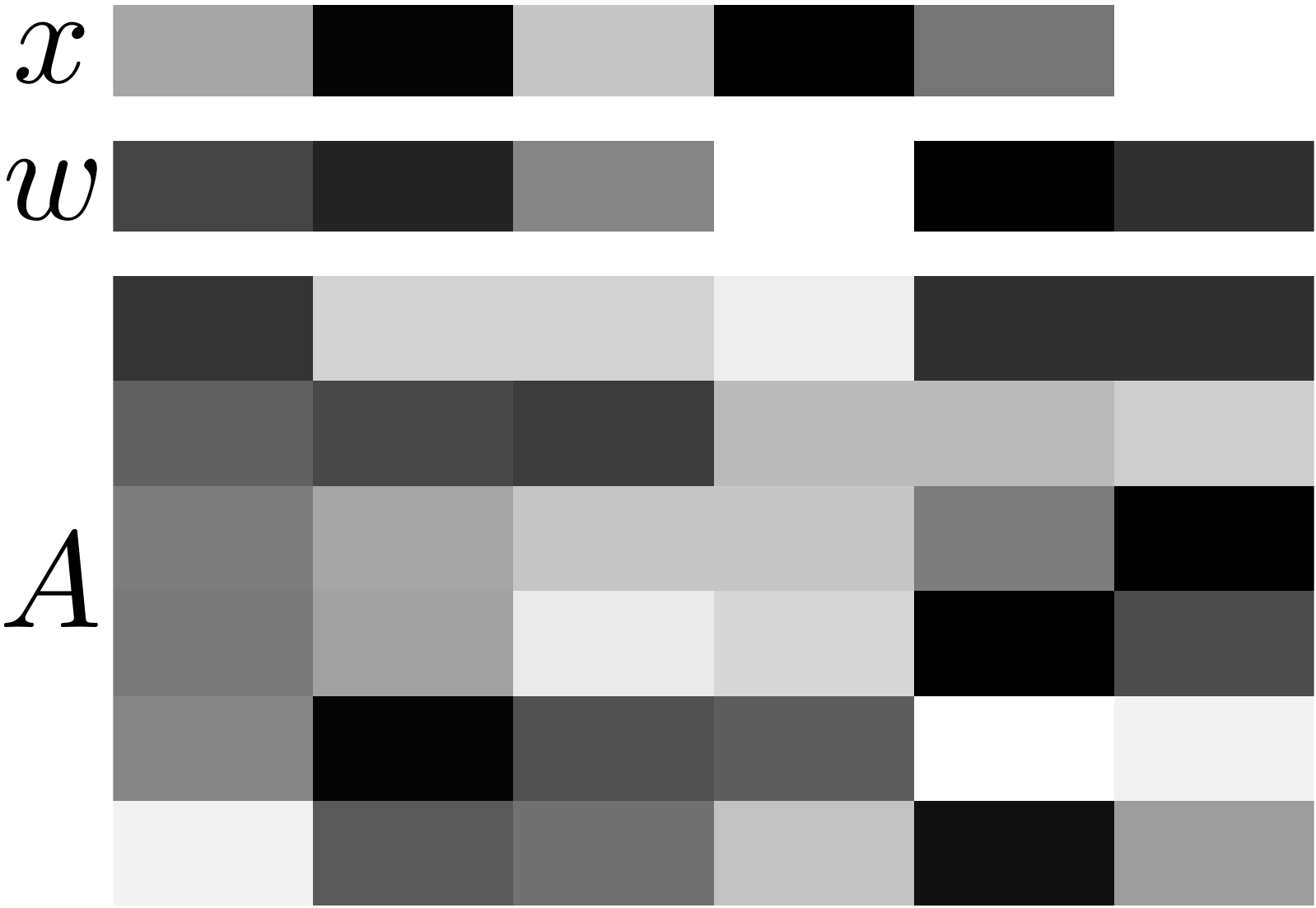}}
\subfigure[no influence dynamics, $t=2$]{ \includegraphics[width=.222\linewidth]{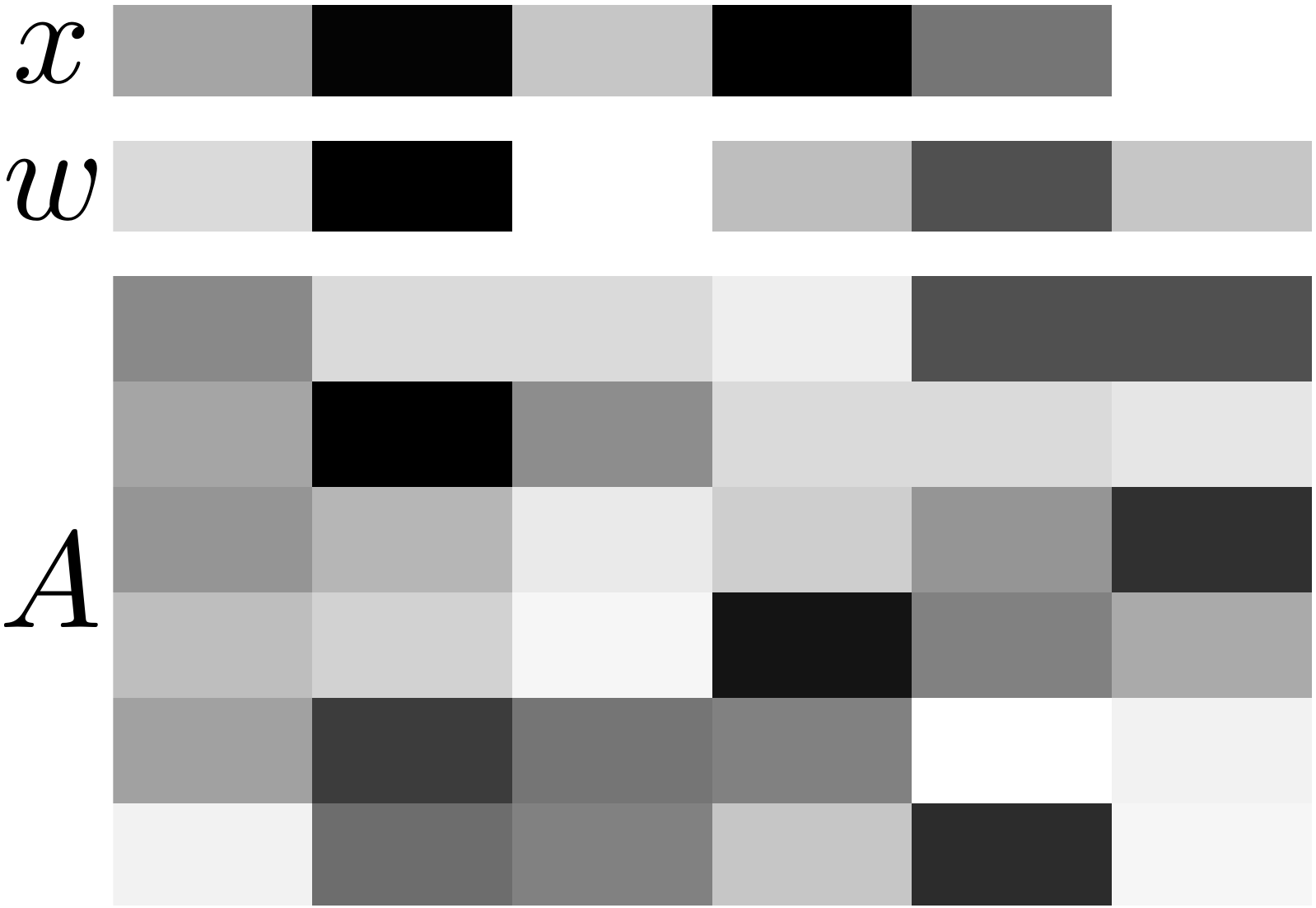}}
\subfigure[no influence dynamics, $t=30$]{ \includegraphics[width=.222\linewidth]{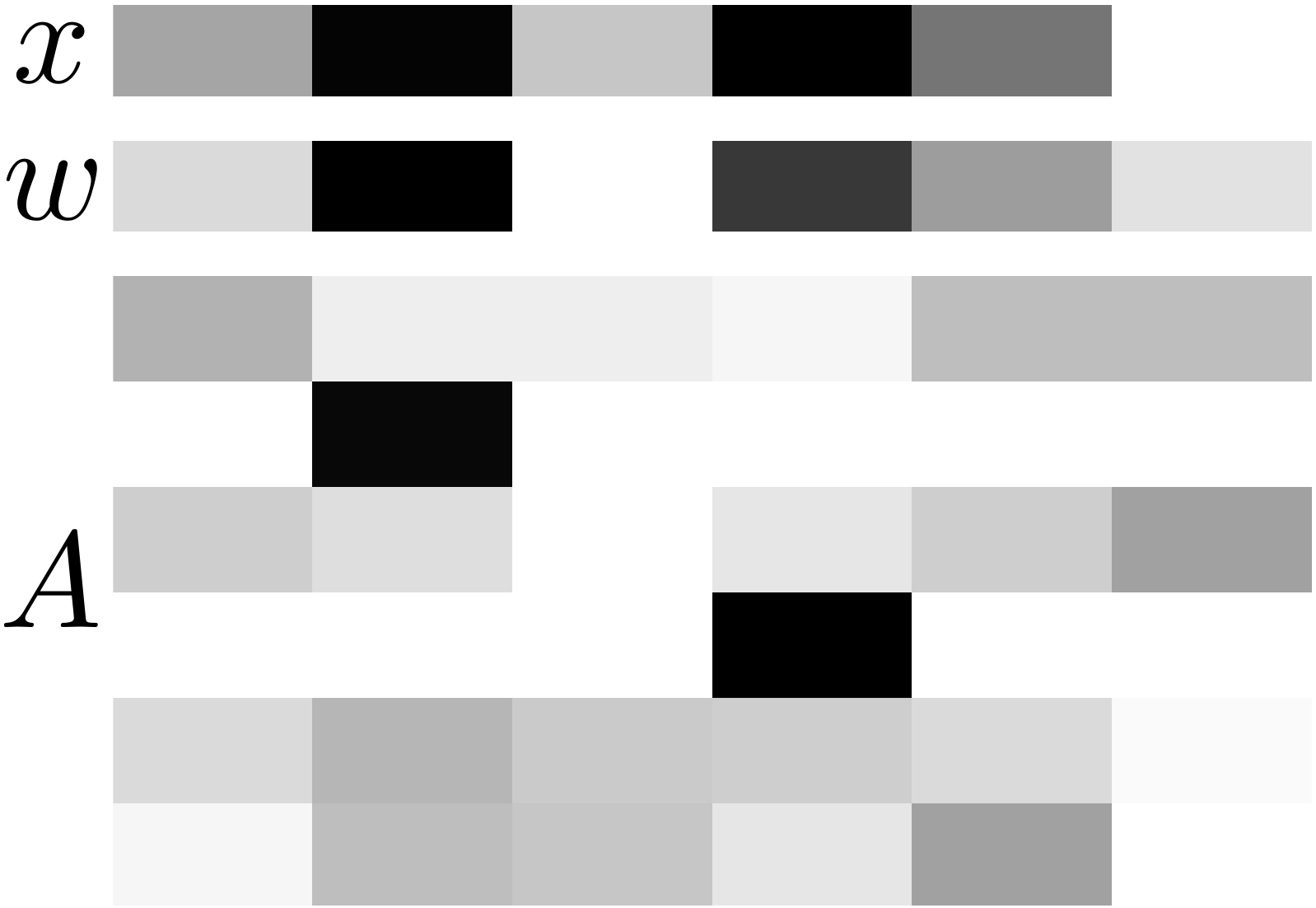}}
\subfigure[no influence dynamics, $t=50$]{ \includegraphics[width=.222\linewidth]{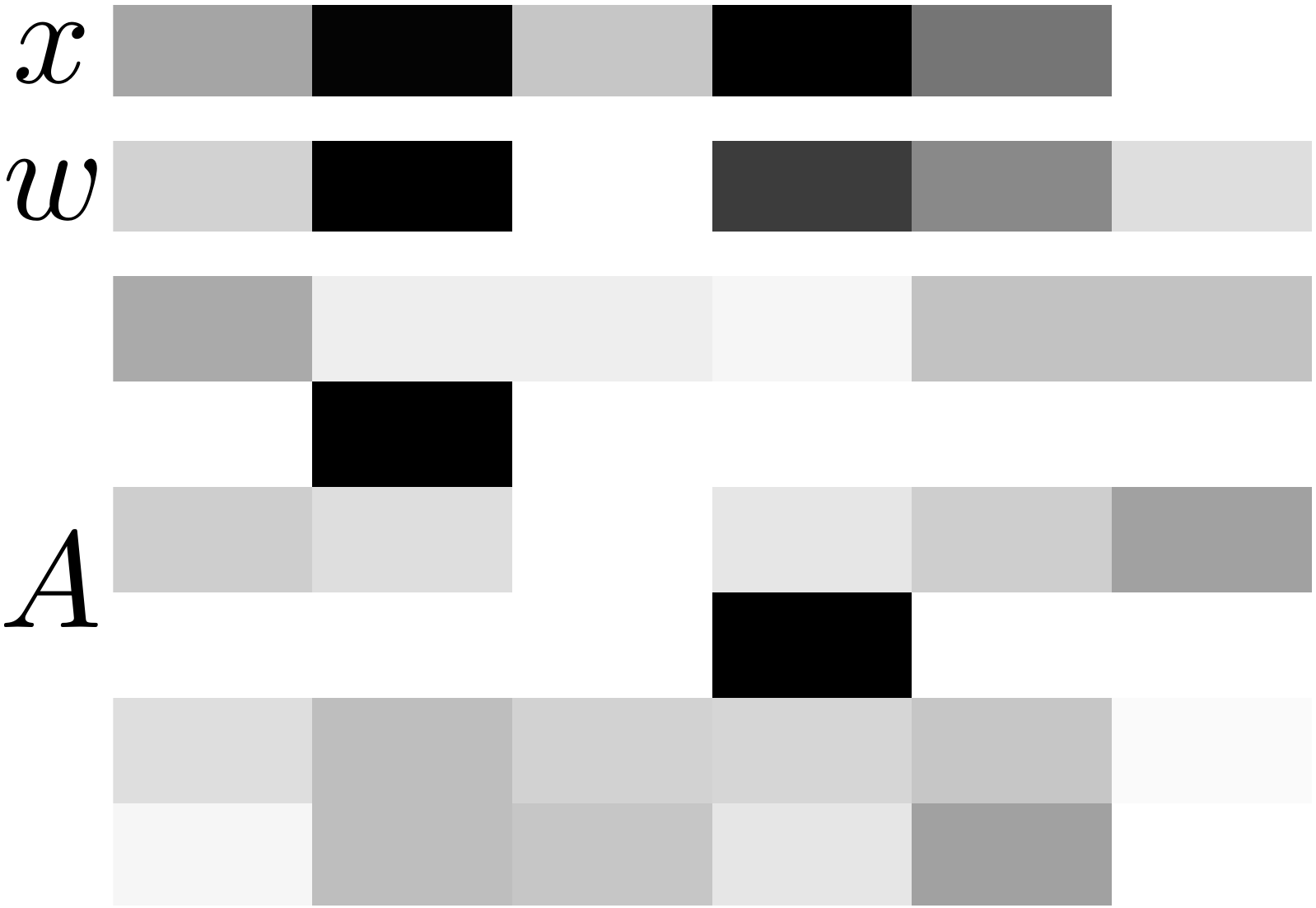}}
\subfigure[with influence dynamics, $t=0$]{ \includegraphics[width=.222\linewidth]{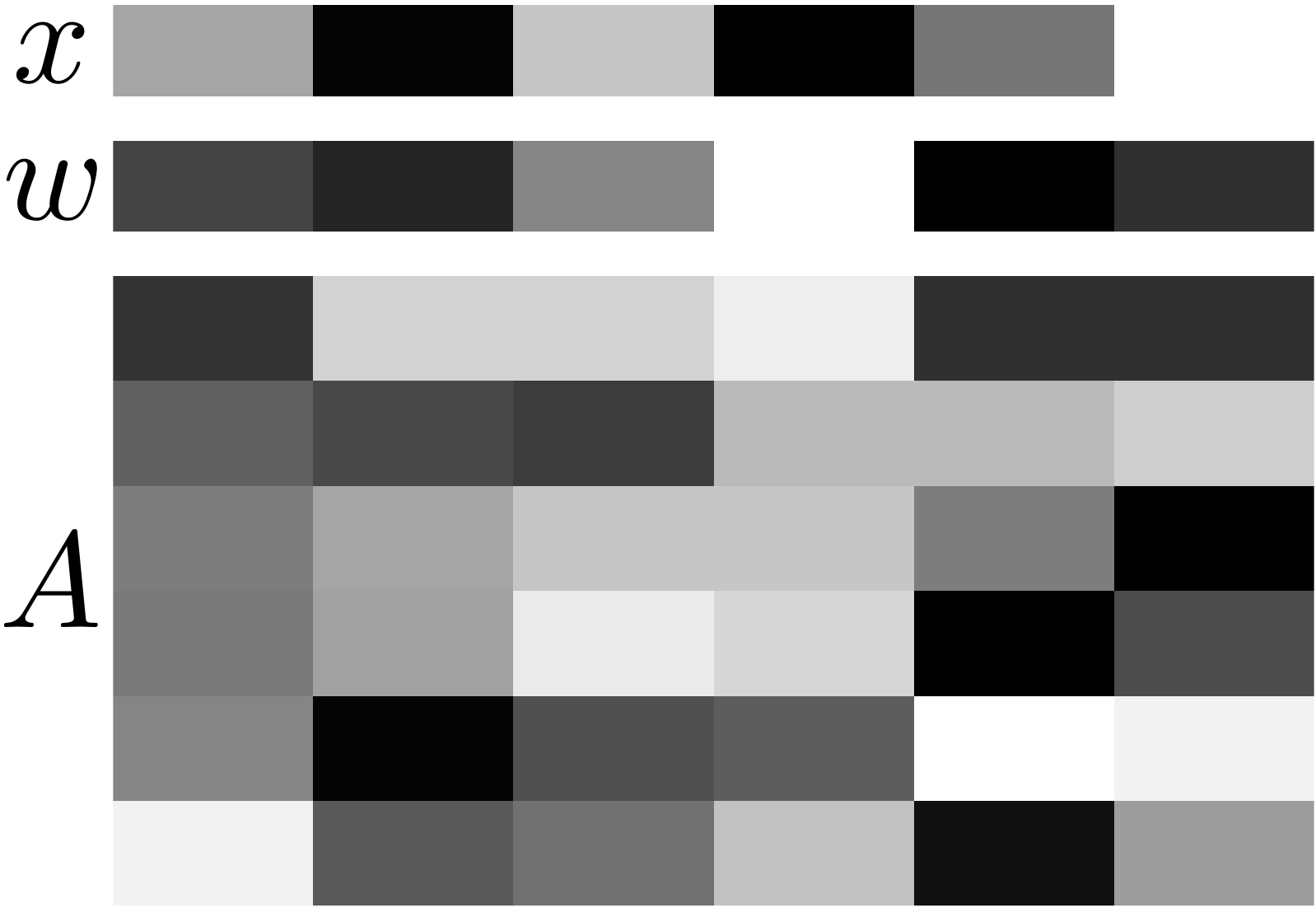}}
\subfigure[with influence dynamics, $t=2$]{ \includegraphics[width=.222\linewidth]{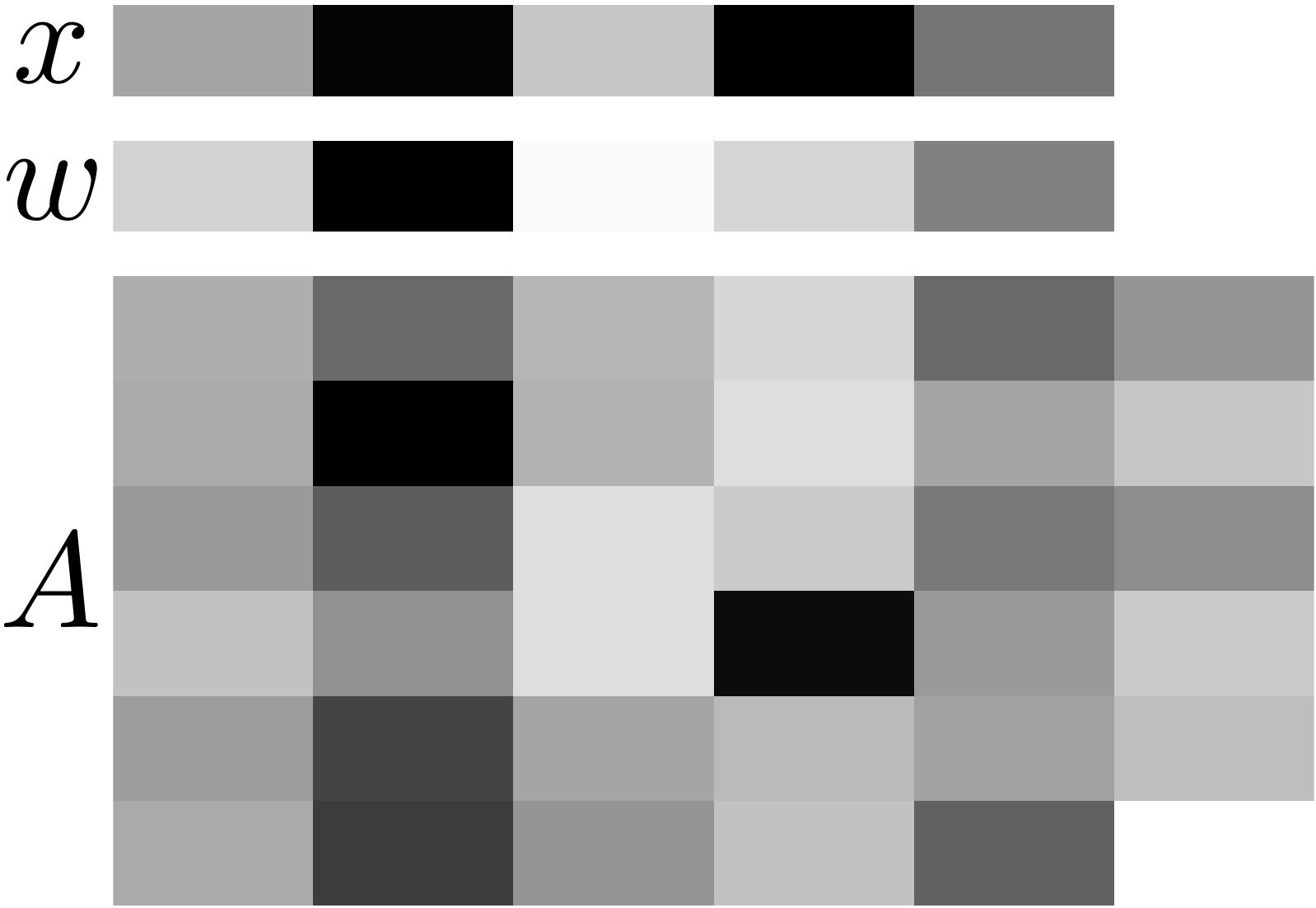}}
\subfigure[with influence dynamics, $t=30$]{ \includegraphics[width=.222\linewidth]{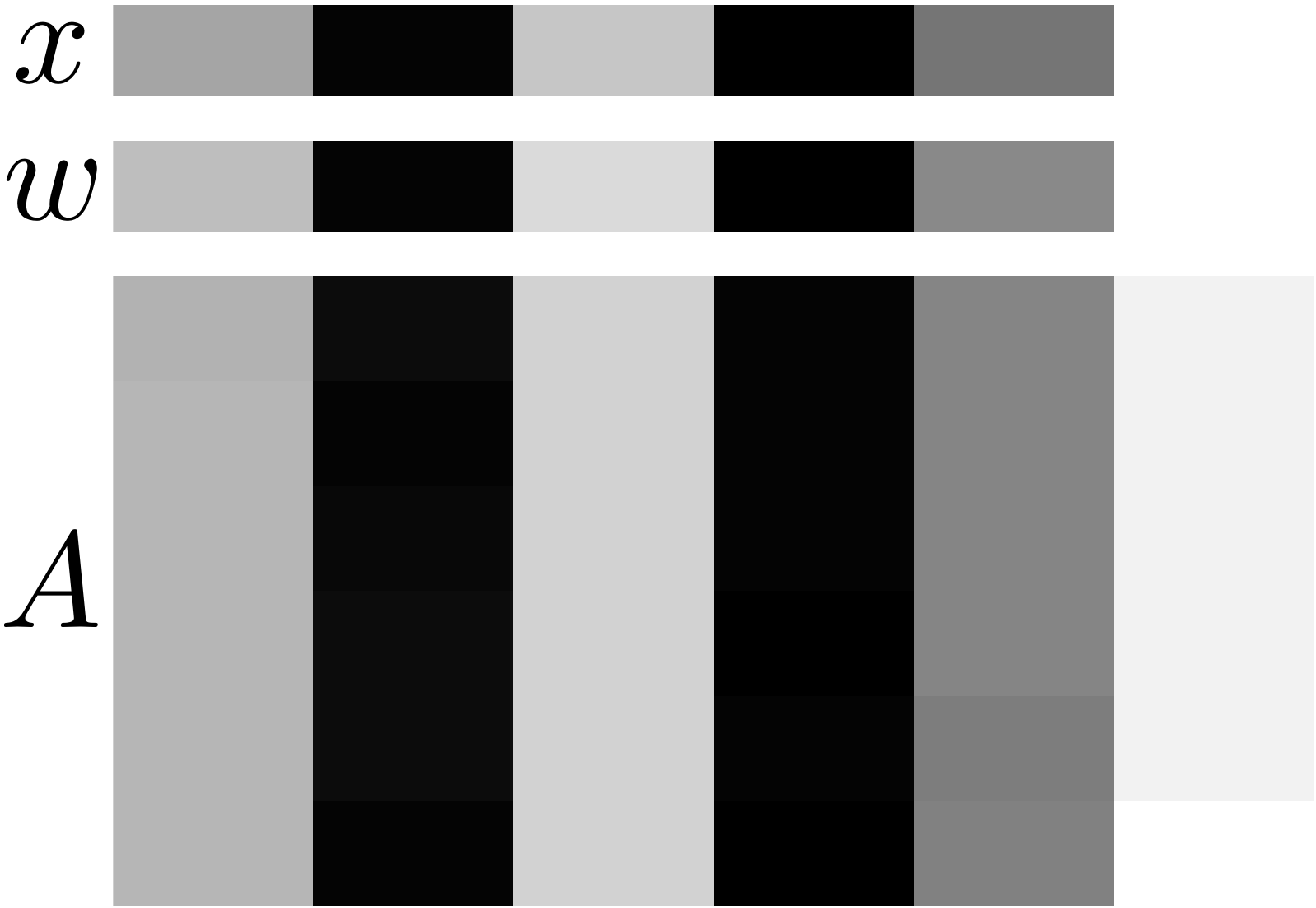}}
\subfigure[with influence dynamics, $t=50$]{ \includegraphics[width=.222\linewidth]{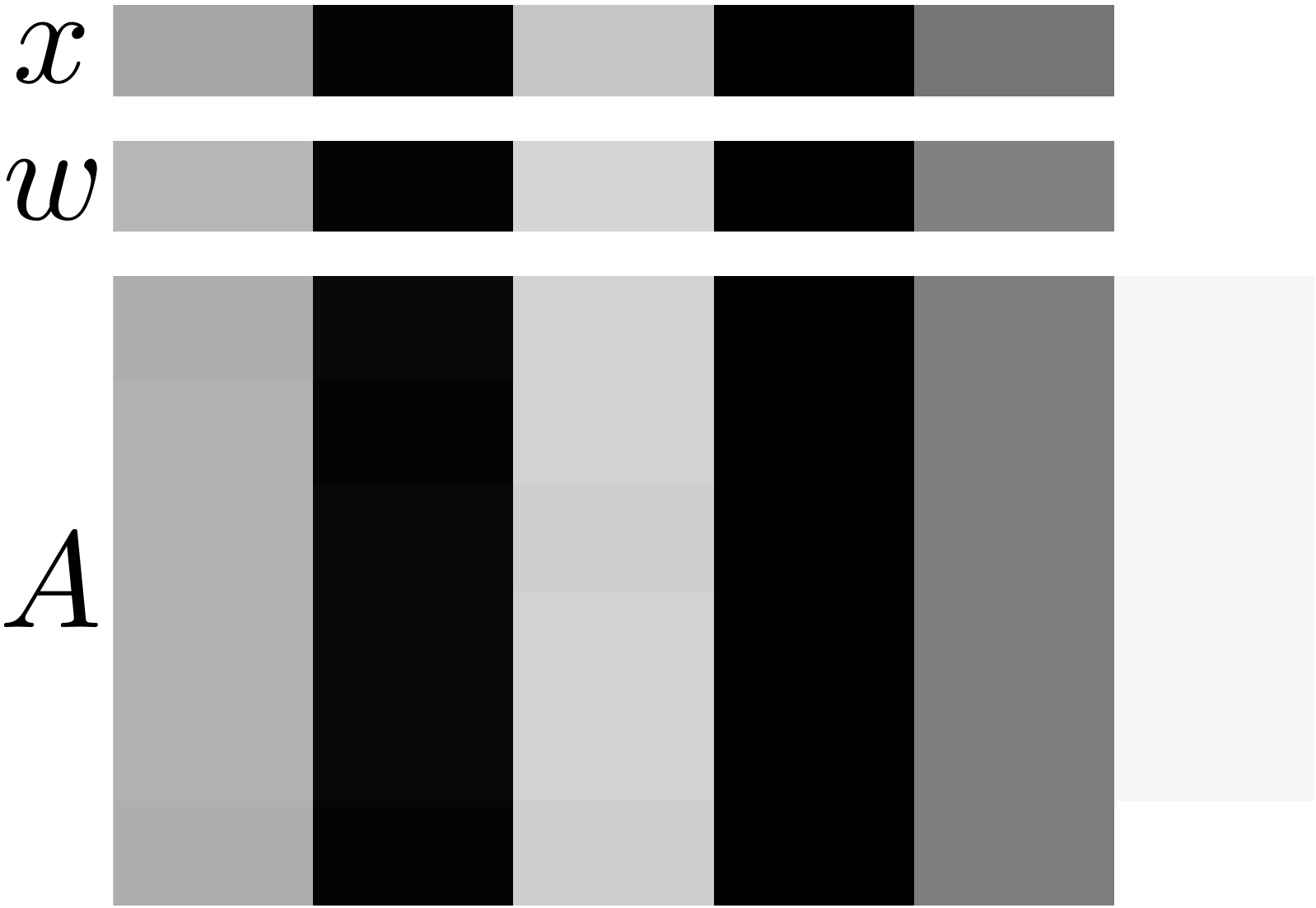}}
\caption{Examples of the assign/appraise (first row) and the assign/appraise/influence (second row) dynamics in which the assignment is based on the individuals' in-degree centrality. The assign/appraise dynamics does not achieve the collective learning, while the assign/appraise/influence dynamics does. }
\label{fig:visualization-assign-column}
\end{center}
\end{figure}

\emph{a) Variation in the assignment rule: task assignment based on degree centrality: }In Assumption~\ref{asmp:assign-rule}, the task assignment is based on the individuals' eigenvector centrality in the appraisal network. If we assume instead that the assignment is based on the individuals' normalized in-degree centrality in the appraisal network, i.e., $\bm{w}(t)=A^{\top}(t)\vectorones[n]/\vectorones[n]^{\top}A(t)\vectorones[n]$, then the numerical simulation, see Figure~\ref{fig:visualization-assign-column}, shows the following results: the team obeying the assign/appraise dynamics does not necessarily achieve collective learning, while the team obeying the assign/appraise/influence dynamics still achieves both collective learning and appraisal consensus. 

\begin{figure}
\begin{center}
\subfigure[no influence dynamics, $t=0$]{ \includegraphics[width=.222\linewidth]{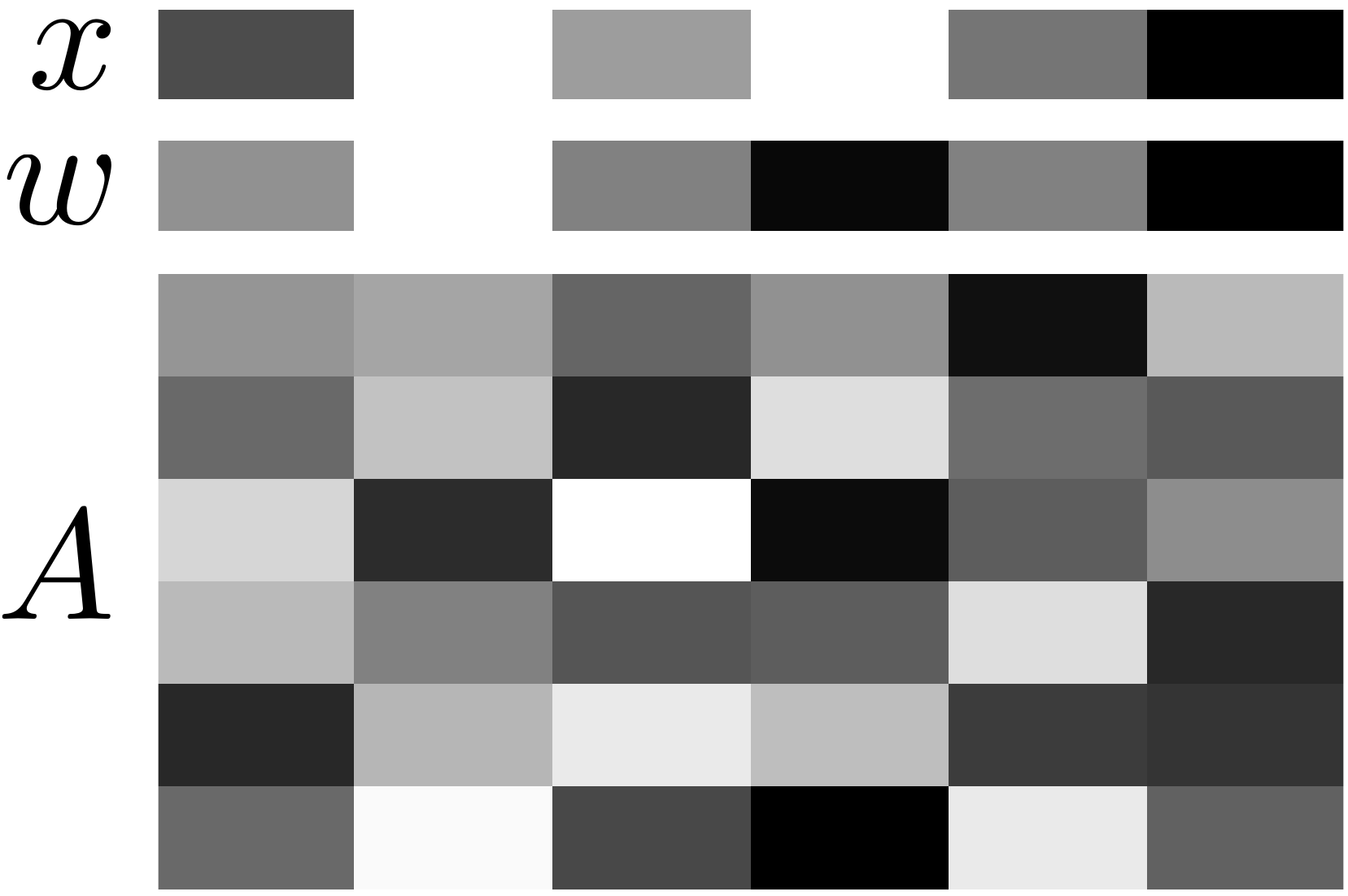}}
\subfigure[no influence dynamics, $t=1$]{ \includegraphics[width=.222\linewidth]{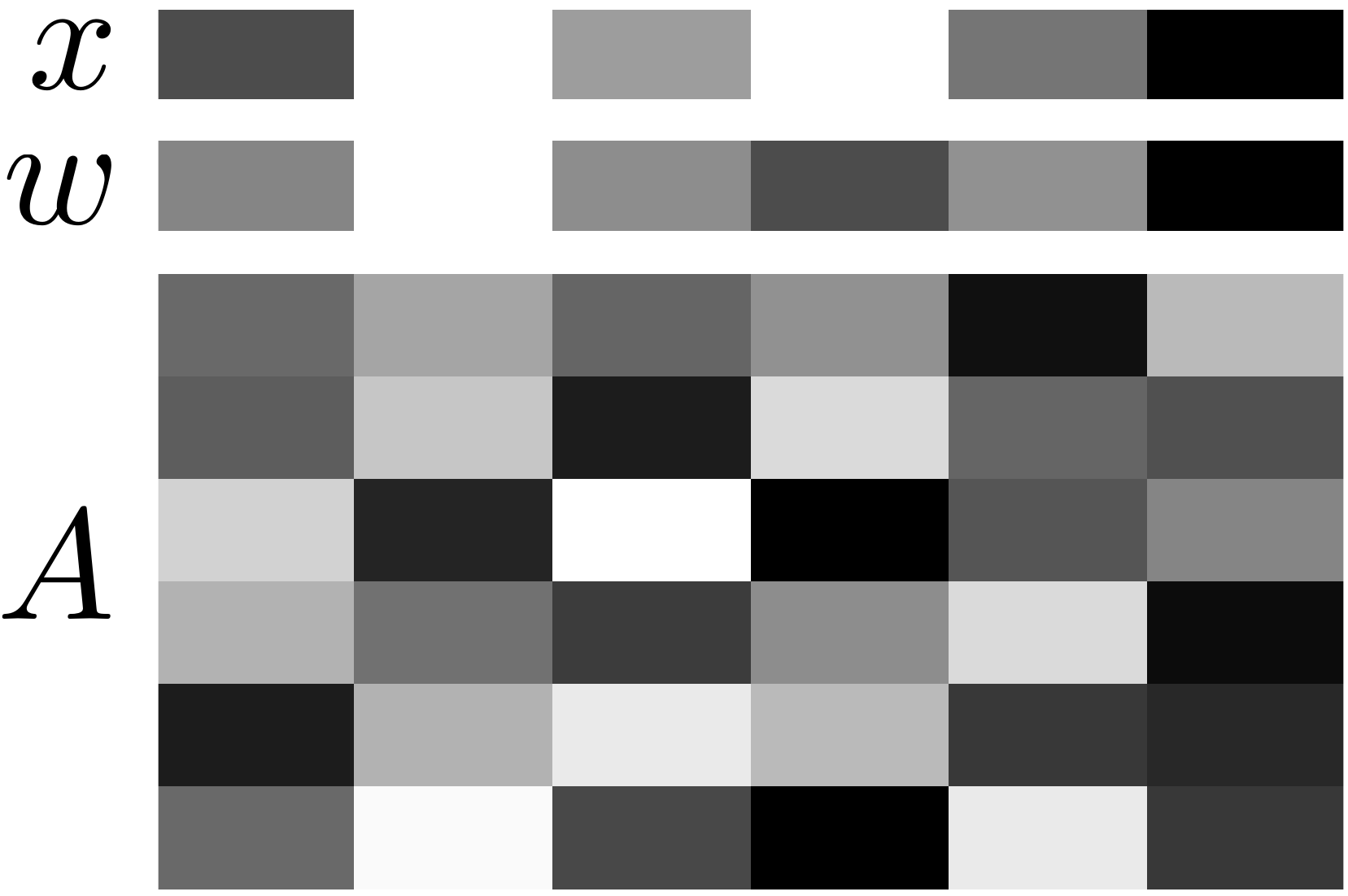}}
\subfigure[no influence dynamics, $t=5$]{ \includegraphics[width=.222\linewidth]{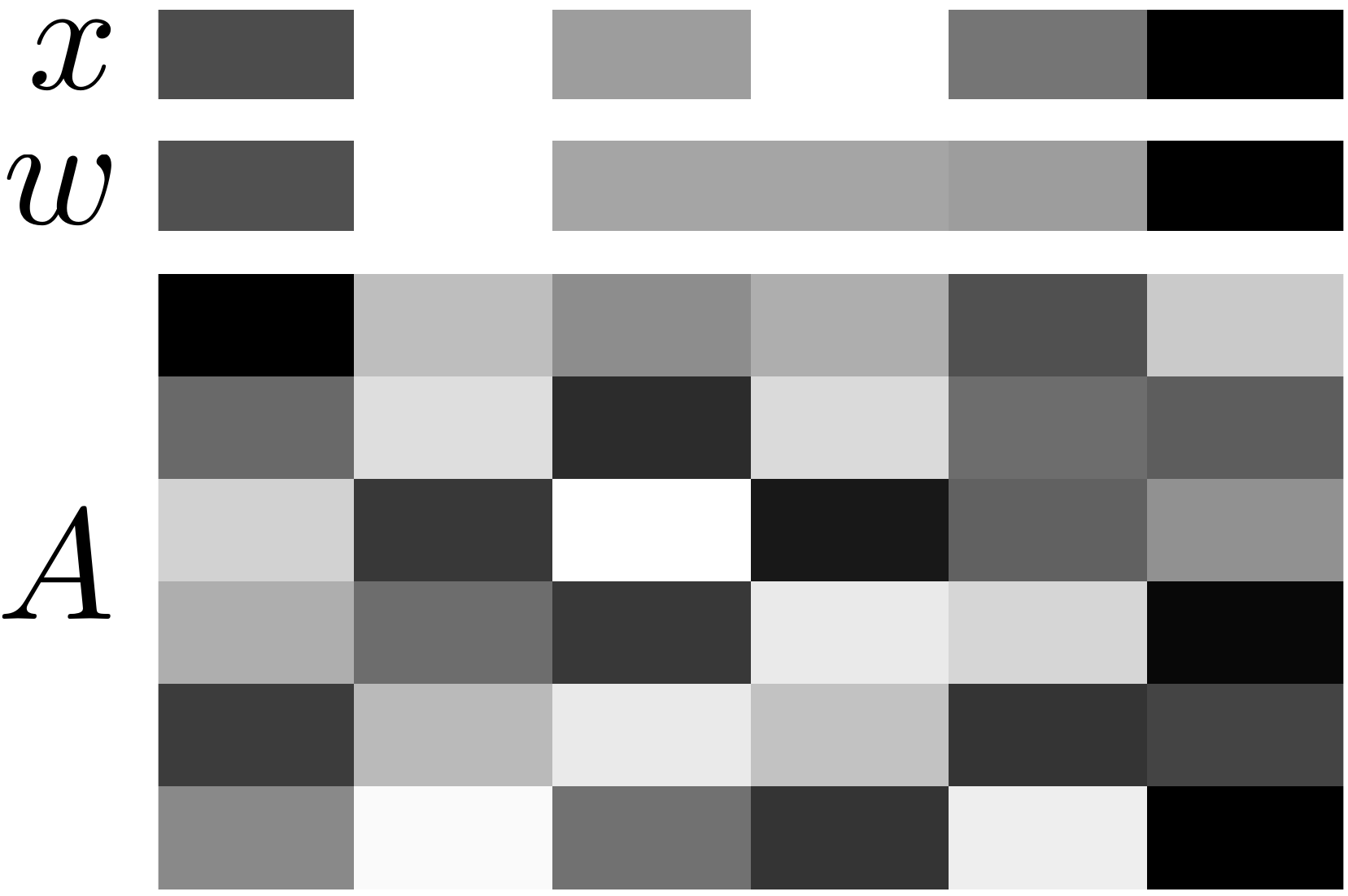}}
\subfigure[no influence dynamics, $t=10$]{ \includegraphics[width=.222\linewidth]{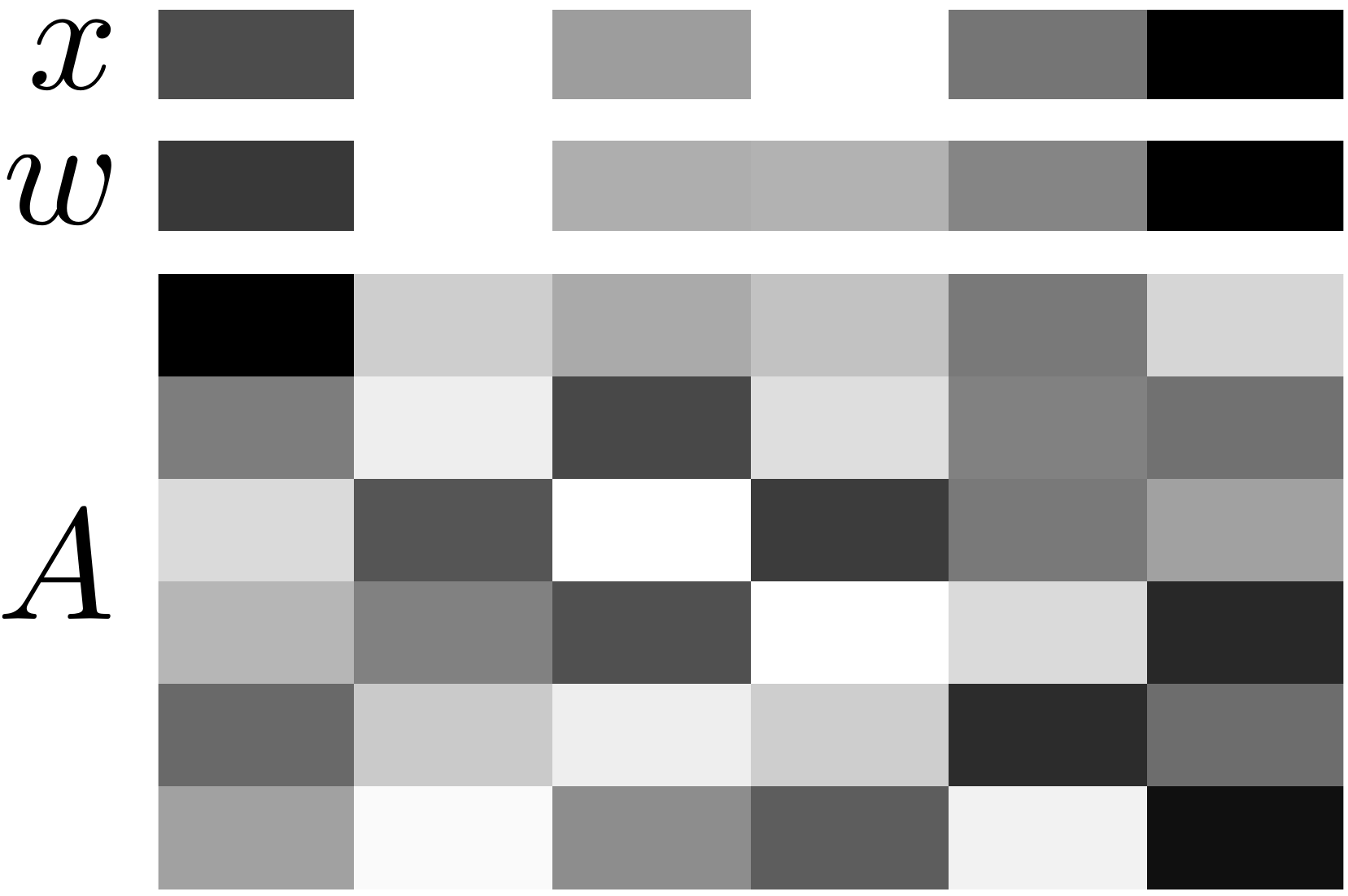}}
\subfigure[with influence dynamics, $t=0$]{ \includegraphics[width=.222\linewidth]{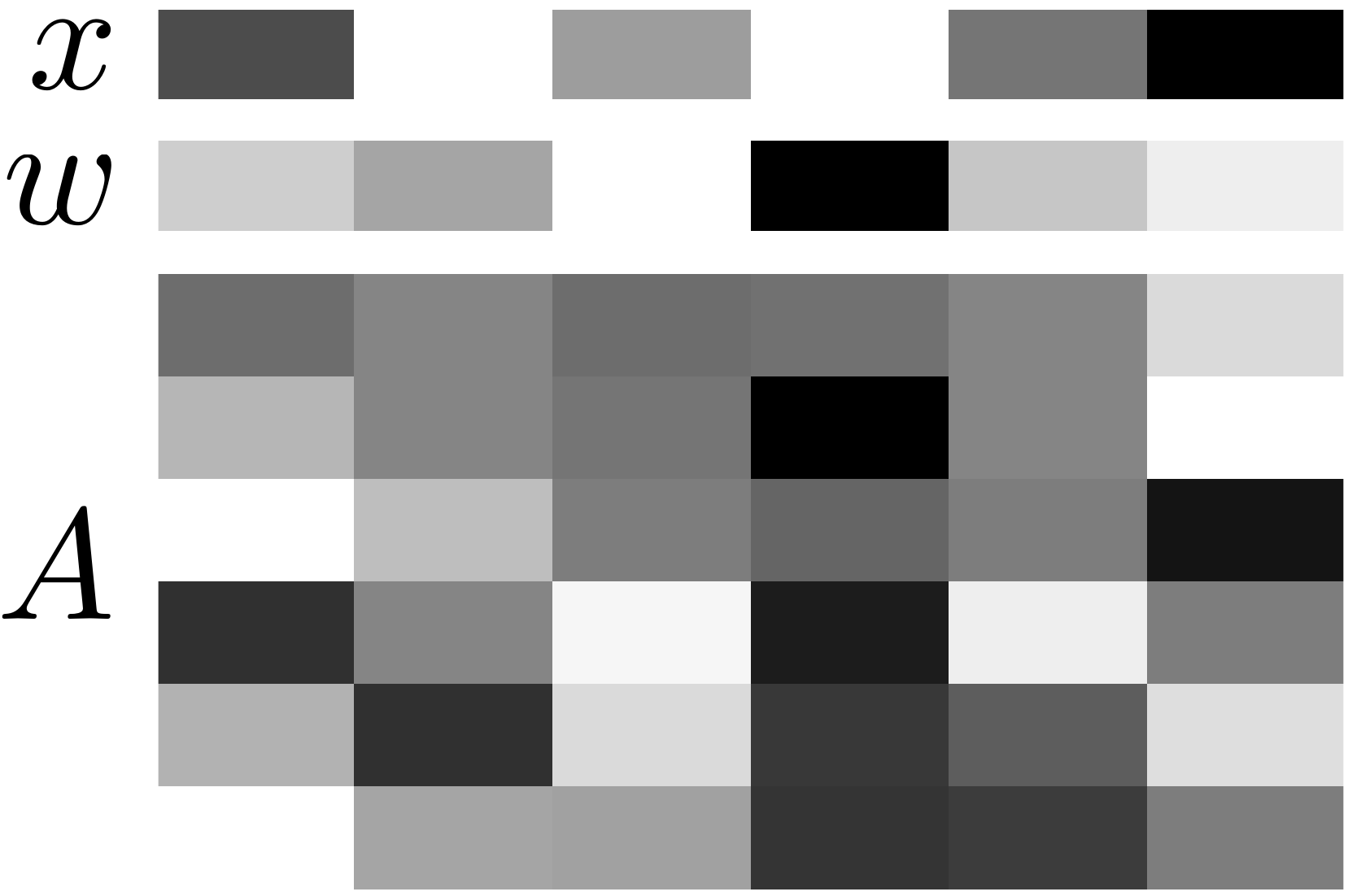}}
\subfigure[with influence dynamics, $t=5$]{ \includegraphics[width=.222\linewidth]{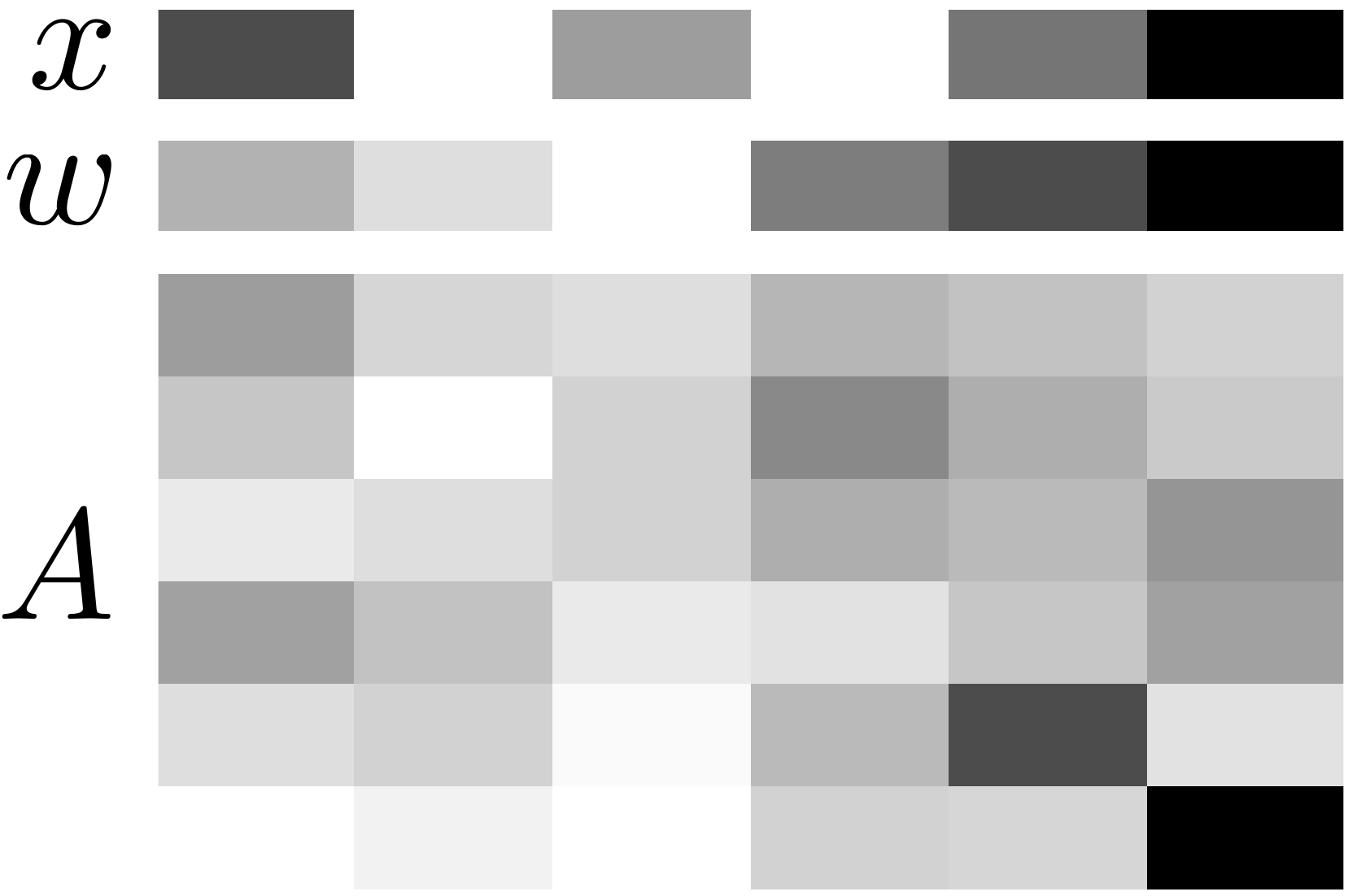}}
\subfigure[with influence dynamics, $t=50$]{ \includegraphics[width=.222\linewidth]{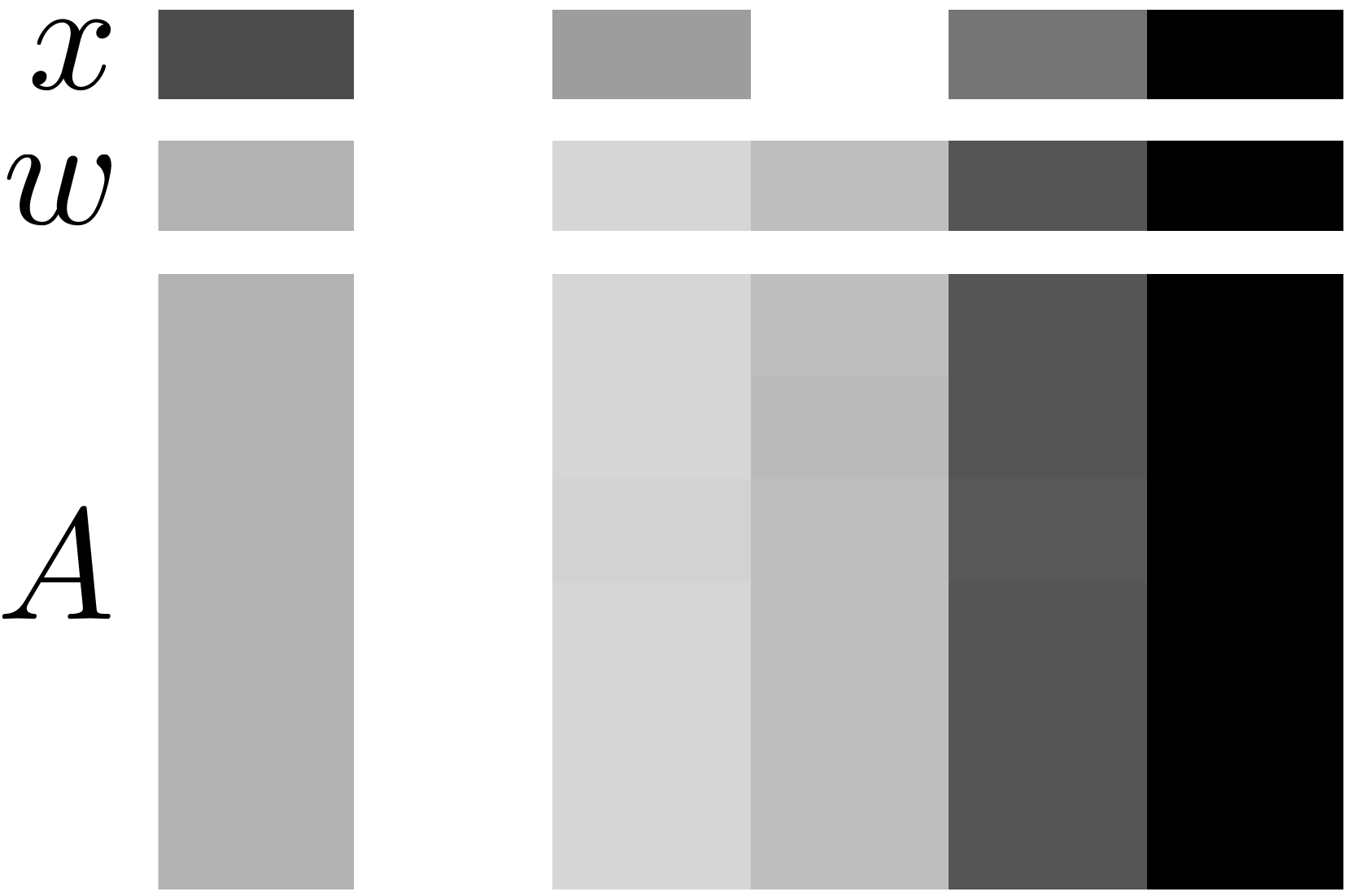}}
\subfigure[with influence dynamics, $t=60$]{ \includegraphics[width=.222\linewidth]{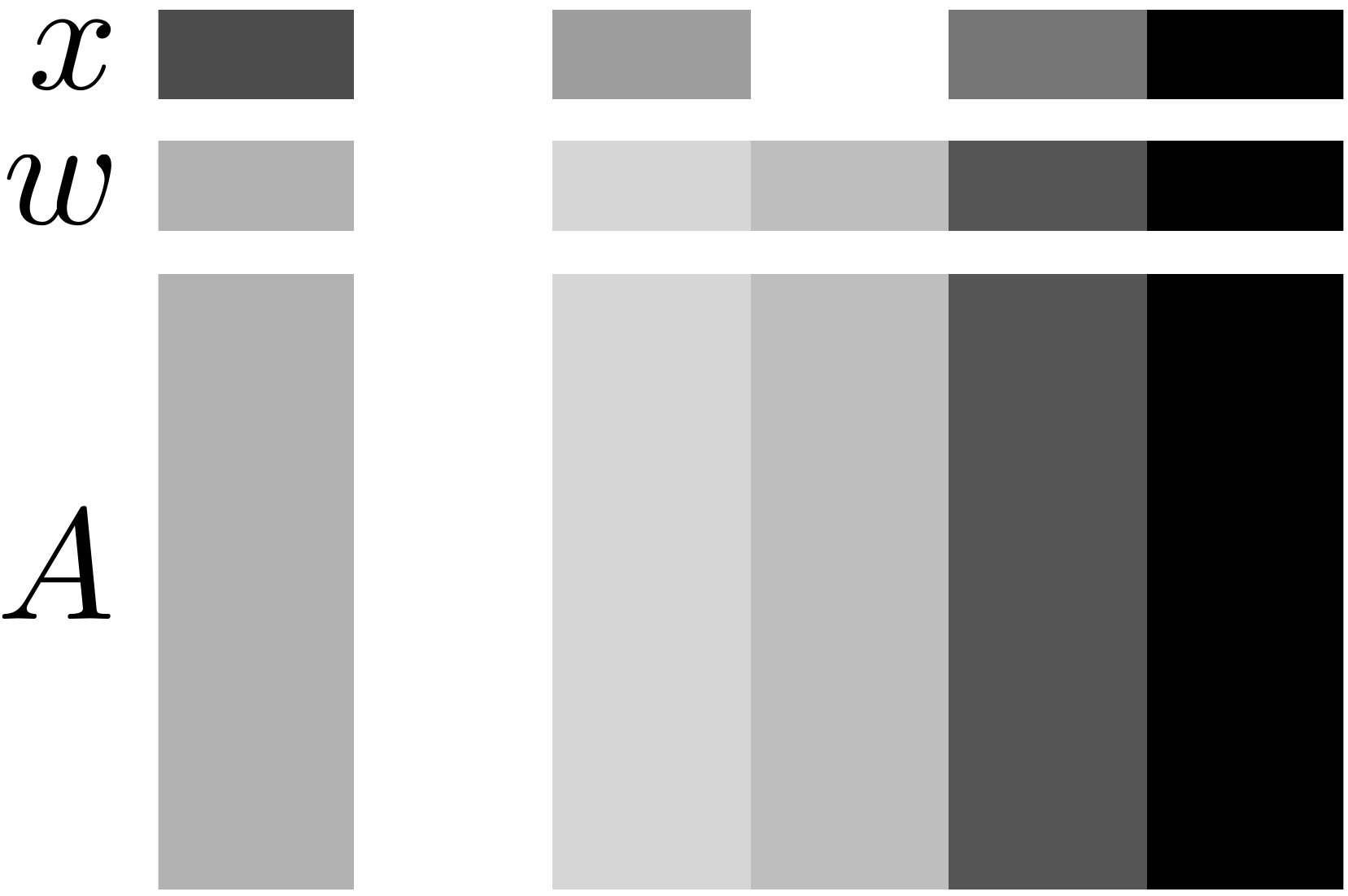}}
\caption{Examples of failure to learn with partial observation for a six-individual team. The figures in the first row correspond to the assign/appraise dynamics, in which the observation network is not strongly connected but contains a globally reachable node. The figures in the second row correspond to the assign/appraise/influence dynamics, in which the observation network does not contain a globally reachable node. In both cases, $A(t)$ converges but $\lim\limits_{t\to +\infty}\bm{w}(t)\neq \bm{x}$. }
\label{fig:visualization-partial-observation-fail}
\end{center}
\end{figure}

\emph{b) Variation in the appraise dynamics: partial observation of performance feedback: }According to Assumption~\ref{asmp:perf-feedback}, the observation network $G(M)$ determines the feedback signals received by each individual. If the observation network does not have the desired connectivity property, the individuals do not have sufficient information to achieve collective learning. Simulation results in Figure~\ref{fig:visualization-partial-observation-fail} shows that, if $G(M)$ is not strongly connected for the assign/appraise dynamics, or if $G(M)$ does not contain a globally reachable node for the assign/appraise/influence dynamics, the team does not necessarily achieve collective learning.

\begin{figure}
\begin{center}
\subfigure[$t=0$]{ \includegraphics[width=.222\linewidth]{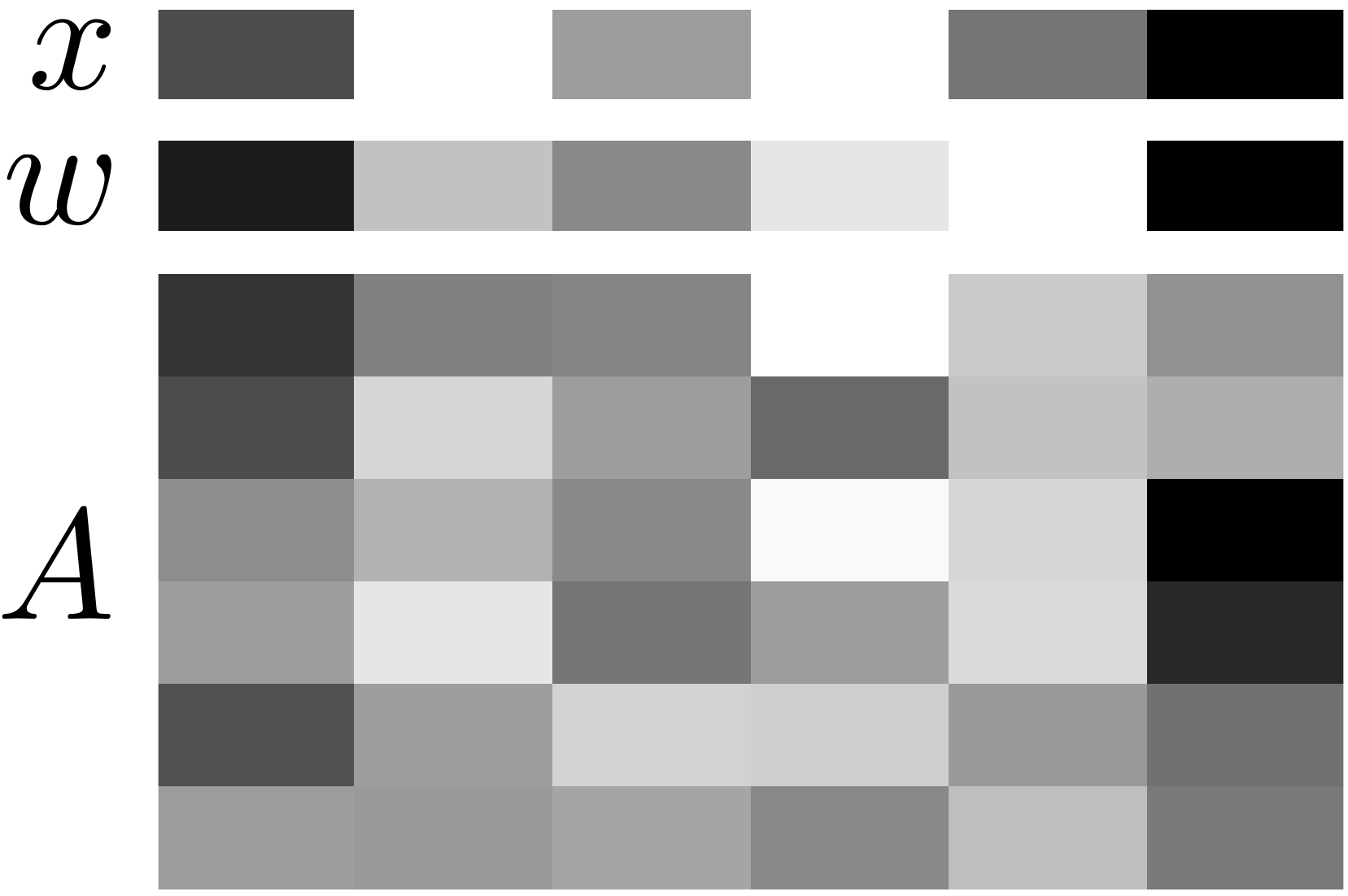}}
\subfigure[$t=1$]{ \includegraphics[width=.222\linewidth]{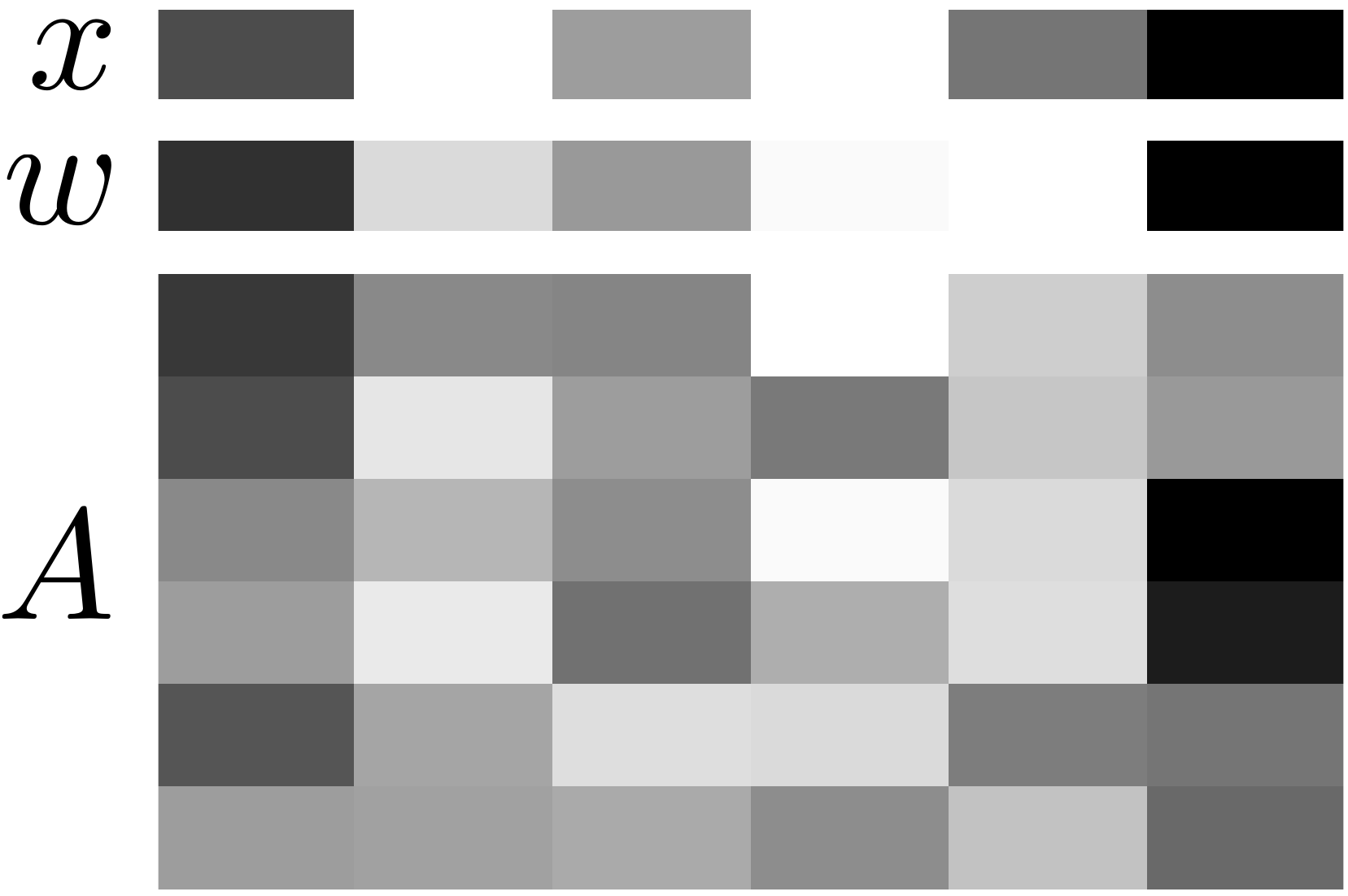}}
\subfigure[$t=5$]{ \includegraphics[width=.222\linewidth]{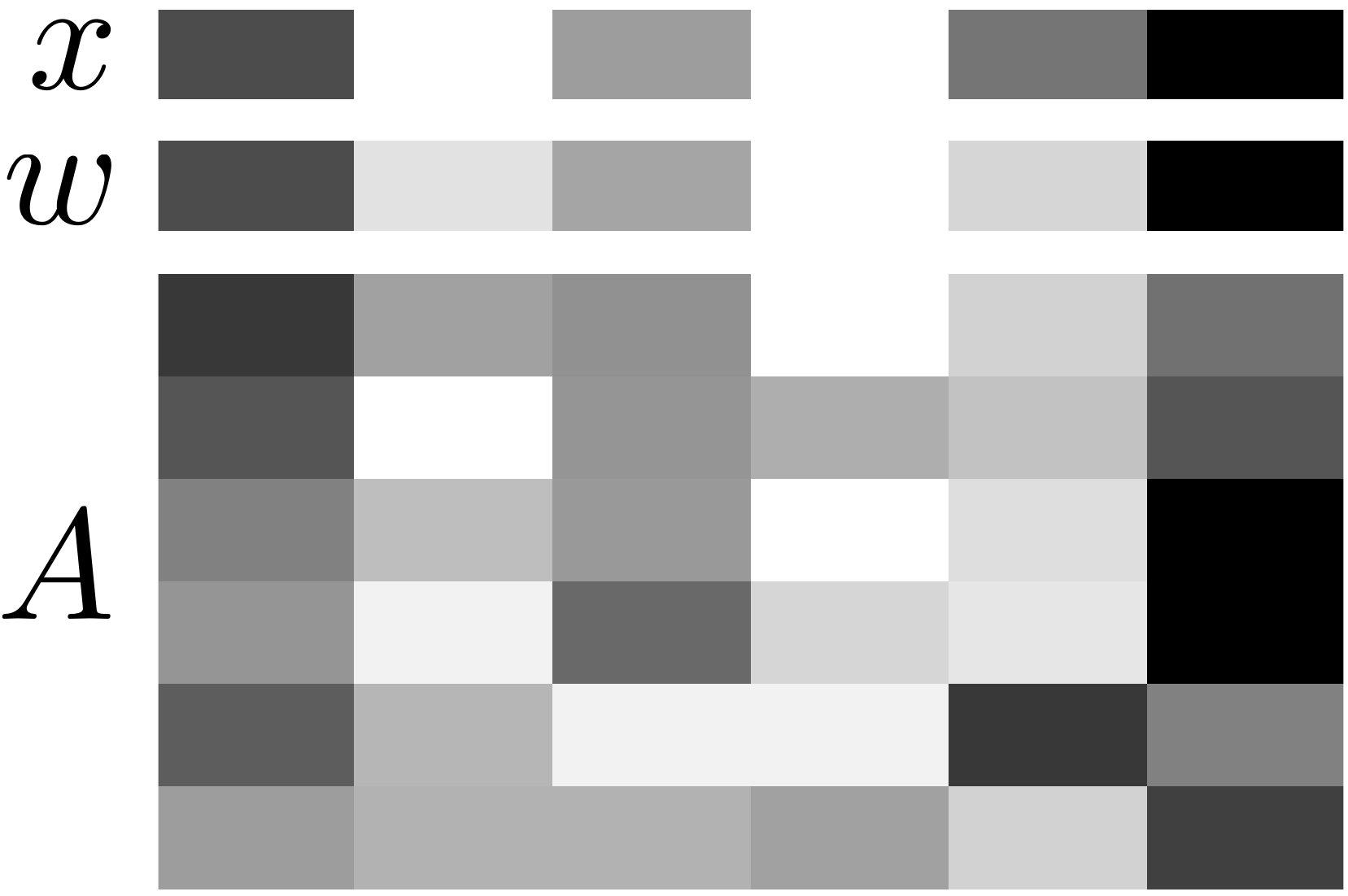}}
\subfigure[$t=10$]{ \includegraphics[width=.222\linewidth]{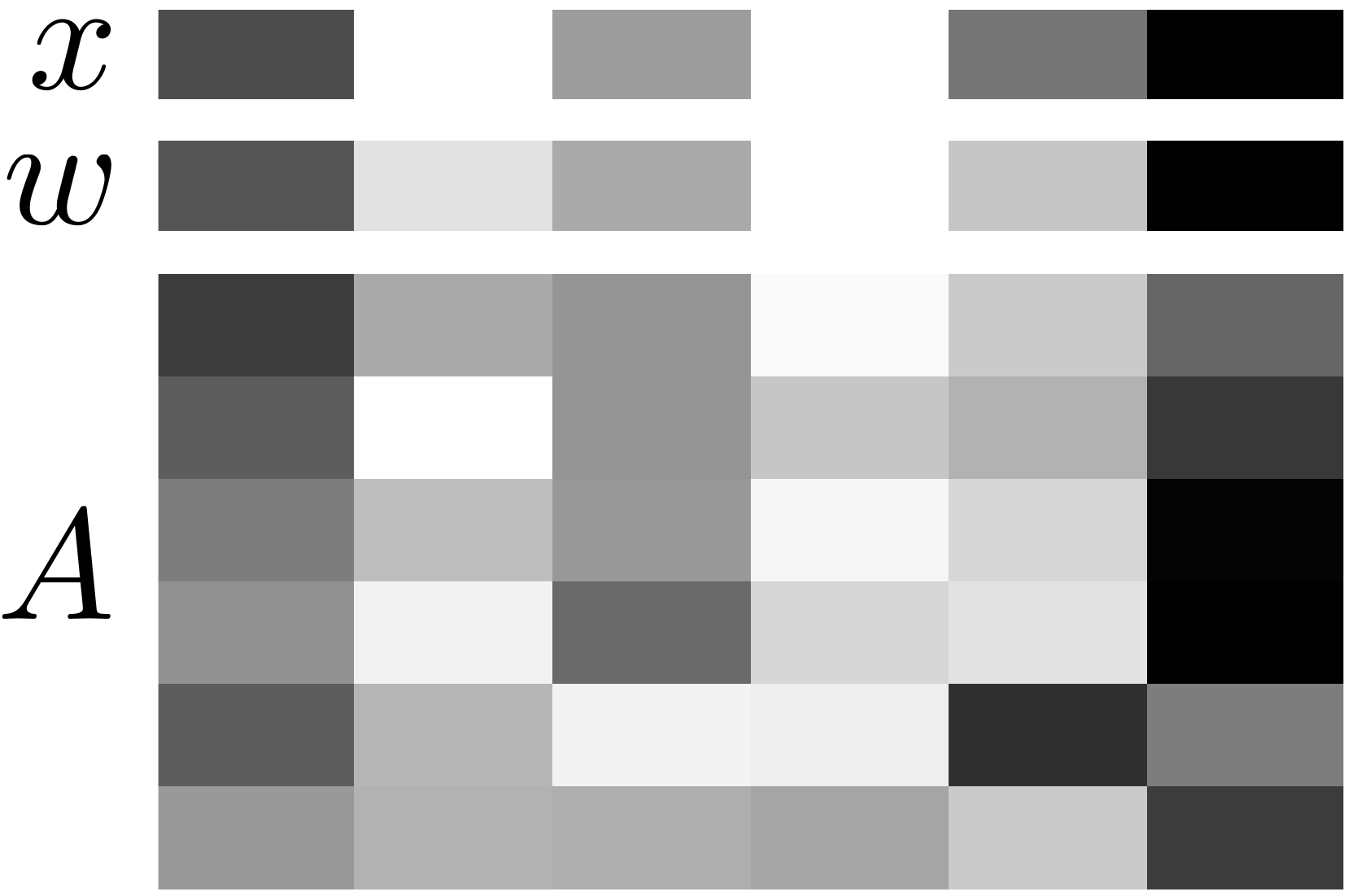}}
\caption{Example of the evolution of $A(t)$ and $\bm{w}(t)$ in the prejudice model with $n=6$. The darker the entry, the higher value it has. The simulation result shows that $A(t)$ converges but $\bm{w}(t)=\vleft\big( A(t) \big)$ does not necessarily converges to $\bm{x}$.}
\label{fig:visualization-FJ}
\end{center}
\end{figure}

\emph{c) Variation in the influence dynamics: prejudice model: }In Assumption~\ref{asmp:inf-dyn}, we assume that the individuals obey the DeGroot opinion dynamics. If we instead adopt the Friedkin-Johnsen opinion dynamics,
given by
\begin{equation*}
\subscr{F}{ave}( A,W ) = -\Lambda ( I_n-W )A + ( I_n-\Lambda)( A(0)-A),
\end{equation*} 
where $\Lambda=\diag(\lambda_1,\dots,\lambda_n)$ and each $\lambda_i$ characterizes individual $i$'s attachment to her initial appraisals. Numerical simulation, see Figure~\ref{fig:visualization-FJ}, shows that the team does not necessarily achieve collective learning. The Friedkin-Johnsen model captures the
social-psychological mechanism in which individuals show an attachment
to their initial opinions. This attachment is a cause of failure of
collective learning. 

\section{Further Discussion and Conclusion}
\subsection{Connections with TMS theory} 
\begin{figure}
\begin{center}
  \subfigure[$e^{-\mathcal{H}_1(\bm{w},\bm{x})}$]
  {\label{fig_evo_H1}\includegraphics[width=.435\linewidth]{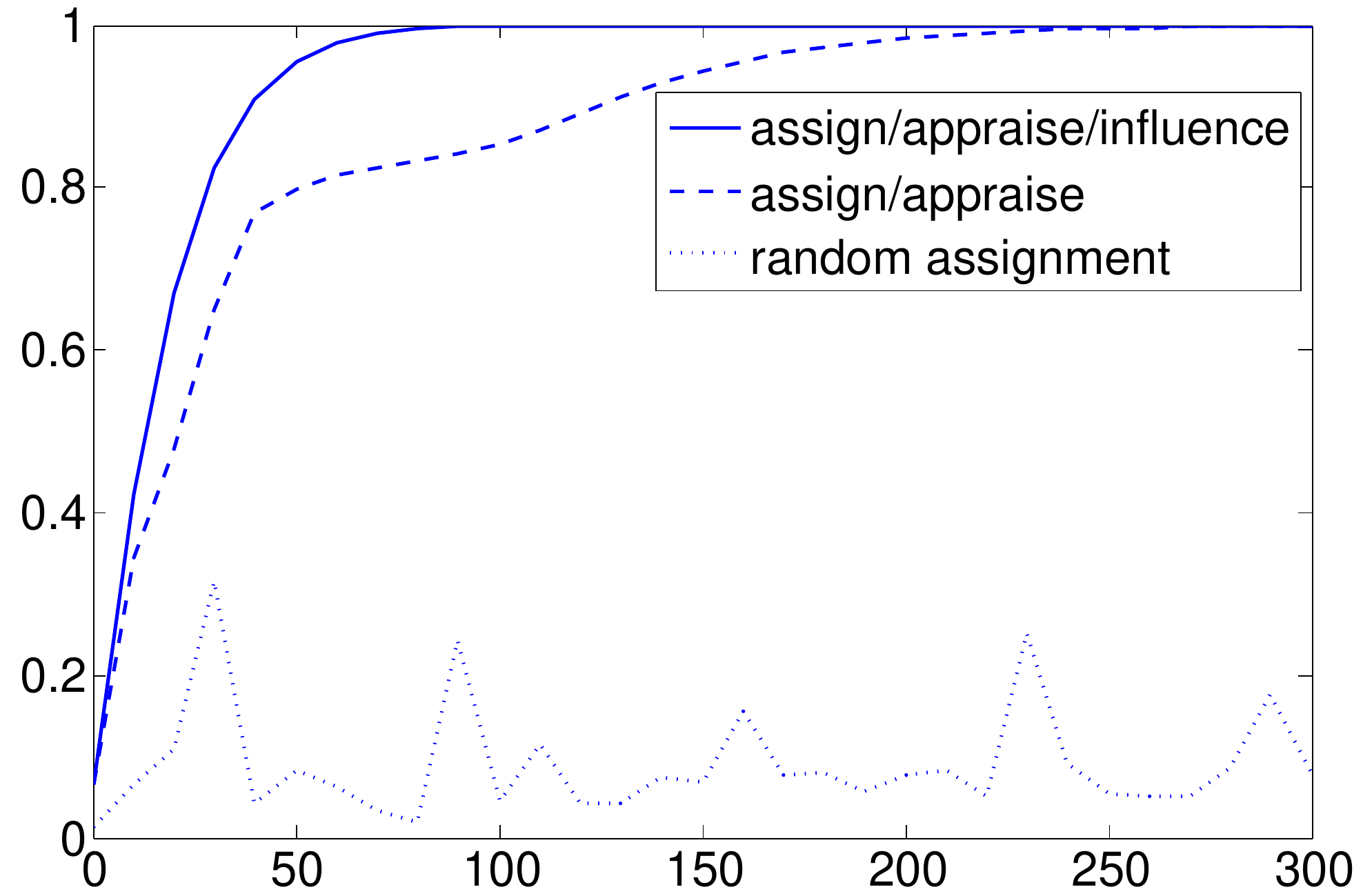}}
  %%  
  %\subfigure[$\frac{\mathcal{H}_2(\bm{w},\bm{x})}{\mathcal{H}_2(\bm{w}^*,\bm{x})}$]
  %{\label{fig_evo_H2}\includegraphics[width=.45\linewidth]{fig_evo_H2_cropped}}
  %%
  \subfigure[Number of non-transitive triads]
  {\label{fig_evo_non_tran_triads}\includegraphics[width=.44\linewidth]{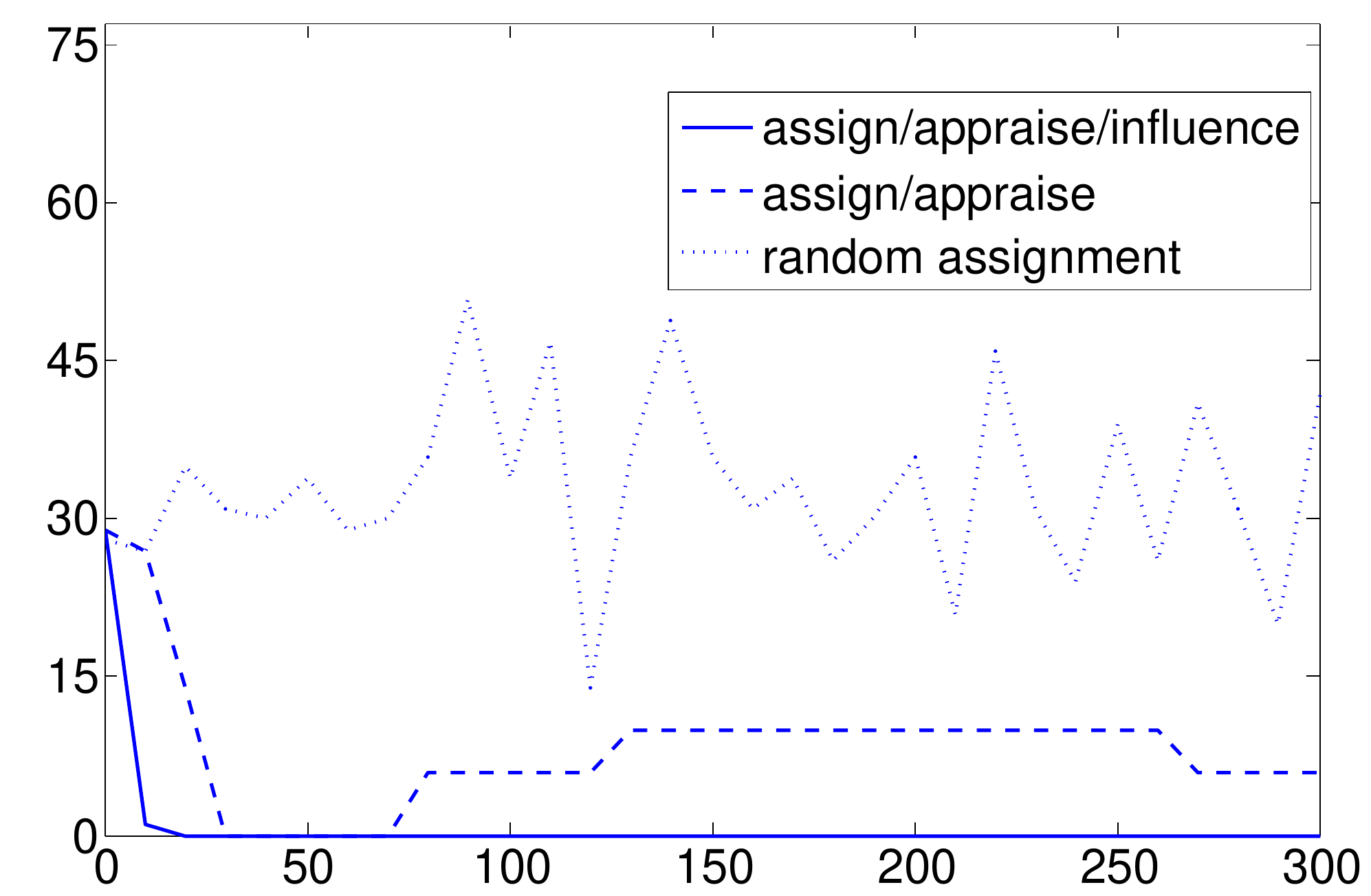}}
  %%
  %\subfigure[Second largest eigenvalue]
  %{\label{fig_evo_lambda2}\includegraphics[width=.45\linewidth]{fig_evo_lambda2_cropped}}
  %%
  \caption{Evolution of the measure of mismatch between assignment and individual skill levels, and the number of non-transitive triads in the comparative appraisal graph. The solid curves represent the team obeying the assign/appraise/influence dynamics. The dash curves represent the team obeying the assign/appraise dynamics. The dotted curves represent the team that randomly assign sub-task workloads.}
\label{fig_evo_team_perf}
\end{center}
\end{figure}

\emph{TMS structure:} As discussed in the introduction, one important
aspect of TMS is the members' shared understanding about who possess
what expertise. For the case of one-dimension skill, TMS structure is
approximately characterized by the appraisal matrix and thus the
development of TMS corresponds to the collective learning on
individuals' true skill levels. Simulation results in
Figure~\ref{fig_evo_team_perf} compare the evolution of some features
among the teams obeying the assign/appraise/influence model, the
assign/appraise model, and the team that randomly assigns the
sub-tasks, respectively. Figure~\ref{fig_evo_H1} shows that, for both
the assign/appraise/influence dynamics and the assign/appraise
dynamics, function $\mathcal{H}_1(\bm{w},\bm{x})$, as the measure of
the mismatch between task assignment and individual skill levels,
converge to $0$, which exhibits the advantage of a developing TMS.

\emph{Transitive triads: } As Palazzolo~\cite{ETP:05}
  points out, transitive triads are indicative of a well-organized TMS. The underlying logic is that inconsistency of interpersonal appraisals lowers the efficiency of locating the expertise and allocating the incoming information. In order to
reveal the evolution of triad transitivity in our models, we define an
unweighted and directed graph, referred to as the \emph{comparative
  appraisal graph} $\widetilde{G}(A)=(V,E)$, with $V=\{1,\dots,n\}$, as follows: for any $i,j\in V$, $(i,j)\in E$ if $a_{ij}\ge a_{ii}$, i.e., if individual $i$ thinks $j$ has no lower skill level than $i$ herself. We adopt the standard notion
of triad transitivity and use the number of non-transitive triads as the indicator of inconsistency in a team. Figure~\ref{fig_evo_non_tran_triads} shows that, the non-transitive triads vanish in the team obeying the assign/appraise/influence dynamics, but persist in the teams obeying the assign/appraise dynamics or just randomly assigning subtasks.
%for the team obeying the assign/appraise/influence dynamics, the number of non-transitive triads tends to $0$ as $t\rightarrow +\infty$, while non-transitive triads persist in the team obeying the assign/appraise dynamics and the team randomly assigning subtasks. 
%Another indicator
%of the team's inner cognitive consistency is the second largest
%eigenvalue (in magnitude) of the influence matrix, which determines
%the convergence rate for DeGroot opinion dynamics. The larger the
%second eigenvalue, the slower the group converges to
%consensus. Figure~\ref{fig_evo_lambda2} reveals that, for the team
%obeying the assign/appraise/influence dynamics, the second largest
%eigenvalue converges to zero. This desirable behavior occurs neither
%in the assign/appraise dynamics nor in the randomly-assigning teams.

\subsection{Minimum condition for convergence to optimal assignment}
Analysis of the asymptotic behavior of assign/appraise dynamics, assign/appraise/influence dynamics and partial observation model can be interpreted as the exploration of the most relaxed condition for the convergence to optimal task assignment, concluded as follows:
\begin{enumerate}
\item Each individual only needs to know, as feedback, the difference between her own performance and the quality of some parts of the entire task, but do not need to know whom they are compared with;
\item The individuals can have heterogeneous but strictly positive sensitivities to the performance feedback;
\item The exchange of opinions on the interpersonal appraisals is not necessary;
\item With opinion exchange, the observation network with one globally reachable node is sufficient for the convergence to optimal assignment;
\item Without opinion exchange, strongly connected observation network is sufficient for the convergence to optimal assignment. 
\end{enumerate} 

\subsection{Conclusion}
This paper proposes a baseline model: the centralized manager dynamics, and two elaborative multi-agent models on team dynamics: the assign/appraise and the assign/appraise/influence dynamics. We reveal insightful connections between our models and the replicator dynamics in evolutionary game theory. For the multi-agents models, the appraisal network is modeled as a team's basic inner structure. the appraisal network generates the team's task assignments, and the mismatch between the assignment and individuals' true skill levels might indicate the level of team performance. By theoretical analysis we investigate the evolution of appraisal network, and relate its asymptotic behavior, i.e., the convergence to optimal assignment and the appraisal consensus, with the individuals' feedback signal structures. We then propose some variations of the baseline models, in which some sociological and psychological
mechanisms, e.g., the assignment by in-degree centrality in the appraisal network, the prejudice in opinion dynamics, and the lack of the desired connectivity property for the observation network, cause the failure of collective learning. In addition, we show that the qualitative predictions made by our models are consistent with TMS theory in organization science.

%This paper proposes three baseline models of team learning: the
%manager dynamics, the assign/appraise dynamics and the
%assign/appraise/influence dynamics. The first model is centralized,
%whereas the second and third are social multi-agent models. We reveal
%insightful connections between our models and the replicator dynamics
%in evolutionary game theory. We analyse collective learning and
%appraisal consensus for the assign/appraise and the
%assign/appraise/influence model and relate the models' asymptotic behavior to the structure of the %observation network.  We then propose some variations of
%the baseline models, in which some sociological and psychological
%mechanisms, e.g., the assignment by in-degree centrality on the appraisal network, the prejudice in opinion dynamics, and the lack of the desired connectivity property for the observation network, cause the failure of collective learning. In addition, we show that the qualitative
%predictions made by our models are consistent with TMS theory in
%organization science.

\begin{appendix}
\subsection{Proof for Theorem~\ref{thm:manager-dyn}}\label{proof:manager-dyn}
The vector form of equation~\eqref{eq:manager-dyn} is written as
\begin{equation}\label{eq:manager-dyn-vector}
\dot{\bm{w}}=\diag(\bm{w})\left( \bm{p}(\bm{w})-\bm{w}^{\top}\bm{p}(\bm{w})\vectorones[n] \right).
\end{equation}
Left multiply both sides by $\vectorones[n]^{\top}$. We get $d(\vectorones[n]^{\top}\bm{w})/dt=0$.

Since the function $f$ is continuously differentiable, the right-hand side of equation~\eqref{eq:manager-dyn-vector} is continuously differentiable and locally Lipschitz in $\intDeltan$. Define 
\begin{equation*}
V(\bm{w})=-\sum_{i=1}^n x_i \log \frac{w_i}{x_i}.
\end{equation*}
We have $V(\bm{w})\ge 0$ for any $\bm{w}\in \Delta_n$, due to the concavity of $\log$ function, and $V(\bm{w})=0$ if and only if $\bm{w}=\bm{x}$. Moreover, since $V(\bm{w})$ is continuously differentiable in $\bm{w}$, the level set $\{\bm{w}\in \intDeltan\,|\,V(\bm{w})=\xi\}$ is a compact subset of $\intDeltan$. Along the trajectory,
\begin{align*}
\frac{dV(\bm{w})}{dt} = -\sum_{i\in \theta_1(\bm{w})}(x_i-w_i)f(x_i/w_i)-\sum_{i\in \theta_2(\bm{w})} (x_i-w_i)f(x_i/w_i)< 0,
\end{align*} 
where $\theta_1(\bm{w})=\{i\,|\,x_i\ge w_i\}$ and $\theta_2(\bm{w})=\{i\,|\,x_i<w_i\}$. This concludes the proof for the invariant set and the asymptotic stability of $\bm{w}^*=\bm{x}$, and one can infer, from the inequality above, that $\bm{w}^*=\bm{x}$ is the ESS for the evolutionary game with the payoff function $\pi_i(\bm{w})=f(x_i/w_i)$. Moreover, since $V(\bm{w})\to +\infty$ as $\bm{w}$ tends to the boundary of $\Delta_n$, the region of attraction is $\intDeltan$.

\subsection{Justifications of Assumption~\ref{asmp:assign-rule} on task assignment}\label{justification:assign-rule}
We provide some justification of Assumption~\ref{justification:assign-rule} that, the task assignment $\bm{w}$ is given by $\bm{w}=\vleft(A)$. Firstly, the entries of $\vleft(A)$
correspond to the individuals' eigenvector centrality in the appraisal
network and thus reflect how much each individual is appraised by the
team. Secondly, there is a natural way in which the interpersonal
interactions lead to the assignment $\bm{w}(t)=\vleft\big( A(t)
\big)$. Assume that for any incoming task, at step $k=0$, each
individual $i$ evenly get $1/n$ workload, and at each step, each
individual $i$ passes $a_{ij}(t)$ of the workload she currently
possesses to each individual $j\in\until{n}$, until the substask
distribution reaches the steady state. Denote by $q_i(k)$ the fraction
of the workload at individual $i$, at step $k$, this workload
distribution process is given by $\bm{q}(k+1)=A(t)^\top \bm{q}(k)$.
%\begin{equation*}
%\bm{q}(k+1)=A(t)^\top \bm{q}(k).
%\end{equation*}
According to Perron-Frobenius theorem, $\bm{q}(k)$
converges to $\vleft\big( A(t) \big)$; Thirdly, our eigenvector
assignment rule has the following natural property: in a team deprived
of performance feedback as the information inflow, the team's task
assignment does not change. In other words, along the
assign/appraise/influence dynamics with
$\frac{1}{\subscr{\tau}{app}}=0$, vector
$\vleft\big(A(t)\big)$ remains unchanged.  These arguments justify
Assumption~\ref{asmp:assign-rule}; recall also
Section~\ref{sec:variations}(a) with a numerical evaluation of a
different assignment rule.

\subsection{Proof for Theorem~\ref{thm:ass/app-finite-time}}\label{proof:ass/app-finite-time}
Before the proof, we state a useful lemma summarized from the argument on Page 62-67 of~\cite{JHW:65}.
\begin{lemma}[Continuity of eigenvalue and eigenvector]\label{lem:continuity-eigenpair}
Suppose $A,B\in \mathbb{R}^{n\times n}$ satisfy $\lvert a_{ij}\rvert <1$ and $\lvert b_{ij}\rvert <1$ for any $i,j\in \{1,\dots,n\}$. For sufficiently small $\epsilon >0$,
%Let $A$ and $B$ be two $n\times n$ matrices satisfying $\lvert a_{ij}\rvert <1$ and $\lvert b_{ij}\rvert <1$ for any $i,j\in \{1,\dots,n\}$. For sufficiently small $\epsilon >0$,
\begin{enumerate}
\item the eigenvalues $\lambda$ and $\lambda ^{'}$ of $A$ and $(A+\epsilon B)$, respectively, can be put in one-to-one correspondence so that $\lvert \lambda ^{'} -\lambda \rvert<2(n+1)^2(n^2\epsilon)^{\frac{1}{n}}$;
\item if $\lambda$ is a simple eigenvalue of $A$, then the corresponding eigenvalue $\lambda(\epsilon)$ of $A+\epsilon B$ satisfies $\lvert \lambda(\epsilon) - \lambda \rvert=O(\epsilon)$;
\item if $\bm{v}$ is an eigenvector of $A$ associated with a simple eigenvalue $\lambda$, then the eigenvector $\bm{v}(\epsilon)$ of $A+\epsilon B$ associated with the corresponding eigenvalue $\lambda(\epsilon)$ satisfies $\lvert v_i(\epsilon)-v_i\rvert=O(\epsilon)$ for any $i\in \{1,\dots,n\}$.
\end{enumerate}
\end{lemma}
\smallskip

\emph{Proof of Theorem~\ref{thm:ass/app-finite-time}:} In this proof, we extend the definition of $\vleft(A)$ to the normalized entry-wise positive left eigenvector, associated with the eigenvalue of $A$ with the largest magnitude, if such an eigenvector exists and is unique. According to Perron-Frobenius theorem and Lemma~\ref{lem:continuity-eigenpair}, vector $\vleft(A)$, as long as well-defined, depends continuously on the entries of $A$. %Therefore, there exists a sufficiently small $\tau>0$ such that $A(t)$ is continuously differentiable to $t$ and $\vleft\big( A(t) \big)$ is well-defined for all $t\in [0,\tau]$. 
Therefore, for system~\eqref{eq:ass/app-dyn-matrix}, there exists a sufficiently small $\tau>0$ such that $A(t)$ and $\bm{w}(t)$ are well-defined and continuously differentiable at any $t\in [0,\tau]$, and, moreover, $p_i\big( \bm{w}(t) \big)-\sum_k m_{ik} p_k\big( \bm{w}(t) \big)$ remains finite. Therefore, for any $t\in [0,\tau]$ and $i,j\in\{1,\dots,n\}$, $a_{ij}(t)>0$ if $a_{ij}(0)>0$; $a_{ij}(t)=0$ if $a_{ij}(0)=0$, and thus $A(t)$ is row-stochastic and primitive for any $t\in [0,\tau]$.

For any $i\in \{1,\dots,n\}$, there exists $k\neq i$ such that $a_{ik}(0)>0$. According to equation~\eqref{eq:app-dyn}, 
%for any $j\in \{1,\dots,n\}\setminus\{i,k\}$,
\begin{equation*}
\frac{da_{ij}(t)}{da_{ik}(t)}=\frac{a_{ij}(t)}{a_{ik}(t)},\text{ }\forall t\in [0,\tau],\text{ }\forall j\in \{1,\dots,n\}\setminus \{i,k\},
\end{equation*}  
which leads to $a_{ij}(t)/a_{ik}(t)=a_{ij}(0)/a_{ik}(0)$.
%\begin{equation}\label{eq_assign_appraise_a_ij_a_ik_constant_ratio}
%\frac{a_{ij}(t)}{a_{ik}(t)}=\frac{a_{ij}(0)}{a_{ik}(0)}.
%\end{equation}
Let $C$ be an $n\times n$ matrix with the entries $c_{ij}$ defined as: (i) $c_{ii}=0$ for any $i\in \{1,\dots,n\}$; (ii) $c_{ij}=a_{ij}(0)\big/\big( 1-a_{ii}(0) \big)$ for any $j\neq i$. One can check that $C$ is row-stochastic and $A(t)$ is given by equation~\eqref{eq:assign/appraise-struc-A(t)}, for any $t\in [0,\tau]$, where $\bm{a}(t)=\big( a_1(t),\dots,a_n(t) \big)^{\top}$ with $a_i(t)=a_{ii}(t)$. Since the digraph, with $C$ as the adjacency matrix, has the same topology with the digraph associated with $A(0)$, matrix $C$ is irreducible and $\bm{c}=\vleft(C)$ is well-defined. 

Since the matrix $A(t)$ has the structure given by~\eqref{eq:assign/appraise-struc-A(t)}, according to Lemma~2.2 in~\cite{PJ-AM-NEF-FB:13d}, for any $t\in [0,\tau]$, 
\begin{equation*}
w_i(t)=\frac{c_i}{1-a_i(t)}\Big / \sum_k\frac{c_k}{1-a_k(t)}.
\end{equation*}

Therefore, for any $t\in [0,\tau]$,
\begin{equation*}
p_i\big( \bm{w}(t) \big)=f\left( \frac{x_i}{c_i}\big( 1-a_i(t) \big)\sum_k w_k(t)\frac{c_k}{1-a_k(t)} \right).
\end{equation*}
According to equation~\eqref{eq:app-dyn}, $\dot{a}_j(t)\le 0$ for any $j\in \argmin_k \frac{x_k}{c_k}\big( 1-a_k(t) \big)$. Therefore, $\argmin_k \frac{x_k}{c_k}\big( 1-a_k(t) \big)$ is increasing, and similarly, $\argmax_k \frac{x_k}{c_k}\big( 1-a_k(t) \big)$ is decreasing with $t$, which implies that, the set
\begin{align*}
\Omega_A \big( A(0) \big)=\Big{\{}A\in \mathbb{R}^{n\times n}\,\Big|\, A=\diag(\bm{a})+(I-\diag(\bm{a}))C,\text{ }\text{ }\, 0\le a_i \le 1-\frac{c_i}{x_i}\min_k \frac{x_k}{c_k}\big( 1-a_{kk}(0) \big),\forall i \Big{\}}
\end{align*}
is a compact positive invariant set for system~\eqref{eq:ass/app-dyn-matrix}, as long as $A(0)$ is row-stochastic, irreducible and has strictly positive diagonal. Moreover, one can check that, for any $A\in \Omega_A\big( A(0) \big)$, $\bm{w}=\vleft(A)$ is well-defined and strictly lower (upper resp.) bounded from $0$ ($1$ resp.). 
%\begin{equation}\label{eq:ass/app-bound-wi}
%0<\frac{c_i}{c_i+\sum_{k\neq i}\mu_k} \le w_i \le \frac{\mu_i}{\mu_i+\sum_{k\neq i} c_k} <1,
%\end{equation}
%for any $i\in \{1,\dots,n\}$, where $\mu_i=x_i\big/\min_l \frac{x_l}{c_l}\big( 1-a_i(0) \big)$, for all $k\in \{1,\dots,n\}$. 
Therefore, the solution $A(t)$ is extensible to all $t\in [0,+\infty)$ and equations~\eqref{eq:assign/appraise-struc-A(t)} and~\eqref{eq:ass/app-reduced} hold for any $t\in [0,+\infty)$. Moreover, since $p_i\big( \bm{w}(t) \big)-\sum_k m_{ik}p_k\big( \bm{w}(t) \big)$ remains bounded, we have $a_{ij}>0$ if $a_{ij}(0)>0$ and $a_{ij}(t)=0$ if $a_{ij}(0)=0$. This concludes the proof for (i) - (iv).

For statement~(v), differentiate both sides of the equation $\bm{w}^{\top}(t)A(t)=\bm{w}^{\top}(t)$ and substitute equation~\eqref{eq:ass/app-dyn-matrix} into the differentiated equation. We obtain
\begin{equation*}
\begin{split}
( \bm{w}^{\top}\!\diag( \bm{p}(\bm{w})-M\bm{p}(\bm{w}))A_d - \frac{d\bm{w}^{\top}}{dt}\!)(I_n-A)=\vectorzeros[n]^{\top},
\end{split}
\end{equation*}
where time index $t$ is omitted for simplicity. Equation~\eqref{eq:ass/app-dyn-w(t)} in~(v) is obtained due to $\bm{w}^{\top}(t)\vectorones[n]=1$.

%\begin{multline*}
%\bigg( \bm{w}^{\top}(t)  \diag\!\Big( \bm{p}\big( \bm{w}(t) \big)-M\bm{p}\big( \bm{w}(t) \big)\Big)\subscr{A}{d}(t)\\
% -\frac{d\bm{w}^{\top}(t)}{dt}\bigg) \big( I_n-A(t) \big)=\vectorzeros[n]^{\top}.
%\end{multline*}

\subsection{Proof for Theorem~\ref{thm:ass-app-asym-behav}}\label{proof:ass-app-asym-behav}
We prove the theorem by analyzing the generalized replicator dynamics~\eqref{eq:ass/app-dyn-w(t)} for $\bm{w}(t)$, and the reduced assign/appraise dynamics~\eqref{eq:ass/app-reduced} for $\bm{a}(t)$, given any constant, normalized and entry-wise positive vector $\bm{c}$. According to equation~\eqref{eq:ass/app-reduced}, the assignment $\bm{w}=\vleft(A)$ can be considered as a function of the self appraisal vector $\bm{a}$, that is, $\bm{w}(t)=\bm{w}\big( \bm{a}(t) \big)$ for any $t\ge 0$. In this proof, let $\bm{\phi}(\bm{a})=\bm{p}\big( \bm{w}(\bm{a}) \big)-M\bm{p}\big( \bm{w}(\bm{a}) \big)$ and denote by $\dist:\mathbb{R}^n\times \mathbb{R}^n\to \mathbb{R}_{\ge 0}$ the distance induced by the $2$-norm in $\mathbb{R}^n$. For any $\bm{x}\in \mathbb{R}^n$ and subset $S$ of $\mathbb{R}^n$, defined $\dist(\bm{x},S)=\inf_{\bm{y}\in S}\dist(\bm{x},\bm{y})$. 

First of all, for any given $\bm{a}(0)\in (0,1)^n$, we know that the set $\Omega$, as defined in Theorem~\ref{thm:ass/app-finite-time}(iv), is a compact positively invariant set for dynamics~\eqref{eq:ass/app-reduced}, and $\bm{w}(t)$ is well-defined and entry-wise strictly lower (upper resp.) bounded from $\vectorzeros[n]$ ($\vectorones[n]$ resp.), for all $t\in [0,+\infty)$.

Secondly, for any $\bm{a}\in \Omega$, define a scalar function
\begin{equation*}
V(\bm{a})=\log \frac{\max_k x_k/w_k(\bm{a})}{\min_k x_k/w_k(\bm{a})},
\end{equation*}
and the following index sets
\begin{align*}
\overline{\theta}(\bm{a}) & = \Big{\{} i\,\Big|\, \exists t_i>0\text{ s.t. }\frac{x_i}{w_i\big( \bm{a}(t) \big)}=\max_k \frac{x_k}{w_k\big( \bm{a}(t) \big)}\text{ for any }t\in [0,t_i],\text{ with }\bm{a}(0)=\bm{a}\Big{\}},\text{ and, }\\
\underline{\theta}(\bm{a}) &  = \Big{\{} j\,\Big|\, \exists t_j>0\text{ s.t. }\frac{x_j}{w_j\big( \bm{a}(t) \big)}=\min_k \frac{x_k}{w_k\big( \bm{a}(t) \big)}\text{ for any }t\in [0,t_j],\text{ with }\bm{a}(0)=\bm{a}\Big{\}}.
\end{align*}
Then the right time derivative of $V\big( \bm{a}(t) \big)$, along the solution $\bm{a}(t)$, is given by
\begin{equation*}
\frac{d^+V\big( \bm{a}(t) \big)}{dt}=a_j(t)\phi_j\big( \bm{a}(t) \big) - a_i(t)\phi_i\big( \bm{a}(t) \big),
\end{equation*}
for any $i\in \overline{\theta}\big( \bm{a}(t) \big)$ and $j\in \underline{\theta}\big( \bm{a}(t) \big)$. Define
\begin{align*}
E & = \big{\{} \bm{a}\in \Omega \,\big|\, a_j\phi_j(\bm{a})-a_i\phi_i(\bm{a})=0\text{for any }i\in \overline{\theta}(\bm{a}), j\in \underline{\theta}(\bm{a}) \big{\}},\\
E_1 & = \big{\{} \bm{a}\in E \,\big|\, \bm{\phi}(\bm{a})=\vectorzeros[n] \big{\}},\\
E_2 & = \big{\{} \bm{a}\in E \,\big|\, \bm{\phi}(\bm{a})\neq \vectorzeros[n] \big{\}}.
\end{align*}
One can check that $E$ and $E_1$ are compact subsets of $\Omega$, $E=E_1\cup E_2$, and $E_1\cap E_2$ is empty. Denote by $\hat{E}$ the largest invariant subset of $E$. Applying the LaSalle Invariance Principle, see Theorem~3 in~\cite{JPL:68}, we have $\dist\big( \bm{a}(t),\hat{E} \big)\to 0$ as $t\to +\infty$. Note that, $\lim\limits_{t\to +\infty}\dist\big( \bm{a}(t),\hat{E} \big)=0$ does not necessarily leads to $\lim\limits_{t\to +\infty}\bm{w}(t)=\bm{x}$. We need to further refine the result.

For set $E_1$, it is straightforward to see that $E_1\in \hat{E}$ and $\bm{w}(\bm{a})=\bm{x}$ for any $\bm{a}\in E_1$. Now we prove by contradiction that, if $E_2\cap \hat{E}$ is not empty, then, for any $\bm{a}\in E_2\cap \hat{E}$, there exists $i\in \overline{\theta}(\bm{a})$ such that $a_i=0$. Suppose $a_i>0$ for any $i\in \overline{\theta}(\bm{a})$. Since the observation network $G(M)$ is strongly connected, there exists a directed path $i,k_1,\dots,k_q,j$ on $G(M)$, where $i\in \overline{\theta}(\bm{a})$ and $j\in \underline{\theta}(\bm{a})$. We have $k_1\in \overline{\theta}(\bm{a})$, otherwise, starting with $\tilde{\bm{a}}(0)=\bm{a}$, there exists sufficiently small $\Delta t>0$ such that $\phi_i\big( \tilde{\bm{a}}(t) \big)>0$ and $\tilde{a}_i(t)>0$, which contradicts the fact that $\bm{a}$ is in the largest invariant set of $E$. Repeating this argument, we have $j\in \overline{\theta}(\bm{a})$, which contradicts $\bm{\phi}(\bm{a})\neq \vectorzeros[n]$. Similarly, we have that, for any $\bm{a}\in E_2\cap \hat{E}$, there exists $j\in \underline{\theta}(\bm{a})$ with $a_j=0$.

If the fixed vectors $\bm{c}$ and $\bm{x}$ satisfy $\bm{c}=\bm{x}$, then there can not exist $\bm{a}\in E_2\cap \hat{E}$ satisfying all the following three properties: i) there exists $i\in \overline{\theta}(\bm{a})$ such that $a_i=0$; ii) there exists $j\in \underline{\theta}(\bm{a})$ such that $a_j=0$; iii) $\bm{\phi}(\bm{a})\neq \vectorzeros[n]$. In this case, $E_2\cap \hat{E}$ is an empty set, which implies that $\bm{a}(t)\to \hat{E}=E_1$ and thus $\bm{w}(t)\to \bm{x}$ as $t\to +\infty$.

Before discussing the case when $\bm{c}\neq \bm{x}$, we present some properties of the individual performance measure:

\emph{P1:} For any $k,l\in \{1,\dots,n\}$, $\frac{x_k}{c_k}(1-a_k) \le \frac{x_l}{c_l}(1-a_l)$ leads to $p_k(\bm{a}) \le p_l(\bm{a})$, and $\frac{x_k}{c_k}(1-a_k) > \frac{x_l}{c_l}(1-a_l)$ leads to $p_k(\bm{a}) > p_l(\bm{a})$;

\emph{P2:} If there exists $\tau \ge 0$ such that $i\in \overline{\theta}\big(\bm{a}(\tau)\big)$ and $a_i(\tau)=0$, then $i\in \overline{\theta}\big(\bm{a}(t)\big)$ for all $t\ge \tau$; 

\emph{P3:} $\bm{p}(\bm{a}(t))$ is finite and strictly bounded from $0$, satisfying 
\begin{equation*}
f\big( \frac{x_i}{c_i} ( 1-\zeta_i(\bm{a}(0))) \big) \le p_i( \bm{a}(t)) \le f\big( \frac{x_i}{c_i}\sum_k \frac{c_k}{\zeta_k(\bm{a}(0))} \big),
\end{equation*}
with $\zeta_i(\bm{a})$ defined in Theorem~\ref{thm:ass/app-finite-time}(iv).

%\begin{enumerate}
%\item For any $k,l\in \{1,\dots,n\}$, $\frac{x_k}{c_k}(1-a_k) \le \frac{x_l}{c_l}(1-a_l)$ leads to $p_k(\bm{a}) \le p_l(\bm{a})$, and $\frac{x_k}{c_k}(1-a_k) > \frac{x_l}{c_l}(1-a_l)$ leads to $p_k(\bm{a}) > p_l(\bm{a})$;
%\begin{align*}
% \frac{x_k}{c_k}(1-a_k) & \le \frac{x_l}{c_l}(1-a_l) \quad \Rightarrow \quad p_k(\bm{a}) \le p_l(\bm{a}),\\
% \frac{x_k}{c_k}(1-a_k) & > \frac{x_l}{c_l}(1-a_l) \quad \Rightarrow \quad p_k(\bm{a}) > p_l(\bm{a});
%\end{align*}
%\item Since $d p_k(\bm{a})/dt \le dp_i(\bm{a})/dt$ for any $k\in \argmax_l \frac{x_l}{c_l}(1-a_l)$ and $i\in \overline{\theta}(\bm{a})$ with $a_i=0$, if there exists $\tau \ge 0$ such that $i\in \overline{\theta}\big(\bm{a}(\tau)\big)$ and $a_i(\tau)=0$, then $i\in \overline{\theta}\big(\bm{a}(t)\big)$ for all $t\ge \tau$;
%\item According to Theorem~\ref{thm:ass/app-finite-time}(iv), $f\Big( \frac{x_i}{c_i} & \big( 1-\zeta_i(\bm{a}(0)) \big) \Big) \le p_i\big( \bm{a}(t) \big) \le f\Big( \frac{x_i}{c_i}\sum_k \frac{c_k}{\zeta_k(\bm{a}(0))} \Big)$, where $\zeta_i(\bm{a})$ is as defined in Theorem~\ref{thm:ass/app-finite-time}(iv). Therefore, $\bm{p}\big( \bm{a}(t) \big)$ is finite, and strictly bounded from $0$ for all $t\ge 0$.
%\end{enumerate}

For the case when $\bm{c}\neq \bm{x}$, consider the partition $\varphi_1,\dots,\varphi_m$ of the index set $\{1,\dots,n\}$, with $m\le n$, satisfying the following two properties:
\begin{enumerate}
\item $x_k/c_k=x_l/c_l$ for any $k,l$ in the same subset $\varphi_r$;
\item $x_k/c_k > x_l/c_l$ for any $k\in \varphi_r$, $l \in \varphi_s$, with $r<s$.
\end{enumerate}
For any $\bm{a}\in E_2\cap \hat{E}$, since there exists $j\in \underline{\theta}(\bm{a})$ with $a_j=0$, we have $\varphi_m \subset \underline{\theta}(\bm{a})$. For any $i\in \cup_{r=1}^{m-1}\varphi_r$, let
\begin{align*}
E_{2,i} = \Big{\{} \bm{a}\in \Omega \,\Big|\, & a_i=0,\text{ }a_j=0\text{ for any }j\in \varphi_m,\\ & 1-\frac{x_i}{c_i}\frac{c_k}{x_k} \le a_k \le 1-\min_{l\in \{1,\dots,n\}} \frac{x_l}{c_l}\frac{c_k}{x_k}\text{ for any }k\in \varphi_1\cup \dots \cup \varphi_{m-1}\setminus \{i\} \Big{\}}.
\end{align*}
With properties P1 and P2 of $\bm{p}(\bm{a})$, for any $\bm{a}\in E_{2,i}$, we have $i\in \overline{\theta}(\bm{a})$ and $a_i=0$. Moreover,
\begin{enumerate}
\item $E_{2,i}\subset \mathbb{R}^n$ is compact for any $i\in \varphi_1 \cup \dots \cup \varphi_{m-1}$;
\item $\cup_{i\in \varphi_1} E_{2,i},\dots,\cup_{i\in \varphi_{m-1}} E_{2,i}$ are disjoint and compact subsets of $\mathbb{R}^n$;
\item $E_2\cap \hat{E}\subset \bigcup_{i\in \varphi_1\cup \dots \cup \varphi_{m-1}} E_{2,i}$.
\end{enumerate}

For any $\bm{a}\in E_2\cap \hat{E}$, since there exists $i\in \overline{\theta}(\bm{a})$ and $j\in \underline{\theta}(\bm{a})$ such that $a_i=a_j=0$, on the observation network $G(M)$, there must exists a path $i,k_1,\dots,k_q$ satisfying: i) $i\in \overline{\theta}(\bm{a})$ and $a_i=0$; ii) $a_{k_q}=0$ and $x_{k_q}/c_{k_q} < x_i/c_i$; iii) $a_{k_l}>0$ for any $l\in \{1,\dots,q-1\}$. Consider the trajectory $\tilde{\bm{a}}(t)$ with $\tilde{\bm{a}}(0)=\bm{a}$, we have 
\begin{align*}
\dot{\tilde{a}}_{k_{q-1}} \ge \tilde{a}_{k_{q-1}}(1-\tilde{a}_{k_{q-1}})\bigg( f\Big( \frac{x_{k_{q-1}}}{c_{k_{q-1}}}(1-\tilde{a}_{k_{q-1}})\sum_{l=1}^n \frac{c_l}{1-\tilde{a}_l} \Big)f\Big( \big( m_{k_{q-1}k_q}\frac{x_{k_q}}{c_{k_q}} + (1-m_{k_{q-1}k_q})\frac{x_i}{c_i} \big)\sum_{l=1}^n \frac{c_l}{1-\tilde{a}_l} \Big) \bigg). 
\end{align*}
The inequality is due to properties P1-P3 of $p_i(\bm{a})$ for $i\in \overline{\theta}(\bm{a})$ with $a_i=0$, and the concavity of the function $f$. Moreover, since $\tilde{a}_{k_{q-1}}$ is strictly bounded from $1$ and $\sum_l c_l/(1-\tilde{a}_l)$ is strictly lower bounded from $0$, there exists $T_{k_{q-1}}(M,\bm{a}(0),\bm{a})>0$ such that
\begin{align*}
p_{k_{q-1}} \big( \tilde{\bm{a}}(t) \big)< \frac{2-m_{k_{q-1}k_q}}{2}p_i\big( \tilde{\bm{a}}(t) \big) + \frac{m_{k_{q-1}k_q}}{2}p_{k_q}\big( \tilde{\bm{a}}(t) \big).
\end{align*}

Applying the same argument to $k_{q-2},\dots,k_1$, we have that, there exists $T_{k_1}(M,\bm{a}(0),\bm{a})>0$ and \\$\eta_{ik_1\dots k_q}(M)\in (0,1)$ such that, for the solution $\tilde{\bm{a}}(t)$ with $\tilde{\bm{a}}(0)=\bm{a}$, 
\begin{align*}
p_{k_1}\big( \tilde{\bm{a}}(t) \big) < \big( 1-\eta_{ik_1\dots k_q}(M) \big)p_i\big( \tilde{\bm{a}}(t) \big)+ \eta_{ik_1\dots k_q} (M)p_{k_q}\big( \tilde{\bm{a}}(t) \big),
\end{align*}
for all $t\ge T_{k_1}(M,\bm{a}(0),\bm{a})$. This inequality implies that,
\begin{align*}
\phi_i\big( \tilde{\bm{a}}(t) \big) & \ge m_{ik_1}\eta_{ik_1\dots k_q}(M)\Big( p_i\big( \tilde{\bm{a}}(t) \big)-p_{k_q}\big( \tilde{\bm{a}}(t) \big) \Big)\\
& \ge m_{ik_1}\eta_{ik_1\dots k_q}(M) f'\left(\frac{x_i}{c_i}\right)\cdot \sum_{l=1}^n \frac{c_l}{1-\zeta_l\big( \bm{a}(0) \big)}\Big( \frac{x_i}{c_i}-\frac{x_{k_q}}{c_{k_q}} \Big)>0.
\end{align*}
Since the choices of $i$ and the paths $i,k_1,\dots,k_q$ are finite, there exists a constant $\eta>0$ such that, for any $\bm{a}\in E_2\cap \hat{E}$, there exists $T\big( \bm{a}(0),\bm{a} \big)>0$ such that, for any $t\ge T\big( \bm{a}(0),\bm{a} \big)>0$, the solution $\tilde{\bm{a}}(t)$, with $\tilde{\bm{a}}(0)=\bm{a}$, satisfies $i\in \overline{\theta}\big( \tilde{\bm{a}}(t) \big)$ and $\phi_i\big( \tilde{\bm{a}}(t) \big)\ge \eta >0$. 

For any $i\in \varphi_1\cup \dots \cup \varphi_{m-1}$, define
\begin{equation*}
\hat{E}_{2,i}=\big{\{} \bm{a}\in E_{2,i}\,\big|\, p_i(\bm{a})-\sum_{k=1}^n m_{ik}p_k(\bm{a})\ge \eta \big{\}}.
\end{equation*}
We have: i) each $\hat{E}_{2,i}$ is a compact subset of $\mathbb{R}^n$; ii) $\cup_{i\in \varphi_1} \hat{E}_{2,i},\dots, \cup_{i\in \varphi_{m-1}} \hat{E}_{2,i}$ are disjoint and compact subsets of $\mathbb{R}^n$. Let $\hat{E}_2=\cup_{r=1}^{m-1}\big( \cup_{r\in \varphi_r} \hat{E}_{2,i} \big)$. For dynamics~\eqref{eq:ass/app-reduced}, due to the continuous dependency on the initial condition, for any $\bm{a}\in (E_2\cap \hat{E})\setminus (\hat{E}_2\cap \hat{E})$, there exists $\delta>0$ such that, for any $\tilde{\bm{a}}(0)\in \mathcal{U}(\bm{a},\delta)\cap(E_2\cap \hat{E})$, where $\mathcal{U}(\bm{a},\delta)=\big{\{} \bm{b}\in \Omega\,\big|\, \dist(\bm{b},\bm{a})\le \delta \big{\}}$, $\tilde{\bm{a}}(t)\in \hat{E}_2\cap \hat{E}$ for sufficiently large $t$. Therefore, $\bm{a}$ can not be an $\omega$-limit point of $\bm{a}(0)$. We thus obtain that, the $\omega$-limit set of $\bm{a}(0)$ is in the set $ E_1\cup (\hat{E}_2\cap \hat{E})$. Moreover, since $E_1,\cup_{i\in \varphi_1}\hat{E}_{2,i},\dots,\cup_{i\in \varphi_{m-1}}\hat{E}_{2,i}$ are disjoints compact subsets of $\mathbb{R}^n$, and the $\omega$-limit set of $\bm{a}(0)$ is connected and compact, $\bm{a}(t)$ can only converge to one of the sets $E_1,\cup_{i\in \varphi_1}\hat{E}_{2,i},\dots,\cup_{i\in \varphi_{m-1}}\hat{E}_{2,i}$. 

Now we prove $\lim_{t\to +\infty} \dist(\bm{a}(t),E_1)=0$ by contradiction. Suppose $\omega\big( \bm{a}(0) \big)\in \cup_{i\in \varphi_r}\hat{E}_{2,i}$ for some $r\in \{1,\dots,m-1\}$. Since each $\hat{E}_{2,i}$ is a compact set, there exists $\epsilon>0$ and $\eta({\epsilon})>0$ such that $\phi_i(\bm{a})\ge \eta(\epsilon)>0$ for any $\bm{a}\in \mathcal{U}(\hat{E}_{2,i},\epsilon)$. For this given $\epsilon>0$, since $\omega\big( \bm{a}(0) \big)\in \cup_{i\in \varphi_r}\hat{E}_{2,i}$ leads to $\dist\big( \bm{a}(t),\cup_{i\in \varphi_r} \hat{E}_{2,i} \big)\to 0$ as $t\to +\infty$, we conclude that, there exists $T>0$ such that, for any $t\ge T$, $\bm{a}(t)\in \cup_{i\in \varphi_r}\mathcal{U}(\hat{E}_{2,i},\epsilon)$. Define $V_r(\bm{a})=\min_{i\in \varphi_r} a_i$, for any $\bm{a}\in \cup_{i\in \varphi_r}\mathcal{U}(\hat{E}_{2,i},\epsilon)$. The function $V_r(\bm{a})$ satisfies that, $V_r(\bm{a})\ge 0$ for any $\bm{a}\in \cup_{i\in \varphi_r}\mathcal{U}(\hat{E}_{2,i},\epsilon)$ and $V_r(\bm{a})=0$ if and only if $\bm{a}\in \cup_{i\in \varphi_r}\hat{E}_{2,i}$. Therefore, $\dist\big( \bm{a}(t),\cup_{i\in \varphi_r}\hat{E}_{2,i} \big)\to 0$ leads to $V_r\big( \bm{a}(t) \big)\to 0$ as $t\to +\infty$. Moreover, since $\bm{a}\in \mathcal{U}(\hat{E}_{2,i},\epsilon)$ for any $i\in\argmin_{k\in \varphi_r}a_k$, we have
\begin{align*}
\frac{d^+V_r\big( \bm{a}(t) \big)}{dt} = \min_{i\in \operatornamewithlimits{argmin}\limits_{k\in \varphi_r} a_k(t)}\dot{a}_i(t) \ge \delta a_i(t)\big( 1-a_i(t) \big).
\end{align*}
According to Theorem~\ref{thm:ass/app-finite-time}(i), for any given $\bm{a}(0)\in (0,1)^n$,
$\bm{a}(t)\in (0,1)^n$ for all $t\ge 0$. Therefore, \\$d^+V_r( \bm{a}(t) )/dt >0$ for all $t\ge T$, which contradicts $\lim_{t\to +\infty}V_r\big( \bm{a}(t) \big)=0$. Therefore, we have \\$\lim_{t\to +\infty} \dist(\bm{a}(t),E_1)=0$ and $\lim_{t\to +\infty} \bm{w}(t)=\bm{x}$.

Since $\dot{A}(t)\to \vectorzeros[n\times n]$ as $\bm{\phi}\big( \bm{a}(t) \big) \to \vectorzeros[n]$, there exists an entry-wise non-negative and irreducible matrix $A^*$, depending on $A(0)$ and satisfying $\vleft(A^*)=\bm{x}$, such that $A(t)\to A^*$ as $t\to +\infty$. This concludes the proof.

\subsection{Proof for Lemma~\ref{lem:ass/app/inf-initial->0}}\label{proof:ass/app/inf-initial->0}
Since $A(0)$ is primitive and row-stochastic, following the same argument in the proof for Theorem~\ref{thm:ass/app-finite-time}(i), we have that, there exists $\Delta \tilde{t}>0$ such that, for any $t\in [0,\Delta \tilde{t}]$: i) $\bm{w}(t)$ is well-defined and $\bm{w}(t)\succ \vectorzeros[n]$; ii) $A(t)$ is bounded, continuously differentiable to $t$, and satisfies $A(t)\vectorones[n]=\vectorones[n]$; iii) $\bm{p}\big( \bm{w}(t) \big)-M\bm{p}\big(  \bm{w}(t)\big)$ is bounded.  Therefore, for any $t\ge 0$, there exists $\mu$, depending on $t$ and $A(0)$, such that $\dot{A}(t)\succeq \frac{1}{\subscr{\tau}{ave}} A^2(t)-(\frac{1}{\subscr{\tau}{ave}}+\mu)A(t)$. 

Consider the equation $\dot{B}(t)=\frac{1}{\subscr{\tau}{ave}} B^2(t)-(\frac{1}{\subscr{\tau}{ave}} + \mu)B(t)$, with $B(0)=A(0)$. According to the comparison theorem, $A(t)\succeq B(t)$ for any $t\ge 0$. Let $\bm{b}_i(t)$ be the $i$-th column of $B(t)$ and let $\bm{y}_k(t)=e^{(\frac{1}{\subscr{\tau}{ave}} + \mu)t}\bm{b}_k(t)$. We obtain $\dot{\bm{y}}_k(t) = \frac{1}{\subscr{\tau}{ave}} B(t)\bm{y}_k(t)$.

Denote by $\Phi(t,0)$ the state transition function for the equation $\dot{\bm{y}}_k(t) = \frac{1}{\subscr{\tau}{ave}} B(t)\bm{y}_k(t)$, which is written as $\Phi(t,0)=I_n + \sum_{k=1}^{\infty} \Phi_k(t)$, where $\Phi_1(t)=\int_0^t B(\tau_1)d\tau_1$ and $\Phi_l(t)=\int_0^t B(\tau_1)\int_0^{\tau_1}\dots B(\tau_{l-1})\int_{0}^{\tau_{l-1}}B(\tau_l)d\tau_l$ for $l\ge 2$.  By computing the MacLaurin expansion for each $\Phi_k(t)$ and summing them together, we obtain that
\begin{equation*}
\begin{split}
\Phi(t,0) =I_n + h_1(t)B(0) + h_2(t)B^2(0)+\dots+h_{n-1}(t)B^{n-1}(0)+O(t^n),
\end{split}
\end{equation*}
where $h_k(t)$ is a polynomial with the form $h_k(t)=\eta_{k,k}t^k + \eta_{k,k+1}t^{k+1}+\dots$, and, moreover, $\eta_{k,k}>0$ for any $k\in \mathbb{N}$. Therefore, for $t$ sufficiently small, we have $h_k(t)>0$ for any $k\in \{1,\dots,n-1\}$. Moreover, since $B^k(0)\succeq \vectorzeros[n\times n]$ for any $k\in \mathbb{N}$ and $B(0)+\dots+B^{n-1}(0)\succ \vectorzeros[n\times n]$, there exists $\Delta t\le \Delta \tilde{t}$ such that $\Phi(t,0)\succ \vectorzeros[n\times n]$ for any $t\in [0,\Delta t]$.

\subsection{Discussion on Conjecture~\ref{conj:ass/app/inf-lower-bound-A(t)}}\label{discussion:conjecture-pos}

The Monte Carlo method~\cite{RT-GC-FD:05} is adopted to estimate the probability that Conjecture~\ref{conj:ass/app/inf-lower-bound-A(t)} holds. For any randomly generated $A(0)\in \intDeltan$, define the random variable $Z:\intDeltan \to \{0,1\}$ as
\begin{enumerate}
\item $Z\big(A(0)\big)=1$ if there exists $a_{\min}>0$ such that $A(t)\succeq a_{min} \vectorones[n]\vectorones[n]^{\top}$ for all $t\in [0,1000]$;
\item $Z\big( A(0) \big)=0$ otherwise. 
\end{enumerate}
Let $p=\mathbb{P}\big[Z\big( A(0) \big)=1\big]$. For $N$ independent random samples $Z_1,\dots,Z_N$, in each of which $A(0)$ is randomly generated in $\intDeltan$, define $\hat{p}_N=\sum_{i=1}^N Z_i/N$. For any accuracy $\epsilon\in (0,1)$ and confidence level $1-\xi \in (0,1)$, $\lvert \hat{p}_N -p \rvert <\epsilon$ with probability greater than $1-\xi$ if 
\begin{equation}\label{eq:Chernoff-bound}
N\ge \frac{1}{2\epsilon^2} \log \frac{2}{\xi}.
\end{equation}
For $\epsilon=\xi=0.01$, the Chernoff bound~\eqref{eq:Chernoff-bound} is satisfied by $N=27000$. We run $27000$ independent MATLAB simulations of the assign.appraise/influence dynamics with $n=7$ and find that $\hat{p}_N=1$. Therefore, for any $A(0)\in \intDeltan$, with $99\%$ confidence level, there is at least $0.99$ probability that $A(t)$ is entry-wise strictly lower bounded from $\vectorzeros[n\times n]$ for all $t\in [0,10000]$.

Moreover, we present in the following lemma a sufficient condition for Conjecture~\ref{conj:ass/app/inf-lower-bound-A(t)} on the initial appraisal matrix $A(0)$ and the parameters $\subscr{\tau}{ave}$, $\subscr{\tau}{app}$.
\begin{lemma}[Strictly positive lower bound of appraisals]\label{lem:ass/app/inf-strictly-pos-entry-sufficient}
Consider the assign/appraise/influence dynamics~\eqref{eq:ass/app/inf-dyn-matrix-form}, based on Assumptions~\ref{asmp:assign-rule}-\ref{asmp:inf-dyn}, with the assignment $\bm{w}(t)$ and performance $\bm{p}(\bm{w})$ as in Assumptions~\ref{asmp-team-task-assign} and~\ref{asmp:indiv-perf} respectively. For any initial appraisal matrix $A(0)$ that is entry-wise positive and row-stochastic, as long as
\begin{equation*}
\frac{\subscr{\tau}{app}}{\subscr{\tau}{ave}}\ge \frac{1-\xi_0}{\xi_0}\left( f\left( \frac{x_{\max}}{\xi_0} \right) - f\left( \frac{x_{\min}}{1-(n-1)\xi_0} \right) \right),
\end{equation*} 
where the constant $\xi_0$ is defined as in Theorem~\ref{thm:ass/app/inf-asym-behav}~(ii), then there exists $a_{\min}>0$ such that $A(t)\succeq a_{\min} \vectorones[n]\vectorones[n]^{\top}$. 
\end{lemma}
\smallskip
\begin{proof}
First of all, by definition we have $w_s(t)=\sum_k w_k(t) a_{ks}(t)$. The right-hand side of this equation is a convex combination of $\{a_{1s}(t),\dots,a_{ns}(t)\}$. Therefore, $\max_k a_{ks}(t)\ge w_s(t)\ge \xi_0$ for all $t\in [0,+\infty)$.

At any time $t\ge 0$, for any pair $(i,j)$ such that $a_{ij}(t)=\min_{k,l}a_{kl}(t)$, the dynamics for $a_{ij}(t)$ is
\begin{align*}
\dot{a}_{ij}(t) =\frac{1}{\subscr{\tau}{ave}} \left( \sum_k a_{ik}(t)a_{kj}(t)-a_{ij}(t) \right)-\frac{1}{\subscr{\tau}{app}} a_{ii}(t)a_{ij}(t)\Big( p_i\big(\bm{w}(t)\big) - \sum_{k=1}^n m_{ik}p_k\big( \bm{w}(t) \big) \Big).
\end{align*} 
For simplicity, in this proof, denote $\phi_i=p_i\big(\bm{w}(t)\big) - \sum_{k=1}^n m_{ik}p_k\big( \bm{w}(t) \big)$. Suppose $a_{mj}(t)=\max_k a_{kj}(t)$. We have
\begin{align*}
\dot{a}_{ij}(t) \ge \frac{1}{\subscr{\tau}{ave}} a_{ij}(t)a_{mj}(t) - \frac{1}{\subscr{\tau}{ave}} a_{ij}^2(t)- \frac{1}{\subscr{\tau}{app}} a_{ii}(t)a_{ij}(t)\phi_i .
\end{align*}
Therefore,
\begin{align*}
\frac{\dot{a}_{ij}}{a_{ij}}  \ge \frac{1}{\subscr{\tau}{ave}} \xi_0  - \frac{1}{\subscr{\tau}{app}} (1-\xi_0) \Big( f\big( \frac{x_{\max}}{\xi_0} \big)- f\big( \frac{x_{\min}}{1-(n-1)\xi_0} \big) \Big).
\end{align*}
%& \ge \frac{1}{\subscr{\tau}{ave}} \xi_0 - \frac{1}{\subscr{\tau}{app}} \left( f\left(\frac{x_i}{w_i(t)}\right) - \sum_k w_k(t)f\left( \frac{x_k}{w_k(t)} \right) \right)\\
The condition on $\frac{1}{\subscr{\tau}{ave}}/\frac{1}{\subscr{\tau}{app}}$ in Lemma~\ref{lem:ass/app/inf-strictly-pos-entry-sufficient} guarantees that $\dot{a}_{ij}(t)\big/a_{ij}(t)$ is positive if $a_{ij}(t)=\min_{k,l}a_{kl}(t)$. This concludes the proof.
\end{proof}

\subsection{Proof for Theorem~\ref{thm:ass/app/inf-asym-behav}}\label{proof:ass/app/inf-asym-behav}
Statement~(i) is proved following the same argument in the proof for Theorem~\ref{thm:ass/app-finite-time}~(i). For any given $A(0)$ that is row-stochastic and entry-wise positive, the closed and bounded invariant set $\Omega$ for $A(t)$ is given by $\Omega=\left\{ A\in \mathbb{R}^{n\times n}\,\big|\,A\succ a_{\min}\vectorones[n]\vectorones[n]^{\top}, A\vectorones[n]=\vectorones[n] \right\}$, where $a_{\min}>0$ is given by Conjecture~\ref{conj:ass/app/inf-lower-bound-A(t)}.

Since $\bm{w}^{\top}(t)\big( A^2(t)-A(t) \big)=\vectorzeros[n]^{\top}$ for all $t\ge 0$, we conclude that, $\bm{w}(t)$ in the assign/appraise/influence dynamics also obeys the generalized replicator dynamics~\eqref{eq:ass/app-dyn-w(t)}. Consider $\bm{w}(t)$ as a function of $A(t)$. Define $\bm{\phi}( A)=\bm{p}\big( \bm{w}(A) \big) - M\bm{p}\big( \bm{w}(A) \big)$ and 
\begin{equation*}
V(A) = \log \frac{\max_k x_k/w_k(A)}{\min_k x_k/w_k(A)}.
\end{equation*}
For any $t\in [0,+\infty)$, there exists $i\in \argmax_k x_k/w_k\big(A(t)\big)$ and $j\in \argmin_k x_k/w_k\big( A(t) \big)$ such that $V\big( A(t) \big)=\log \Big( x_i w_j\big( A(t) \big)\big/ x_j w_i\big( A(t) \big) \Big)$, and $\frac{d^+ V( A )}{dt} = a_{jj}\phi_j(A) - a_{ii}\phi_i(A) \le 0$.
%\begin{equation*}
%\frac{d^+ V( A )}{dt} = a_{jj}\phi_j(A) - a_{ii}\phi_i(A) \le 0,
%\end{equation*}
 Therefore, $V\big( A(t) \big)$ is non-increasing with $t$, which in turn implies
\begin{equation*}
\frac{x_i}{x_j} \frac{w_j(t)}{w_i(t)} \le \frac{\max_k x_k/w_k(0)}{\min_k x_k/w_k(0)}=\gamma_0,
\end{equation*}  
for any $i,j\in \{1,\dots,n\}$. This inequality, combined with the fact that $\sum_k w_k(t)=1$ for any $t\ge 0$, leads to the inequalities in statement~(ii).

Similar to the proof for Theorem~\ref{thm:ass-app-asym-behav}, define 
\begin{align*}
\overline{\theta}(A) & = \Big{\{} i\,\Big|\, \exists\, t_i>0\text{ s.t. } \frac{x_i}{w_i\big(A(t)\big)}=\max_k \frac{x_k}{w_k\big( A(t) \big)}\text{ for any }t\in [0,t_i]\text{ with }A(0)=A \Big{\}},\\
\underline{\theta}(A) & = \Big{\{} j\,\Big|\, \exists\, t_j>0\text{ s.t. } \frac{x_j}{w_j\big(A(t)\big)}=\min_k \frac{x_k}{w_k\big( A(t) \big)} \text{ for any }t\in [0,t_j]\text{ with }A(0)=A \Big{\}},
\end{align*}
and let $E=\big{\{} A\in \Omega \,\big|\, d^+V(A)/dt=0  \big{\}}$. For any $A\in E$, since $A\succeq a_{\min}\vectorones[n]\vectorones[n]^{\top}$, we have $\phi_i(A)=\phi_j(A)=0$ for any $i\in \overline{\theta}(A)$ and $j\in \underline{\theta}(A)$. Suppose individual $s$ is a globally reachable node in the observation network. There exists a directed path $i,k_1,\dots,k_q,s$. Without loss of generality, suppose $q\ge 1$. For any $A$ in the largest invariant subset of $E$, we have $k_1\in \overline{\theta}(A)$ and therefore $\phi_{k_1}(A)=0$. This iteration of argument leads to $s\in \overline{\theta}(A)$. Following the same line of argument, we have $s\in \underline{\theta}(A)$. Therefore, for any given $A(0)\succ \vectorzeros[n\times n]$ that is row-stochastic, the solution $A(t)$ converges to $\hat{E}=\{ A\in \Omega \,|\, \bm{\phi}(A)=\vectorzeros[n] \}=\{ A\in \Omega \,|\, \vleft(A)=\bm{x} \}$. 

Let $\tilde{A}=\max_j \big( \max_k a_{kj} - \min_{k} a_{kj} \big)$. One can check that $d^+\tilde{V}(A)/dt$ along the dynamics~\eqref{eq:ass/app/inf-dyn-matrix-form} is a continuous function of $A$ for any $A\in \Omega$. Define $\hat{E}_{\epsilon/2}=\big{\{} A\in \hat{E} \,\big|\, \lVert A-\vectorones[n]\bm{x}^{\top} \rVert_2 \ge \epsilon/2 \big{\}}$. Since $\hat{E}$ is compact, $\hat{E}_{\epsilon/2}$ is also a compact set. For any $A\in \hat{E}_{\epsilon/2}$, since $d^+ \tilde{V}(A)/dt$ is strictly negative and depends continuously on $A$, there exists a neighborhood $\mathcal{U}(A,r_A)=\{\tilde{A}\in \Omega \,|\, \lVert \tilde{A}-A \rVert_2\le r_A\}$ such that $d^+ \tilde{V}(\tilde{A})/dt<0$ for any $\tilde{A}\in \mathcal{U}(A,r_A)$. Due to the compactness of $\hat{E}_{\epsilon/2}$ and according to the Heine-Borel finite cover theorem, there exists $K\in \mathbb{N}$ and $\{A_k,r_k\}_{k\in \{1,\dots,K\}}$, where $A_k\in \hat{E}_{\epsilon/2}$ and $r_k>0$ for any $k\in \{1,\dots,K\}$, such that $\hat{E}_{\epsilon/2} \subset \cup_{k=1}^K \mathcal{U}(A_k,r_k)$.

Define the distance $\dist: \mathbb{R}^n \times \mathbb{R}^n \to \mathbb{R}_{\ge 0}$ as in the proof for Theorem~\ref{thm:ass-app-asym-behav}. Let $\delta =\min \{r_1,\dots,r_k,\epsilon/2\}$ and
\begin{align*}
B_1 & = \big{\{} A\in \Omega \,\big|\, \dist(A,\hat{E})\le \delta, \dist(A,\hat{E}_{\epsilon/2})> \delta \big{\}},\\
B_2 & = \big{\{} A\in \Omega \,\big|\, \dist(A,\hat{E})\le \delta, \dist(A,\hat{E}_{\epsilon/2})\le \delta \big{\}}.
\end{align*}
We have $B_1\cap B_2$ is empty. For any $A\in B_1$, since $\dist(A,\hat{E})\le \delta$, $\dist(A,\hat{E}_{\epsilon/2})>\delta$, there exists $\tilde{A}\in \hat{E}_{\epsilon/2}$ such that $\dist(A,\tilde{A}) \le \delta$. Since $\dist( \tilde{A},\vectorones[n]\bm{x}^{\top})<\epsilon/2$, we have $\dist(A,\vectorones[n]\bm{x}^{\top})\le \dist(A,\tilde{A})+\dist(\tilde{A},\vectorones[n]\bm{x}^{\top})<\epsilon$. Therefore, $B_1\subset \mathcal{U}(\vectorones[n]\bm{x}^{\top},\epsilon)$. Moreover, since $B_2$ is compact, $\tilde{V}(A)$ is lower bounded and $d^+ \tilde{V}(A)/dt$ is strictly upper bounded from $0$ in $B_2$. Since $\lim_{t\to +\infty}\dist(A(t),\hat{E})=0$, there exists $t_0>0$ such that $A(t)\in B_1\cup B_2$ for any $t\ge 0$. Therefore, for any $t\ge t_0$, there exists $t_1\ge t$ such that $A(t_1)\in B_1$. This argument is valid for any $\epsilon>0$, which implies that $\vectorones[n]\bm{x}^{\top}$ is an $\omega$-limit point for any given $A(0)$.

Since $\hat{E}$ is compact, $\dist(A,\tilde{E})$ is strictly positive. Since $\lim_{t\to +\infty} \dist \big( A(t),\hat{E} \big)=0$, any $A\in \Omega\setminus \hat{E}$ can not be an $\omega$-limit point of $A(0)$. For any $A\in \hat{E}\setminus \{\vectorones[n]\bm{x}^{\top}\}$, since the solution passing through $A$ asymptotically converges to $\vectorones[n]\bm{x}^{\top}$, $A\in \hat{E}\setminus \{\vectorones[n]\bm{x}^{\top}\}$ can not be an $\omega$-limit point of $A(0)$ either. Therefore, the $\omega$-limit set of $A(0)$ is $\{ \vectorones[n]\bm{x}^{\top} \}$. This concludes the proof.

\end{appendix}
\bibliographystyle{plain}
\bibliography{alias,FB,Main}

\end{document}